\documentclass[aps,pre,twocolumn,superscriptaddress,floatfix,notitlepage,10pt]{revtex4-2}
\usepackage{etoolbox}
\usepackage[utf8]{inputenc}
\usepackage{amsmath,amsfonts,amssymb,amsthm,bm,color,dsfont,graphicx,psfrag,mathtools,cancel}

\usepackage{pgfplots}
\pgfplotsset{compat=1.17}
\usepackage{xcolor}
\usepackage{soul}

\usepackage[export]{adjustbox}
\usepackage{natbib}
\usepackage{graphicx}
\usepackage[hybrid]{markdown}
\usepackage{mathrsfs}

\usepackage{blkarray}
\usepackage{multirow}
\usepackage[caption = false]{subfig}
\usepackage{hyperref}

\usepackage{stmaryrd}
\usepackage{tikz-cd}
\usepackage{comment}
\usepackage{listings}

\usepackage[normalem]{ulem}

\newcommand{\arcin}[6]{#1_{\mathsf{#2}_{#3}\mathsf{#4}_{#5}}^{\mathsf{i}_{#6}}}
\newcommand{\arc}[5]{#1_{\mathsf{#2}_{#3}\mathsf{#4}_{#5}}}

\newcommand{\numberneighborsMB}[2]{\widetilde{N}_{\mathsf{#1#2}}}
\newcommand{\numberneighbors}[2]{\widetilde{N}_{\mathsf{#1#2}_{0}}}

\newcommand{\arcprime}[2]{#1_{\mathsf{#2}_{0}}}

\newcommand{\transitionprob}[3]{%
    \if\relax\detokenize{#3}\relax
        {#1^{\mathsf{#2}}}
    \else
        {#1^{\mathsf{#2}}(#3)}
    \fi
}

\newcommand{\rhoOne}[1]{\rho^{\mathsf{#1}}_1}

\begin{document}

\title{A Minimal Network of Brain Dynamics:\\Hierarchy of Approximations to Quasi-critical Neural Network Dynamics}
\author{Jeremy B. Goetz}
\thanks{These authors contributed equally.}
\author{Naruepon Weerawongphrom }
\thanks{These authors contributed equally.}
\affiliation{\mbox{Department of Physics, Indiana University, Bloomington, Indiana 47405, USA}}
\author{\\Rashid V. Williams-García}
\affiliation{\mbox{Institut Denis Poisson, Parc de Grandmont, Tours 37200 FRANCE}}
\author{John M. Beggs}
\author{Gerardo Ortiz}
\affiliation{\mbox{Department of Physics, Indiana University, Bloomington, Indiana 47405, USA}}

\date{\today}

\begin{abstract}
We present an interacting model of neural network dynamics that incorporates key biological features, including multiple forms of inhibitory interactions. We develop a hierarchy of analytical mean-field approximations to characterize nonequilibrium phase transitions between ordered, disordered, and chaotic regimes, complemented by a detailed stability analysis. We show that inhibition generically enhances the stability of network dynamics. The model is consistent with the quasi-criticality hypothesis, exhibiting regions of maximal dynamical susceptibility and mutual information, modulated by the strength of external stimuli. We further demonstrate that, at the mean-field level, the critical transition belongs to the mean-field directed percolation universality class, in agreement with prior experimental and theoretical studies. More broadly, our framework may offer insights into neurological disorders, with the unstable regime exhibiting chaotic dynamics that may be associated with epileptic seizures.

\end{abstract}
\maketitle

\section{Introduction}

The human brain is a highly intricate open system with nonequilibrium stochastic dynamics, numerous neurotransmitters, and hundreds of different types of neurons. In total, it is estimated that there are $20\times10^{9}$ neurons with $150\times10^{12}$ connections in the neocortex \cite{numberofneurons,PAKKENBERG200395}. The cortex is the most evolutionarily recent component of the mammalian brain and is widely regarded as central to higher cognition, as well as to sensory processing and motor function \cite{Cortex_Function}. Given its importance, researchers have developed numerous biological neural network models to study the cortex. These models typically rely on either continuous partial differential equations or discrete cellular automata. Broadly, they fall into two categories: highly detailed models that resist analytical treatment, and simplified minimal approaches. Mean-field (MF) approximations of these neural network models are analytically tractable but oversimplify the dynamics, failing to capture essential spatiotemporal fluctuations. In general, they cannot reproduce phenomena such as avalanches—bursts of activity separated by quiescent periods—that are commonly observed in experiments. 

Stochastic branching-like processes provide the basis for a class of simplified minimal models of neuronal activity. A representative model from this class is the cortical branching model (CBM) \cite{williams2014quasicritical}, which exhibits avalanche dynamics and an analytically tractable dynamical MF approximation. Branching-like network models \cite{Beggs11167,PhysRevLett.94.058101,Kinouchi2006,williams2014quasicritical,zierenberg_description_2020,zeraati_topology-dependent_2024,PhysRevLett.94.058101, 10.1371/journal.pcbi.1000271,Kello2013} have successfully reproduced several critical properties and avalanche phenomena. However, these models typically lack an explicit representation of inhibitory neurons, which constitute approximately 20\% of all neurons in the cortex \cite{Meinecke1987}. Inhibitory neurons are believed to play a crucial role in stabilizing network dynamics while enabling complex information processing \cite{Importance_of_Inhbitory_Neurons}. 

The seminal work of Wilson and Cowan \cite{WilsonCowan} underpins many modern models that incorporate and explore the role of inhibitory neurons \cite{Shew2011,Munoz2022}. However, these models are intrinsically rate-based, or  MF, in nature, which limits their ability to capture key features of neuronal dynamics. In particular, they cannot reproduce experimentally observed avalanches and typically neglect neuronal refractory periods.

This paper aims to develop a minimal network model that addresses these common shortcomings while remaining consistent with experimental observations. In particular, we focus on capturing two key features: (1) that living neuronal networks exhibit many signatures of operation near criticality, and (2) that they are not truly critical, but instead reside in a quasi-critical regime. We now elaborate on these two observations.

Regarding operation near criticality, we begin with well-established experimental evidence of neuronal avalanches. These emergent degrees of freedom exhibit apparent scale-free distributions of avalanche sizes and durations, scaling relations among critical exponents, and universal scaling functions (i.e., effective shape collapse) \cite{Beggs11167,RevModPhys.90.031001,Fosque2021,PhysRevE.108.034110,PhysRevLett.122.208101,PhysRevLett.108.208102}. Scale-free behavior is typically associated with optimal information processing properties \cite{Beggs11167}, including divergent dynamical susceptibility \cite{williams2014quasicritical}, maximal mutual information \cite{williams2014quasicritical}, enhanced computational power \cite{bertschinger_real-time_2004} and capacity \cite{PhysRevLett.94.058101,Shew2011,safavi_signatures_2024}, optimal dynamical range \cite{Kinouchi2006}, enhanced learning capabilities \cite{de_arcangelis_learning_2010,zeraati_self-organization_2021}, and is often taken as a hallmark of universal critical behavior \cite{Shew2011,Fagerholm2016}. Exponent scaling relations have been studied across a wide range of experimental systems, including anesthetized rats, monkeys, freely moving mice, and ex vivo turtles \cite{Fontenele2019}. However, these same experiments also reveal variability in the measured critical exponents for a given system over time. This suggests that, while neuronal activity operates close to criticality, it does not exhibit strict universality \cite{Beggs11167,PhysRevLett.108.208102,10.1371/journal.pcbi.1000314,PASQUALE20081354,tmtctPlenz}. Indeed, experimental data are inconsistent with true criticality \cite{williams2014quasicritical,Fosque2021,fosque_quasicriticality_2022,
zeraati_self-organization_2021}.

This apparent lack of universality raises an important question: is there an emergent organizing principle that can account for these seemingly inconsistent experimental observations? The brain is continuously driven by inputs from other regions and the external environment, which tend to push any self-organizing dynamics away from a true critical point. The quasi-criticality hypothesis addresses this by explicitly incorporating the effects of external stimuli \cite{williams2014quasicritical}. According to this framework, the cortex operates in a region of parameter space close to a surface of maximal dynamical susceptibility and mutual information—the nonequilibrium Widom hypersurface—whose location depends on the strength of the external drive \cite{williams2014quasicritical}. Numerical simulations, together with experimental observations, support this picture and reproduce effective scaling-law behavior \cite{Fosque2021}. This, in turn, provides a way to quantify how far the system deviates from ideal critical scaling \cite{williams2014quasicritical,Fosque2021}.

To build on the experimental and theoretical evidence for quasi-criticality, we introduce the generalized cortical branching model (GCBM), an extension of the CBM that incorporates inhibition. The GCBM is a functional network closely related to a nonequilibrium many-body stochastic cellular automaton. It explicitly tracks the state of each neuron—quiescent, active, or refractory—as well as the propagation of activity between neurons, without assuming a regular lattice structure or a predefined spatial metric. Despite its simplicity, the GCBM captures the essential features of neural dynamics, thereby enhancing the predictive power of the quasi-criticality hypothesis. The inclusion of inhibitory neurons provides additional control parameters that allow the system to be tuned toward or away from the quasi-critical regime. This flexibility enables the exploration of a broader range of dynamical behaviors and offers insight into how and why the cortex regulates its proximity to quasi-criticality. 

Due to its exponential scaling with the number of neurons, the many-body GCBM is computationally challenging, limiting its resolution and precluding exact analytical treatment. To overcome this limitation, we develop a hierarchy of MF approximations that remain analytically tractable. This approach reduces the complexity of the system to a set of coupled, autonomous, nonlinear discrete dynamical maps, whose dimensionality is determined by the integer-valued refractory periods.

More generally, this hierarchical MF framework applies to arbitrary functional networks. The different levels of approximation can be represented graphically in terms of network motifs—basic connectivity patterns, such as interactions between excitatory and inhibitory neurons. By replication and translation, these motifs generate larger, more complex networks, much like a simple tile can tessellate a plane. In this sense, motifs serve as the fundamental building blocks of the network.

Given a set of such motifs, MF equations can be systematically constructed using a ``dictionary'' supplemented by substitution rules that encode the underlying dynamics. While less systematic than the framework developed here, early steps toward incorporating inhibitory neurons into MF descriptions were presented in Mark Moore’s thesis \cite{moore2018inhibition}.

These MF equations allow for the examination of the roles of inhibitory neurons in modulating the dynamics and assessing the model’s consistency with the quasi-criticality hypothesis. As we will see, the inclusion of inhibitory neurons increases the system's stability by shifting the phase boundaries to more stable regions, also reducing the peak susceptibility as the strength of inhibition increases, which are experimentally testable prediction. In the absence of external stimuli, i.e., where spontaneous activity is absent, we observe a second-order phase transition between disordered and ordered regimes, marked by a divergence in susceptibility at the critical point —behavior consistent with purely excitatory models. Notably, the model falls within the directed percolation universality class, in agreement with the excitatory-only CBM and other models \cite{williams2014quasicritical, Kinouchi2006, girardi-schappo_synaptic_2020, Munoz2022}. We also characterize the previously labeled quasiperiodic phase as marginally stable, leading to a period-doubling route to chaos. The transitions through marginally stable, periodic orbits, and the period-doubling route to chaos can be related to epileptic seizures as explored in chaos theory and neural dynamics studies \cite{Low_Dimensional_Chaos, Chaos_Theory_and_Epilepsy, Sackellares2000EPILEPSYW,PANAHI2019395}.

The rest of this paper is organized as follows: Starting with Section \ref{section2}, we show the equations of the many-body GCBM to derive the MF approximations. Section \ref{section3} provides the steps for developing the hierarchy of MF approximations, which enables a concise method for writing solutions for every motif. Section \ref{section4} provides examples of how to solve MF equations for a simple and a more complex network. In Section \ref{section5}, we will analyze how adding inhibitory neurons affects the dynamics, analyze phase diagrams, compute the dynamical susceptibility with variations of inhibitory strengths, discuss the universality class of the model for no external driver, the physical and non-physical chaotic regimes of the model, discuss paths to chaos, and explore forcing inhibitory neurons to be periodic and its effect on excitatory neurons. We conclude in Section \ref{section6} with a summary of the main results, where we highlight the central findings and their implications. A notation guide can be found in Appendix \ref{App:Notation Guide}, where terms are organized by order of appearance. Several appendices provide additional details and technical derivations that supplement the calculations presented in the main text.

\section{Generalized Cortical Branching Model}
\label{section2}

\subsection{Branching dynamics}

The GCBM simulates biological spiking activity using a branching process model. Briefly, this model consists of nodes (neurons), connected to other nodes in a network that can fire in response to inputs. Unlike leaky integrate and fire models that are common in computational neuroscience, this model does not have nonlinear thresholds for its neurons. Rather, the neurons will fire if they receive net positive inputs and they are not in the refractory state. 

We will now describe the model in more detail. The model consists of excitatory and inhibitory neurons, located at the node or vertex of the network, labeled $\mathsf{e}$ and $\mathsf{i}$, respectively, while $\mathsf{n}\in\{\mathsf{e},\mathsf{i}\}$ denotes a generic neuron. The total number of neurons is given by $N=\arc{N}{e}{}{}{}+\arc{N}{i}{}{}{}$, where $\arc{N}{e}{}{}{}$ and $\arc{N}{i}{}{}{}$ denote the numbers of excitatory and inhibitory neurons, respectively.
Excitatory and inhibitory neurons are labeled by $\mathsf{e}_{\mu}$ and $\mathsf{i}_{\nu}$ where index $\mu=1,...,N_{\mathsf{e}}$ and $\nu=1,...,N_{\mathsf{i}}$. The subscripts $\mu,\nu$ are used to distinguish excitatory and inhibitory neurons, respectively. 

Neurons are connected via fixed transmission channels.  
All neurons start in a resting state but can become active spontaneously or due to signals from presynaptic neighbors.  
Active neurons propagate signals probabilistically to their postsynaptic targets at the next (discrete) time step.  Neuron states are updated synchronously, progressing through active, refractory, and resting phases according to well-defined transition probabilities, a key difference from the family of integrate-and-fire models \cite{hodgkin_quantitative_1952}. 
Our framework captures the branching nature of spike propagation in a minimal yet biologically motivated manner.  
Detailed descriptions of the model components and update rules follow, along with an illustrative example.

\subsection{Neural configuration space}


The state of each neuron $\mathsf{n}$ at time $t$ is described by a dynamical state variable $\arc{z}{n}{}{}{}(t)\in\mathcal{S}_{\mathsf{n}}$, with values in the set $\mathcal{S}_{\mathsf{n}}=\{0,1,.
	..,\tau_{\mathsf{n}}\}$, where $\arc{\tau}{n}{}{}{}$ is the integer refractory period. 
Following activation of a neuron {\sf n} from state $z_{\mathsf{n}}=0$ to state $z_{\mathsf{n}}=1$, the neuron's state evolves cyclically through the refractory states ($\arc{z}{n}{}{}{}=\{2,...,\arc{\tau}{n}{}{}{}\}$) starting from 2 and ending at $\tau_{\mathsf{n}}$ before returning to $z_{\mathsf{n}}=0$. At time step $t = 0$, all neurons are initialized in the resting (quiescent) state, denoted by $z_{\mathsf{n}} = 0$.

Therefore, at any time step $t$ the state of the neural network can be represented by a sequence of individual neuron states
\[ z(t)=\left (z_{\mathsf{e}_{1}}(t),z_{\mathsf{e}_{2}}(t),.
	..,z_{\mathsf{e}_{N_{\mathsf{e}}}}(t),z_{\mathsf{i}_{1}}(t),z_{\mathsf{i}_{2}}(t),...,z_{\mathsf{i}_{N_{\mathsf{i}}}}(t) \right).
\]
The neural network state configuration space is 
\[C=\{z(t)\ | \ z_{\mathsf{e}_{\mu}}(t)\in\mathcal{S}_{\mathsf{e}}\ \mbox{and }z_{\mathsf{i}_{\nu}}(t)\in\mathcal{S}_{\mathsf{i}}\}.\]

In the GCBM, neurons interact 
through the transmission and processing of signals, where the processing dynamics are governed by the activation (threshold) functions defined in Eqs.~\eqref{def:Ufn_e} and \eqref{def:Ufn_i}.  A neuronal connection (interaction) establishes a transmission channel, labeled as 
$\mathsf{nn}'$, meaning $(\mathsf{n} \rightarrow \mathsf{n}')$,  represented by a directed link or arc. Those transmission channels are fixed at the start of the simulation, as we show next.

\subsection{Types of interactions and states of arcs}
\label{subsec:Interaction-probability}

\begin{figure}
	\includegraphics[width=.95 \columnwidth]{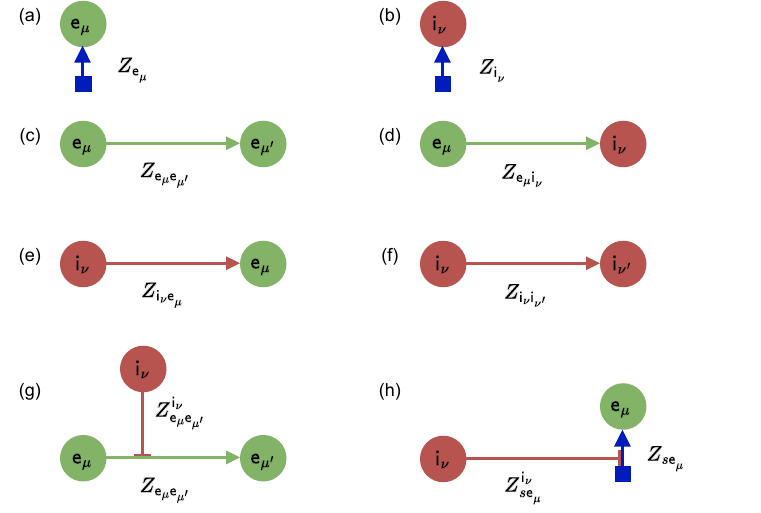}

	\caption{Eight types of interactions defined in the GCBM.
		Panels illustrate: {(a--b)} Interactions between a neuron and an external source.
		{(c--f)}
		Interactions among neurons.
		{(g)}
		Inhibition of the excitatory-to-excitatory $\mathrm{E}\shortrightarrow\mathrm{E}$ channel.
		{(h)}
		Inhibition on the spontaneous activation of an excitatory neuron.
		Arcs are represented by arrows from the source to the target neuron.
		Hyperarcs are depicted as arrows with a flat line from an inhibitory neuron to another arc.
		Red-colored nodes and arcs denote inhibition; green nodes and arcs denote excitation.
		The blue box and arrow represent external spontaneous activation.
	}
	\label{fig:All-interactions}
\end{figure}

The GCBM is composed of eight types of interactions shown in Fig.~\ref{fig:All-interactions}. These interactions are meant to capture biologically relevant neuronal processes, like: excitatory neurons activating other neurons, inhibitory neurons regulating other neurons, axons from a neuron either activating or attenuating other axons, and neurons spontaneously firing. The corresponding states of the arcs at any time step $t$ are represented as
\[Z=\left (Z_{\mathsf{e}_{\mu}},Z_{\mathsf{i}_{\nu}},\arc{Z}{e}{\mu}{e}{\mu'},\arc{Z}{e}{\mu}{i}{\nu},\arc{Z}{i}{\nu}{e}{\mu},\arc{Z}{i}{\nu}{i}{\nu'},\arcin{Z}{e}{\mu}{e}{\mu'}{\nu},\arcin{Z}{}{}{e}{\mu}{\nu}\right ),\]
and each state is represented by a Bernoulli random variable (active [1] and inactive [0]) with a given probability of success (see Section \ref{interactionprob}). The subscripts $\mu,\nu$ (and ${\sf n}$) indicate source or target neuron.
A superscript represents a hyperarc inhibition. 

The first group of interactions, Fig.~\ref{fig:All-interactions}~(a–b), accounts for the experimental observation that neurons can spike spontaneously. In the GCBM, this is modeled by external sources that deliver a spontaneous activation signal $Z_{\mathsf{n}}$ to each neuron at every timestep with probability $p_{s{\sf n}}$. This spontaneous activation acts as a driving external input and is responsible for initiating avalanches when all neurons are in the resting state.
The notation for the spontaneous activation probability, $p_{s{\sf n}}$, is a deliberate exception: because this signal originates from an external source, the subscript is followed by $\mathsf{e}_{\mu}$ or $\mathsf{i}_{\nu}$ to indicate spontaneous activation of excitatory or inhibitory neurons, respectively.

The second group, Fig.~\ref{fig:All-interactions} (c-f), accounts for the fact that chemical and electrical signals from a presynaptic neuron can alter the membrane potential of a target neuron, thereby modulating the state transition probability of a resting neuron to the active state (see Section \ref{interactionprob}).
These four interactions E-E, E-I, I-E, and I-I are described by states of the arcs $Z_{\mathsf{e}_{\mu}\mathsf{e}_{\mu'}},Z_{\mathsf{e}_{\mu}\mathsf{i}_{\nu}},Z_{\mathsf{i}_{\nu}\mathsf{e}_{\mu}}$, and $Z_{\mathsf{i}_{\nu}\mathsf{i}_{\nu'}}$.
The associate transmission probability are $P_{\mathsf{e}_{\mu}\mathsf{e}_{\mu'}},P_{\mathsf{e}_{\mu}\mathsf{i}_{\nu}},P_{\mathsf{i}_{\nu}\mathsf{e}_{\mu}}$, and $P_{\mathsf{i}_{\nu}\mathsf{i}_{\nu'}}$, respectively.

The third group accounts for the unique functional roles of inhibitory neurons.
An inhibitory neuron can inhibit a signal on a specific dendritic branch \citep{blot_ultra-rapid_2014}.
This inhibition targets on transmission channel $Z_{\mathsf{e}_{\mu}\mathsf{e}_{\mu'}}$ by source inhibitory neuron $\mathsf{i}_{\nu}$, thus it is denoted by $Z_{\mathsf{e}_{\mu}\mathsf{e}_{\mu'}}^{\mathsf{i}_{\nu}}$with a transmission probability  $P_{\mathsf{e}_{\mu}\mathsf{e}_{\mu'}}^{\mathsf{i}_{\nu}}$.
Technically, this interaction is known as a hyperarc \citep{bretto_hypergraph_2013}.
They are visualized by a line with flat head ending on another arc in Fig.
\ref{fig:All-interactions} (g).
Another type of inhibitory interaction captures the experimental observation that inhibitory neurons can suppress spontaneous activation \citep{freeman_suppression_2002}.
It is denoted by $Z_{\mathsf{e}_{\mu}}^{\mathsf{i}_{\nu}}$ with a transmission probability $P_{\mathsf{e}_{\mu}}^{\mathsf{i}_{\nu}}$ and is shown in Fig.~\ref{fig:All-interactions}~(h).

In the present study, we focus exclusively on a model where inhibitory neurons consistently inhibit as can be appreciated in Fig. \ref{fig:All-interactions} (f), where the arrow is red-colored. The more general case is explored in a companion  paper \citep{weerawongphrom_minimal_2025}.

Signal transmission across these arcs and hyperarcs occurs with a success probability that we next describe. 



\subsection{Transmission probability and network connectivity}
\label{interactionprob}

Altogether, the transmission probabilities are represented as 
\[P=\Big ( p_{s\mathsf{e}_{\mu}},p_{s\mathsf{i}_{\nu}},\arc{P}{e}{\mu}{e}{\mu'},\arc{P}{e}{\mu}{i}{\nu},\arc{P}{i}{\nu}{e}{\mu},\arc{P}{i}{\nu}{i}{\nu'},\arcin{P}{e}{\mu}{e}{\mu'}{\nu},\arcin{P}{}{}{e}{\mu}{\nu}\Big ),\]
and follow a labeling pattern parallel to that of the transmission states.
Each arc state is modeled as a Bernoulli random variable with success probability $P$, conditional on the source node being active. We refer to $P$  as the transmission probability. The GCBM thus defines a random process on a directed hypergraph, where arcs and hyperarcs encode the complete set of neural interactions.

Network connectivity critically influences the branching dynamics observed in the model. The dynamics of the GCBM across several classes of network architectures,  capable of reproducing spike rasters from laboratory experiments,  were explored in a companion paper \citep{weerawongphrom_minimal_2025}. Key features of these networks served as foundational elements in the construction of the MF equations.

\begin{figure}
	\includegraphics[width=8.5cm]{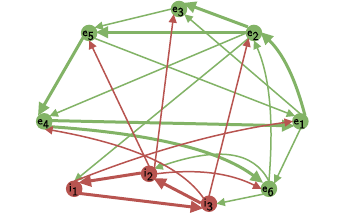}

	\caption{Network topology of the many-body model suitable for a MF approximation. The in-degree $k_{\sf n n'}$ is uniform across all nodes.
	For E-E interaction, $k_{\mathsf{ee}}=2$ and bias parameter $B>0$ (incoming arcs have different weights indicated by thickness).
	One transmission channel is more likely to be activated than another. While $k_{\mathsf{ei}}=k_{\mathsf{ie}}=k_{\mathsf{ii}}=1$, each type has different branching parameter $\kappa_{\sf nn'}$.
	}
	\label{fig:NetworkMFT}
\end{figure}

The first feature is the in-degree,  $k_{\sf n n'}$, defined as the number of incoming arcs to a node $\mathsf{n'}$.
There are four types of $k_{\sf n n'}$ corresponding to node-to-node interaction: $k_{\mathsf{ee}}$, $k_{\mathsf{ie}}$, $k_{\mathsf{ei}}$, and $k_{\mathsf{ii}}$.
While the in-degree $k_{\sf n n'}$ may vary across interaction types in the many-body model, the MF approximation assumes a fixed $k_{\sf n n'}$ for each interaction type.
It is assumed that such a network would more closely approximate translational invariance 
(see Fig.~\ref{fig:NetworkMFT}).
The second feature is the distribution of the connection weights. Experimental evidence \citep{Fosque2021,fosque_quasicriticality_2022} suggests an exponential distribution of the in-degree weight distribution.
Let $\mathrm{r}=1,.
	..,k_{\mathsf{nn'}}$
be the ranking order of the relative weight profile $p_{r}^{\sf n n'}$, then
\begin{equation}
	p_{\mathrm{r}}^{\sf n n'}=\frac{e^{-B\mathrm{r}}}{\sum_{\mathrm{r}=1}^{k_{\mathsf{nn'}}}e^{-B\mathrm{r}}}\label{eq:pnij_RelativeWeight}
\end{equation}
where $B$ is the bias parameter.
If $B=0$, all incoming arcs have the same weight.
If $B\gg1$, the network is essentially reduced to $k_{\mathsf{nn'}}=1$.
The profile of the relative strength of all node-to-node arcs is assumed to have the same form.
We will be using the bias parameter $B=0.5$ throughout the paper for illustrative purposes.

The last feature of the network is
the branching parameter $\kappa_{\mathsf{nn'}}$ (${\sf n},{\sf n'}\in\{{\sf e, i}\}$) that defines the transmission probabilities as:  
$P_{\mathsf{nn'}}(\mathrm{r})=\kappa_{\mathsf{nn'}}p_{\mathrm{r}}^{\sf n n'}$. 
We restrict each $\kappa_{\sf n n'}$ to the range {[}0,$\arc{\kappa}{n}{}{n'}{}^{\sf max}${]}, where the upper bound is given by 
\begin{eqnarray}
 \arc{\kappa}{n}{}{n'}{}^{\sf max}= e^{B}\sum_{\mathrm{r}=1}^{k_{\mathsf{nn}}}e^{-B\mathrm{r}} .
 \label{kappamax}
\end{eqnarray}
If $\kappa_{\mathsf{nn'}}$ exceeds the set maximum value, the model is no longer in a physical regime since the probability of activation of a neighborhood neuron is not restricted to the range of {[}0,1{]}.

Once transmission outcomes are realized, postsynaptic neurons evaluate the resulting arc states using an activation function to determine whether they will transition to the active state, as we next illustrate. 

\subsection{Activation functions}
\label{activationfunctions}

The way a neuron processes signals is encoded in the activation function $U$. 
Define a neighborhood $\mathcal{N}_{\mathsf{n}}$ of postsynaptic neurons $\mathsf{n}$ as the set of the number of excitatory neurons $\widetilde N_{\mathsf{e}\mathsf{n}}$ and the number of inhibitory neurons $\widetilde N_{\mathsf{i}\mathsf{n}}$ inhibitory neurons that have an arc (or hyperarc) terminating at $\mathsf{n}$. 
$\tilde{Z}_{\mathsf{e}_{\mu'}}$($\tilde{Z}_{\mathsf{i}_{\nu'}}$) sequence of all arcs' states of the neighborhood of excitatory (inhibitory) neurons $\mathsf{e}$ ($\mathsf{i}$),
\begin{align*}
	\tilde{Z}_{\mathsf{e}_{\mu'}} & =\left ( Z_{\mathsf{e}_{\mu'}},
	Z_{\mathsf{e}_{\mu}\mathsf{e}_{\mu'}},Z_{\mathsf{i}_{\nu}\mathsf{e}_{\mu'}},Z_{\mathsf{e}_{\mu}\mathsf{e}_{\mu'}}^{\mathsf{i}_{\nu}},Z_{\mathsf{e}_{\mu'}}^{\mathsf{i}_{\nu}}\right ) \\ \tilde{Z}_{\mathsf{i}_{\nu'}} & =\left (Z_{\mathsf{i}_{\nu'}},Z_{\mathsf{e}_{\mu}\mathsf{i}_{\nu'}},Z_{\mathsf{i}_{\nu}\mathsf{i}_{\nu'}} \right ),
    \end{align*} 
    where indices of the source neurons belong to $\mathcal{N}_{\mathsf{n}'}$.

For the simplest case, where hyperarc interactions are not present, an excitatory neuron is modeled to spike when it receives more excitatory than inhibitory signals.
That is, in simple informative form,
\begin{align}
	U(\tilde Z_{\mathsf{e'}}) & =\Theta\left(Z_{\mathsf{e'}}+\sum_{{\sf e}\in \mathcal{N}_{\mathsf{e}'}} Z_{\mathsf{ee'}}-\sum_{{\sf i}\in \mathcal{N}_{\mathsf{e}'}} Z_{\mathsf{ie'}}\right) ,
\label{def:Ufn_e}
\end{align} 
    where $\Theta$ is a step function with $\Theta(0)=0$ and the summations run over all the neighborhood.
Similarly, the activation function for inhibitory neurons is 
\begin{align}
	U(\tilde Z_{\mathsf{i'}}) & =\Theta\left(Z_{\mathsf{i'}}+\sum_{{\sf e}\in \mathcal{N}_{\mathsf{i}'}} Z_{\mathsf{ei'}}-\sum_{{\sf i}\in \mathcal{N}_{\mathsf{i}'}} Z_{\mathsf{ii'}}\right) .
\label{def:Ufn_i}\end{align}

However, inhibitory neurons are more diverse than excitatory neurons.
Although inhibitory neighbor activity consistently decreases the likelihood of excitatory neuron firing, experimental evidence shows an opposite effect in target inhibitory neurons.
Some biological mechanisms for inhibitory neurons to effectively excite other inhibitory neurons are gap junctions \citep{skinner_bursting_1999,gibson_two_1999}, disinhibition \citep{ye_inhibitory_2017}, and an increase in reverse potential \citep{ye_inhibitory_2017}.
In such a situation one needs to replace Eq. \eqref{def:Ufn_i} by  $U(\tilde Z_{\mathsf{i'}})=\Theta\left(Z_{\mathsf{i'}}+\sum Z_{\mathsf{ei'}} + \sum Z_{\mathsf{ii'}}\right)$.


To capture the physics of hyperarc inhibition, the activation function must be extended to 
\begin{align}
	U(\tilde{Z}_{\sf e'})= & \Theta\left(\sum_{{\sf e}\in \mathcal{N}_{\mathsf{e}'}}\,Z_{\mathsf{e}\mathsf{e}'}\cdot\prod_{{\sf i}\in \mathcal{N}_{\mathsf{e}'}}(1-Z_{\mathsf{e}\mathsf{e}'}^{\mathsf{i}})-\sum_{{\sf i}\in \mathcal{N}_{\mathsf{e}'}} Z_{\mathsf{i}\mathsf{e}'} \right.+
	\nonumber                                  \\& \left. Z_{\mathsf{e}'}\cdot\prod_{{\sf i}\in \mathcal{N}_{\mathsf{e}'}}(1-Z_{\mathsf{e}'}^{\mathsf{i}})\right),
    \label{def:Ufn_MarkThesis}
    \end{align}
where the summations and products are taken over all arcs in the neighborhood of neuron $\mathsf{e}_{\mu}$. If any of the hyperarcs is activated, the signal on the target arc is suppressed. 

The activation function takes arc states as input and determines whether a neuron in the resting state transitions to the active state. An output of one indicates that the neuron becomes active. 
Since individual arc and hyperarc states are stochastic variables, the output of the activation function is also a Bernoulli random variable. In the following section, we compute the probability that the activation function gives a value of one, that is, the transition probability $W_{01}^{\sf n}$.

\subsection{Transition probability}
\label{transitionprob}

To compute the transition probability of the target neuron $\mathsf{n'}$,  $W^\mathsf{n'}_{01}$,  as a function of the states $\tilde{z}_{\sf n'}$ of its neighboring neurons,
\begin{equation}
\tilde{z}_{\mathsf{n'}}(t)=(\dots,z_{\mathsf{e}_{\mu}},\ldots;\dots,z_{\mathsf{i_{\nu}}},\ldots),\quad\mathsf{e_{\mu},}\mathsf{i_{\nu}}\in\mathcal{N}_{\mathsf{n}'} ,
\label{eq:NodeState_Neighbor}
\end{equation}
one has to evaluate the expression
\begin{equation}
	W_{01}^{\mathsf{n'}}(\tilde{z}_{\mathsf{n'}}(t))=\sum_{\forall\tilde{Z}_{\mathsf{n'}}}P(\tilde{Z}_{\mathsf{n'}}|\tilde z_{\mathsf{n'}}(t)) \ U(\tilde{Z}_{\mathsf{n'}}),\label{eq:transition_probability} 
    \end{equation}
which casts $W_{01}^{\mathsf{n'}}$ in terms of the states of the neighboring neurons at a given time step $t$. 
Since $\tilde{Z}_{\mathsf{n'}}$ are independent random variables, the  conditional probabilities become
\begin{align}
P & (\tilde{Z}_{\mathsf{n'}}|\tilde{z}_{\mathsf{n'}}(t))=P(Z_{\mathsf{n'}})\cdot\prod_{\forall\mathsf{n}\in\mathcal{N}_{\mathsf{n'}}}P(Z_{\mathsf{nn'}}|z_{\mathsf{n}}(t))\label{eq:JointCondProb_ArcState}\nonumber  \\
\cdot & \prod_{\forall\mathsf{i}\in\mathcal{N}_{\mathsf{{n}'}}}P(Z_{\mathsf{e{n}'}}^{\mathsf{i}}|z_{\mathsf{i}}(t))\cdot\prod_{\forall\mathsf{i}\in\mathcal{N}_{\mathsf{{n}'}}}P(Z_{\mathsf{{n}'}}^{\mathsf{i}}|z_{\mathsf{i}}(t)).
\end{align}
Each of these probabilities can in turn be expressed as follows
\begin{align}
    P(Z_{\mathsf{n'}}(t)) & =p_{s\mathsf{n'}},\nonumber \\
    P(Z_{\mathsf{nn'}}(t)|z_{\mathsf{n}}(t)) & =(1-\delta_{1,z_{\mathsf{n}}(t)})(1-Z_{\mathsf{nn'}}(t))\nonumber \\
    +\delta_{1,z_{\mathsf{n}}(t)} & \left((1-Z_{\mathsf{nn'}}(t))(1-P_{\mathsf{nn'}})+Z_{\mathsf{nn'}}(t)P_{\mathsf{nn'}}\right),\nonumber \\
    P(Z_{\mathsf{ee'}}^{\mathsf{i}}(t)|z_{\mathsf{i}}(t)) & =(1-\delta_{1,z_{\mathsf{i}}(t)})(1-Z_{\mathsf{ee'}}^{\mathsf{i}}(t))\nonumber \\
    +\delta_{1,z_{\mathsf{i}}(t)} & \left((1-Z_{\mathsf{ee'}}^{\mathsf{i}}(t))(1-P_{\mathsf{ee'}}^{\mathsf{i}})+Z_{\mathsf{ee'}}^{\mathsf{i}}(t)P_{\mathsf{ee'}}^{\mathsf{i}}\right),\nonumber \\
    P(Z_{\mathsf{e'}}^{\mathsf{i}}(t)|z_{\mathsf{i}}(t)) & =(1-\delta_{1,z_{\mathsf{i}}(t)})(1-Z_{\mathsf{e'}}^{\mathsf{i}}(t))\nonumber \\
    +\delta_{1,z_{\mathsf{i}}(t)} & \left((1-Z_{\mathsf{e'}}^{\mathsf{i}}(t))(1-P_{\mathsf{e'}}^{\mathsf{i}})+Z_{\mathsf{e'}}^{\mathsf{i}}(t)P_{\mathsf{e'}}^{\mathsf{i}}\right). 
    \label{eq:ArcProbabilityFxnSite}
\end{align}

\begin{figure}
\includegraphics[width=8cm]{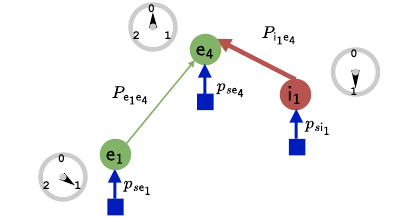} \caption{ Neighborhood of excitatory neuron $\mathsf{e}_{4}$. Green nodes represent excitatory neurons and red nodes represent inhibitory neurons. Directed arrows indicate synaptic interactions: green for excitation and red for inhibition. The layout includes all neurons and interactions directly linked to $\mathsf{e}_{4}$, capturing its environment. The clocks in each of the nodes represent its state $z_{\sf n}$. The arcs are labeled by the probability of activation.
	}
	\label{fig:ExampleWe01}
\end{figure}

For the MF analysis in Section~\ref{section3}, we find that computationally the only transition probability that must be evaluated explicitly is the special case in which all neighboring neurons are active, i.e., $\tilde{z}_{\mathsf{n}}(t)=(1,\dots;1,\dots)$, which is labeled as $\overline{W_{01}^{\mathsf{n}}}$. To proceed, we construct the corresponding event space necessary for computing $\overline{W_{01}^{\mathsf{n}}}$. 

We first define the event $Z_{\mathsf{n'}}(t) = 1  (0)$ to indicate that the directed edge from the spontaneous source to neuron $\mathsf{n'}$ is active (inactive) at time $t$. Next, consider the set of directed edge states $Z_{\mathsf{nn'}}(t)$. We define $Z_{\mathsf{nn'}}(t) = 1 (0)$ to indicate that the directed edge from neuron $\mathsf{n}$ to neuron $\mathsf{n'}$ is active (inactive). Similarly, the events $Z_{\mathsf{ee'}}^{\mathsf{i}}(t) = 1  (0)$ and $Z_{\mathsf{e'}}^{\mathsf{i}}(t) = 1  (0)$ denote that the corresponding hyperarc originating from inhibitory neuron $\mathsf{i}$ is active (inactive).

Since excitatory signals can be canceled by arc-to-arc inhibition as specified by the activation function, we define the complete set of events that actually contribute to the activation of neuron $\mathsf{n}$ at time $t$ as follows
\begin{align}
	A_{0}^{\mathsf{e}}(t) & =(Z_{\mathsf{e}}(t)=1)\bigcap_{\mathsf{i_{\nu}}\in\mathcal{N}_{\mathsf{e}}}(Z_{\mathsf{e}}^{\mathsf{i_{\nu}}}(t)=0),\nonumber \\ A_{0}^{\mathsf{i}}(t) & =(Z_{\mathsf{i}}(t)=1),\nonumber \\ A_{\mu}^{\mathsf{e}}(t) & =(Z_{\mathsf{e_{\mu}e}}(t)=1)\bigcap_{\mathsf{i_{\nu}}\in\mathcal{N}_{\mathsf{e}}}(Z_{\mathsf{e_{\mu}e}}^{\mathsf{i_{\nu}}}(t)=0),\nonumber \\ A_{\mu}^{\mathsf{i}}(t) & =(Z_{\mathsf{e_{\mu}i}}(t)=1),\nonumber \\ B_{\nu}^{\mathsf{n}}(t) & =(Z_{\mathsf{i_{\nu}n}}(t)=1)  ,
	   \label{eq:AB_events}
\end{align}
where $\mu\in\{1,\dots,\tilde{N}_{\mathsf{en}}\}$  ($\nu\in\{1,\dots,\tilde{N}_{\mathsf{in}}\}$) 
indexes the excitatory (inhibitory) neighbors of $\mathsf{n}$.

The probabilities of the events $A_{0}^{\mathsf{n}}(t)$, $A_{\mu}^{\mathsf{n}}(t)$ and $B_{\nu}^{\mathsf{n}}(t)$ conditioned on  all (source) neighboring neurons being
in the active state, are
given by
\begin{align}
	P(A_{0}^{\mathsf{e}}(t)\vert\tilde{z}_{\mathsf{n}}=(1,\dots;1,\dots))    & =p_{s\mathsf{e}}\prod_{\mathsf{i_{\nu}}\in\mathcal{N}_{\mathsf{e}}}(1-P_{\mathsf{e}}^{\mathsf{i_{\nu}}}) , \nonumber              \\
	P(A_{0}^{\mathsf{i}}(t)\vert\tilde{z}_{\mathsf{n}}=(1,\dots;1,\dots))    & =p_{s\mathsf{i}}  ,                                                                                                            \nonumber \\
	P(A_{\mu}^{\mathsf{e}}(t)\vert\tilde{z}_{\mathsf{n}}=(1,\dots;1,\dots))  & =P_{\mathsf{e_{\mu}e}}\prod_{\mathsf{i_{\nu}}\in\mathcal{N}_{\mathsf{e}}}(1-P_{\mathsf{e_{\mu}e}}^{\mathsf{i_{\nu}}}) , \nonumber \\
	P(A_{\mu}^{\mathsf{i}}(t)\vert\tilde{z}_{\mathsf{n}}=(1,\dots;1,\dots))  & =P_{\mathsf{e_{\mu}i}} ,                                                                                                        \nonumber \\
	P(B_{\nu}^{\mathsf{n}}(t)|\tilde{z}_{\mathsf{n}}=(1,\dots;1,\dots)) & =P_{\mathsf{i_{\nu}n}} ,
    \label{eq:Probability_events}
\end{align}
which we will next compactly denote as $\overline{P}(A_{0}^{\mathsf{n}})$, $\overline{P}(A_{\mu}^{\mathsf{n}})$, and $\overline{P}(B_{\nu}^{\mathsf{n}})$. 

With the events $A_{0}^{\mathsf{n}}(t)$, $A_{\mu}^{\mathsf{n}}(t)$ and $B_{\nu}^{\mathsf{n}}(t)$ and their corresponding conditional probabilities defined, we next introduce  $E_{m}^{\mathsf{n}}$ as the probability that at least $m$ of the events $A_{0}^{\mathsf{n}}(t),\dots,A_{\tilde{N}_{\mathsf{en}}}^{\mathsf{n}}(t)$ occur simultaneously.
The probability $E_{m}^{\mathsf{n}}$ can be computed using a generalization of the inclusion–exclusion principle
\begin{align}
	E_{m}^{\mathsf{n}} & =\sum_{r=m}^{\tilde{N}_{\mathsf{en}}+1}(-1)^{r-m}\binom{r-1}{m-1}\,S_{r}^{\mathsf{en}},\ \ m=1,\ldots,\tilde{N}_{\mathsf{en}}+1, \label{eq:E_m_term}
\end{align}
where


\begin{align}
	S_{r}^{\mathsf{en}} & :=\sum_{\ell_r=1}^{L_r} \overline{P}(\bigcap_{j\in J_{\ell_r}}A_{j}^{\mathsf{n}}),\label{eq:S_e_term} \end{align} 
with $J_{\ell_r}\subseteq \{0,\ldots,\tilde{N}_{\mathsf{en}}\}$ denoting a subset of cardinality $r$.
The index $\ell_r$ runs over all such subsets, of which there are $L_r=\binom{\tilde{N}_{\mathsf{en}}+1}{r}$.
The events $A_{0}^{\mathsf{n}}$ and $A_{\mu}^{\mathsf{n}}$ are assumed to be independent.

Similarly, we define $I_{[m]}^{\mathsf{n}}$ as the
probability that exactly $m$ of the events $B_{1}^{\mathsf{n}}(t),\dots,B_{\tilde{N}_{\mathsf{in}}}^{\mathsf{n}}(t)$
simultaneously occur,
\begin{align}
	I_{[m]}^{\mathsf{n}} & =\sum_{r=m}^{\tilde{N}_{\mathsf{in}}}(-1)^{r-m}\binom{r}{m}\,S_{r}^{\mathsf{in}},\ \ m=0,\ldots,\tilde{N}_{\mathsf{in}},
	\label{eq:I_m_term}
\end{align}
where
\begin{align}
	S_{r}^{\mathsf{in}} & :=\sum_{\ell_r=1}^{M_r} \overline{P}(\bigcap_{j\in K_{\ell_r}}B_{j}^{\mathsf{n}}),\label{eq:S_i_term} \end{align} 
with $K_{\ell_r}\subseteq \{1,\ldots,\tilde{N}_{\mathsf{in}}\}$ denoting a subset of cardinality $r$.
The index $\ell_r$ runs over all such subsets, of which there are $M_r=\binom{\tilde{N}_\mathsf{in}}{r}$.
Note that $S_{0}^{\mathsf{in}}=1$. 
The events $B_{\nu}^{\mathsf{n}}$ are also assumed to be independent.
The probability of at least $m$ of the events occurring simultaneously, or the probability of exactly $m$ of the events occurring simultaneously, is calculated from a generalization of the inclusion-exclusion principle known as the Schuette-Nesbitt formula \cite{Feller1968}.
\begin{table}[htb]
	\begin{tabular}{|c|c|c|}
		\hline
		$z_{\mathsf{e}_{1}}$                     & $z_{\mathsf{i}_{1}}$ & $W_{01}^{\mathsf{e}_{4}}$\tabularnewline
		\hline
		\hline
		$\cancel{1}$                             & $\cancel{1}$         & $p_{s\mathsf{e}_{4}}$\tabularnewline
		\hline
		$\cancel{1}$                             & 1                    & $p_{s{\sf e}_{4}}(1-P_{\mathsf{i}_1 {\sf e}_{4}})$\tabularnewline
		\hline
		1                                        & $\cancel{1}$         & $p_{s{\sf e}_{4}}+P_{\mathsf{e}_1{\sf e}_{4}}-p_{s{\sf e}_{4}}
		P_{\mathsf{e}_1{\sf e}_{4}}$\tabularnewline \hline 1 & 1                    & $(p_{s{\sf e}_{4}}+P_{\mathsf{e}_1{\sf e}_{4}}-p_{s{\sf e}_{4}}P_{\mathsf{e}_1{\sf e}_{4}})(1-P_{\mathsf{i}_1{\sf e}_{4}})+p_{s{\sf e}_{4}}P_{\mathsf{e}_1{\sf e}_{4}}P_{\mathsf{i}_1{\sf e}_{4}}$ \tabularnewline \hline\end{tabular}
\caption{Transition probabilities,  $W_{01}^{\mathsf{e}_{4}}$, of a node transitioning from state 0 to state 1 under various neighbor conditions. The transition probability increases when an excitatory neighbor is active and decreases when an inhibitory neighbor is active. The symbol $\cancel{1}$ represents a node that is not active, i.e., 0,2,...,$\tau_{\mathsf{n}}$, and $\tau_{\mathsf{n}}$ denotes the time constant for the node’s state evolution. There is always a chance of spontaneous activation even when the neighbors' states are not active.}
\label{tab:ExampleWe01}
\end{table}

To compute $\overline{W_{01}^{\mathsf{n}}}$, we consider the activation function $U$ in Eq.~\eqref{eq:transition_probability}, which includes only terms for which the number of active excitatory arcs (excluding those subject to arc-to-arc inhibition) exceeds the number of active inhibitory arcs by at least one, thereby contributing to activation of the target neuron.
By accounting for the possible number of $\tilde{N}_{\mathsf{en}}$
and $\tilde{N}_{\mathsf{in}}$, the general expression of $\overline{W_{01}^{\mathsf{n}}}$
is


\begin{equation}
	\begin{aligned}\overline{W_{01}^{\mathsf{n}}}=W_{01}^{\mathsf{n}}((1,\ldots;1,\ldots))=\sum_{m=1}^{\tilde{N}_{\mathsf{in}}+1}E^{\sf n}_{m}I^{\sf n}_{[m-1]} .
	\end{aligned}
	\label{eq:Highest_W}
\end{equation}
\subsection{Neuronal Dynamics}

Having established the components necessary to define the stochastic dynamics of the GCBM, we now proceed with a systematic description. Each neuron has its own probability $W^{\mathsf{n}}_{01}$ of transitioning from the quiescent state (0) at time $t$ to the active state (1) at time $t + 1$. Only neurons in the quiescent state integrate inputs and may spike with probability $W^{\mathsf{n}}_{01}$. This process is illustrated in the following diagram:
\begin{equation} 
\begin{tikzcd}[cramped,row sep=small,column sep=small]
		0 \arrow[r,"W_{01}^\mathsf{n}"] \arrow[loop left,"1-W_{01}^\mathsf{n}",below] &
		1 \arrow[r] &
		2 \arrow[r] & ... \arrow[r] & \tau_{\mathsf{n}}
		\arrow[llll,bend left]
	\end{tikzcd} .
    \label{graphdynamics}
\end{equation}
The only stochastic component in the dynamics is the transition probability $W^{\mathsf{n}}_{01}$, through which neurons may be activated by incoming signals. Once activated, a neuron enters a dormant phase lasting $\tau_{\mathsf{n}}$ time steps, representing the biological refractory period. Neurons in the active or refractory states evolve deterministically in a fixed cycle and do not process incoming inputs during this time.

In summary, the neuronal dynamics of the GCBM is described by the
following algorithm:
\begin{enumerate}
	\item \textit{Initialization}:
All neuronal interactions are defined by a fixed hypergraph—whose topology plays a critical role \cite{weerawongphrom_minimal_2025}— as described in Section \ref{subsec:Interaction-probability}. The initial state of each neuron is set to $z_{\mathsf{n}}(t = 0) = 0$.
	\item \textit{Drive}: At each time step $t$, every neuron ${\sf n}'$ receives a spontaneous signal with probability $p_{s\mathsf{n'}}$ (Section \ref{interactionprob}).
	\item \textit{Relaxation}: If a presynaptic neuron $\mathsf{n}$ is in the active state, $z_{\mathsf{n}}(t) = 1$, it transmits a set of signals $Z$ through all its outgoing channels with transmission probability $P$, as described in Section \ref{interactionprob}.
	      \begin{enumerate}
		      \item If the postsynaptic neuron $\mathsf{n'}$ is in the quiescent state, $z_{\mathsf{n}'}(t) = 0$, it may transition to the active state, $z_{\mathsf{n}'}(t+1) = 1$, by processing incoming signals and evaluating the activation function $U$, as described in Section \ref{activationfunctions}. With $P$ and $U$ evaluated for each target neuron $\mathsf{n'}$, we then compute the transition probability $W_{01}^{\mathsf{n'}}$, as detailed in Section \ref{transitionprob}.
		      \item If the postsynaptic neuron $\mathsf{n'}$ is in a state $z_{\mathsf{n}'}(t) \neq 0$, it progresses to the next refractory state until it returns to the quiescent state. This evolution is governed by
$$ \hspace*{1cm}
z_{\mathsf{n}'}(t+1) = \left( z_{\mathsf{n}'}(t) + 1 \right) \bmod (\tau_{\mathsf{n}'} + 1),
$$
as illustrated in the diagram \eqref{graphdynamics}.
	      \end{enumerate}
	\item \textit{Iteration}: Start the next time step: Return to step 2.
\end{enumerate}

\subsection{Dynamics Characterization}
The density of active excitatory nodes has previously been identified as the order parameter of the CBM \cite{williams2014quasicritical}. For the GCBM, we introduce the density of active excitatory and inhibitory nodes:
\begin{align}
	\rho_{1}^{\mathsf{e}}(t) & =\frac{1}{N_{\mathsf{e}}}\sum_{\mu=1}^{N_{\mathsf{e}}}\delta_{z_{\mathsf{e}_{\mu}}(t),1}   \ , \ \ \ \                 	\rho_{1}^{\mathsf{i}}(t)  =\frac{1}{N_{\mathsf{i}}}\sum_{\nu=1}^{N_{\mathsf{i}}}\delta_{z_{\mathsf{i}_{\nu}}(t),1} .\label{eq:many_body_average_activity}
\end{align}
The time average case is
\begin{eqnarray}
    \Bar{\rho}^{\sf{n}}_{1}=\langle\rhoOne{n}(t)\rangle_{t}=\frac{1}{N_{T}}\sum_{t=1}^{N_{T}}\rhoOne{n}(t) ,
\end{eqnarray}
where $N_{T}$ is the number of iteration time steps of the map. The zero-field dynamical susceptibility $\chi_{\mathsf{n}}$ corresponds to the fluctuations
\begin{equation}
\text{\ensuremath{\chi_{\mathsf{n}}}}=N_{\mathsf{n}}\left(\left\langle \rhoOne{n}(t))^{2}\right\rangle _{t}-(\Bar{\rho}^{\sf{n}}_{1})^{2}\right)\label{eq:def_Suscept}
\end{equation}
and quantifies the dynamical response of the system.

\section{Hierarchy of Mean-Field Approximations}
\label{section3}
Now that we have established the many-body formulation of the GCBM, we aim to develop a semi-analytical method to probe its dynamics. The GCBM has been shown to reproduce avalanche behavior consistent with experimental observations \cite{Fosque2021, weerawongphrom_minimal_2025, fosque_quasicriticality_2022}.
As noted earlier, the GCBM falls within the NP class of problems, meaning that many-body simulations of the GCBM would require exponentially more computations as more nodes are added, limiting the potential for numerical examination and insights
The goal of the MF approximation is to derive expressions for the fraction of neurons in a given state $j$ at time step $t$, within a framework that generalizes across a range of network topologies. 
This framework relies on a series of assumptions that simplify the underlying network dynamics.

First, we enforce translational invariance to reduce the complex interactions in a network of $N$ neurons to a tractable system involving a single representative excitatory neuron, $\mathsf{e}_{0}$, and a single representative inhibitory neuron, $\mathsf{i}_{0}$. 
In general, a motif may contain multiple representative neurons of each type. 
For pedagogical clarity, however, we restrict our examples to a single representative neuron of each type.
This choice of using only a single representative neuron ensures that the number of neighbors coincides with the in-degree, so that $\numberneighborsMB{n}{n'} = k_{\mathsf{n n'}}$.
We also note that this MF approximation does not assume all-to-all connectivity.
The state of the representative neurons is $z_{\mathsf{n}_{0}}(t)$.
The neighborhoods of these representative neurons are characterized by the distribution of neuron states and the connectivity structure of the many-body network.
By varying the form of these effective interactions, the model can represent the collective behavior across a hierarchy of network topologies.
Second, we assume that the dynamical states of neighboring neurons, denoted $\arcprime{\tilde{z}}{n}(t)$, are statistically independent and identically distributed (i.i.d.).
This is a feature of the MF approximation; the full many-body GCBM, in contrast, does exhibit correlations among presynaptic neurons. 
Finally, we assume the probability of finding the representative neuron $\mathsf{n}_{0}$ in state $j$ corresponds to the density of neurons in state $j$ in the full many-body system.

Under these assumptions, the MF approximation replaces the many-body network with a {\it motif} whose dynamics are governed by coupled nonlinear maps.
This approach offers a tractable semi-analytical framework for studying network stability, the influence of inhibitory neurons, the model’s consistency with the quasi-critical hypothesis, and the onset of chaotic dynamics.

\begin{figure}
	\includegraphics[width=7cm]{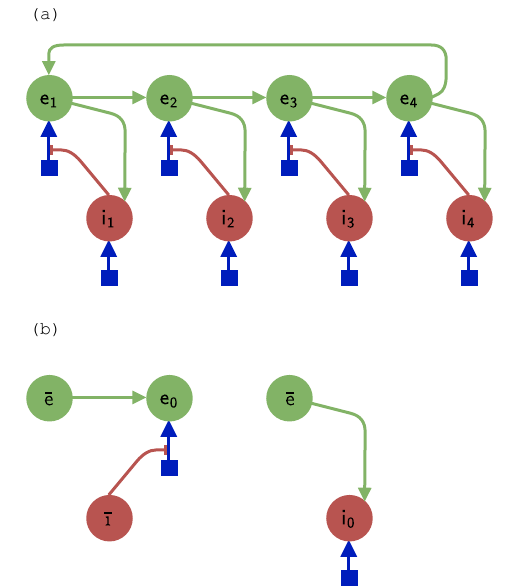} 
    \caption{
    An example network and its subgraphs in the GCBM. 
    {(a)}: A translationally invariant network where every neuron has the same local presynaptic environment. 
    Each excitatory neuron receives one incoming excitatory arc, one inhibitory arc, and one external spontaneous activation arc. 
    Each inhibitory neuron receives one excitatory arc and one external spontaneous activation arc.
    {(b)}: The subgraphs used in the MF approximation.
    We have $\numberneighbors{e}{e}=\numberneighbors{e}{i}=\numberneighbors{i}{e}=1$, and $\numberneighbors{i}{i}=0$.
    Here, $\mathsf{e}_{0}$ denotes the representative excitatory neuron and $\mathsf{i}_{0}$ the representative inhibitory neuron, while
    $\mathsf{\bar e}$ and $\mathsf{\bar {i}}$ represent the mean excitatory and inhibitory neighbors, respectively.
    Blue arrows denote external spontaneous activation. 
    }
	\label{fig:net_to_centralSite}
\end{figure}

\subsection{Motifs and representative neurons}

\begin{figure}
\includegraphics[width=0.85\columnwidth]{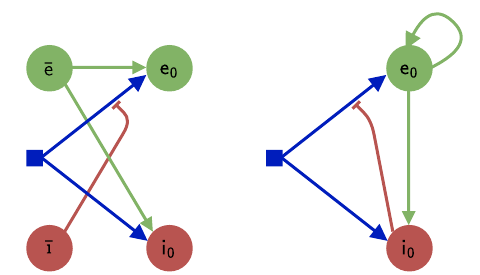} \caption{Example motif of the network in Fig.\textbf{~\ref{fig:net_to_centralSite}.}
		This is an example of how we form the motif from the network subgraphs.
		This example includes the inhibition of a directed edge of the external spontaneous activation.
		The motif has all the mean neighborhood connections on the left and the representative neurons on the right, showing the progression of time from left to right.
		A MF diagram is also shown on the right to emphasize that there are only two representative neurons in the MF approximation.
		We will primarily use the {\it unfold} diagram on the left to clearly show the neighborhood of the representative neurons. We have $\numberneighbors{e}{e}=\numberneighbors{e}{i}=\numberneighbors{i}{e}=1$, and $\numberneighbors{i}{i}=0$.
    Here, $\mathsf{e}_{0}$ denotes the representative excitatory neuron and $\mathsf{i}_{0}$ the representative inhibitory neuron, while
    $\mathsf{\bar e}$ and $\mathsf{\bar {i}}$ represent the mean excitatory and inhibitory neighbors, respectively.
    Blue arrows denote external spontaneous activation. 
	}

	\label{fig:motif}
\end{figure}

An example of a spatially symmetric network composed of \(N_{\mathsf{e}} + N_{\mathsf{i}}\) neurons, to which the MF approximation can be applied, is shown in Fig.~\ref{fig:net_to_centralSite} (a).
This panel consists of excitatory and inhibitory neurons with their interactions. Figure \ref{fig:net_to_centralSite} (b) illustrates representative subgraphs that capture the interactions experienced by a neuron in this network.
The MF approximation reduces the full network to two representative neurons, one excitatory neuron, $\mathsf{e}_{0}$, and one inhibitory neuron, $\mathsf{i}_{0}$, each interacting with a set of MF neighbors.
Specifically, $\mathsf{e}_{0}$ receives input from one spontaneous source, one excitatory neighbor, and one inhibitory neighbor. 
In contrast, $\mathsf{i}_{0}$ receives input from one spontaneous source and one excitatory neighbor. 
These neighbors are denoted by $\mathsf{\bar n}$ to indicate that they represent the average effective environment rather than specific neurons.
The dynamics are defined solely on the representative neurons $\mathsf{e}_{0}$ and $\mathsf{i}_{0}$. 
These two subgraphs are then merged into a single motif, shown in Fig.~\ref{fig:motif}, where both $\mathsf{e}_{0}$ and $\mathsf{i}_{0}$ share the same environment.

\subsection{Dynamical maps}

The MF approximation expresses excitatory and inhibitory neuron state densities at time $t+1$ as functions of those at time $t$.
The only nontrivial state densities are the probability of activation of excitatory and inhibitory neurons, respectively, due to their stochastic dynamics.

To derive the dynamical maps of these nontrivial terms, consider a representative neuron $\mathsf{n}_{0}$ and its mean neighborhood. 
Let the \textit{joint probability} of  neuron's state $\arcprime{z}{n}(t)$ at time $t$ and $t+1$, along with the state of its neighborhood $\arcprime{\tilde{z}}{n}$, be denoted by $P(\arcprime{z}{n}(t+1),\arcprime{z}{n}(t),\arcprime{\tilde{z}}{n}(t))$. We are interested in $P(\arcprime{z}{n}(t+1)=1)$. Therefore, from the law of total probability
\begin{align}
	P(\arcprime{z}{n}(t+1)=1) & \nonumber \\ & \hspace{-2.5cm} =\sum_{\forall\arcprime{\tilde{z}}{n}(t)}\sum_{j=0}^{\tau_{\sf n}}P(\arcprime{z}{n}(t+1)=1,\arcprime{z}{n}(t)=j,\arcprime{\tilde{z}}{n}(t)) ,
\label{eq:P_z_tf_TotalProb} 
    \end{align} 
where summation runs over all possible states of the neighborhood.  From the GCBM transition rules, only $\arcprime{z}{n}(t)=0$ can transit to $\arcprime{z}{n}(t+1)=1$. Thus
\begin{align}
\hspace*{-0.5cm}P(\arcprime{z}{n}(t+1)=1)& \nonumber \\ & \hspace{-2.5cm} =\sum_{\forall\arcprime{\tilde{z}}{n}(t)} P(\arcprime{z}{n}(t+1)=1,\arcprime{z}{n}(t)=0,\arcprime{\tilde{z}}{n}(t)) .
\end{align}
This can be evaluated using the conditional probability
\begin{align}
\hspace*{-0.5cm}	P(\arcprime{z}{n}(t+1)=1) \nonumber \\
	&\hspace*{-2.5cm}=\sum_{\forall\arcprime{\tilde{z}}{n}(t)} W^\mathsf{n_0}_{01}(\arcprime{\tilde{z}}{n}(t)) \cdot P(\arcprime{z}{n}(t)=0, \arcprime{\tilde{z}}{n}(t)) ,\label{eq:P_z_tf_CondProb}
\end{align}
where $W^\mathsf{n_0}_{01}(\arcprime{\tilde{z}}{n}(t))=  P(\arcprime{z}{n}(t+1)=1 \! \mid \! \arcprime{z}{n}(t)=0, \arcprime{\tilde{z}}{n}(t))$.

In the many-body problem, the joint probability $P(z_{\mathsf{n}_{0}}(t)=0, \tilde z_{\mathsf{n}_{0}}(t))$ can be very complex.  
Using our second primary assumption that neuron states are \textit{independent random variables},  
we approximate this joint probability as the product of individual probabilities
\begin{align}
	P&(z_{\mathsf{n}_{0}}(t)=0,\tilde z_{{\mathsf{n}_{0}}}(t))\\ \nonumber &=P(z_{\mathsf{n}_{0}}(t)=0)\cdot\prod_{\mu=1}^{\numberneighbors{e}{n}}P(\tilde{z}_{\mathsf{e}_{\mu}}(t))\cdot\prod_{\nu=1}^{\numberneighbors{i}{n}}P(\tilde{z}_{\mathsf{i}_{\nu}} (t)) .
    \label{eq:assume_independent}
\end{align}

Moreover, it is assumed that the state of all excitatory and inhibitory neurons is drawn from an \textit{identical probability mass distribution}. That is
\begin{equation}
	P(z_{{\sf n}}(t)=m)=P(z_{{\sf n'}}(t)=m),\quad\forall m\in\mathcal{S}_{\sf n} .
    \label{eq:assume_identical} 
    \end{equation} 
To simplify notation, let $x_m(t)$ denote the probability that an excitatory neuron $\mathsf{e}_{\mu}$ is in state $m$ at time step $t$. (The neuron index is omitted due to the assumption of an identical mass distribution across excitatory neurons.)
Similarly, let $y_n(t)$ represent the probability that an inhibitory neuron $\mathsf{i}_{\nu}$ is in state $n$ at time step $t$
\begin{align}
	x_{m}(t) & \triangleq P(z_{\mathsf{e}_{\mu}}(t)=m)\quad m\in\mathcal{S}_{\mathsf{e}}\label{eq:def_xy}, \nonumber\\
	y_{n}(t) & \triangleq P(z_{\mathsf{i}_{\nu}}(t)=n)\quad n\in\mathcal{S}_{\mathsf{i}}.
\end{align}
In other words, we impose that the representative neurons and their neighbors are statistically indistinguishable, sharing the same state distribution. 
This is the MF approximation, leading to a set of dynamical \textit{self-consistent equations} that we show next.

The MF equations for $m=1$ are given by the dynamical maps for $\mathsf{e_{0}}$ and $\mathsf{i}_{0}$ as follows

\begin{align}
   &x_{1}(t+1)=F \cdot {x_{0}(t)},   \label{eq:MFT_x1}\\ &F=\sum_{\forall\tilde{z}_{\mathsf{e}_{0}}(t)} W_{01}^{\mathsf{e_0}}(\tilde{z}_{\mathsf{e}_{0}}(t)) \cdot \prod_{\mu=1}^{\numberneighbors{e}{n}} x_{z_{\mathsf{e}_{\mu}}}(t) \cdot \prod_{\nu=1}^{\numberneighbors{i}{n}} y_{z_{\mathsf{i}_{\nu}}}(t) \label{eq:MFT_F},
\end{align}
\begin{align}
&y_{1}(t+1) =G \cdot {y_{0}(t)},  \label{eq:MFT_y1}\\ &G=\sum_{\forall\tilde{z}_{\mathsf{i}_{0}}(t)} W_{01}^{\mathsf{i_0}}(\tilde{z}_{\mathsf{i}_{0}}(t)) \cdot\prod_{\mu=1}^{\numberneighbors{e}{n}} x_{z_{\mathsf{e}_{\mu}}}(t) \cdot \prod_{\nu=1}^{\numberneighbors{i}{n}} y_{z_{\mathsf{i}_{\nu}}}(t) \label{eq:MFT_G},
\end{align}
where $x_1(t)$ ($y_1(t)$) represents the MF approximation to $\rho_{1}^{\mathsf{e}}(t)$ ($\rho_{1}^{\mathsf{i}}(t)$). These maps relate the probability that the representative neuron is active at time $t+1$ to its own state and the states of its neighbors at time $t$.

To complete the set of MF equations, the refractory periods must be taken into account. Once a neuron enters a refractory state, it progresses deterministically—advancing one step at a time—until it reaches the final refractory state. It then transitions to the quiescent state (0) in the subsequent time step
\begin{align}
	x_{m}(t+1)= & x_{m-1}(t),\ 2\leq m\leq\tau_{\mathsf{e}},\ t\geq1\nonumber           \\
	y_{n}(t+1)= & y_{n-1}(t),\ 2\leq n\leq\tau_{\mathsf{i}},\ t\geq1\label{eq:MFT_xiyi}
\end{align}
Thus, there are $\tau_{\mathsf{e}} + \tau_{\mathsf{i}}$ dynamical variables, each subject to specific constraints. Since $x_m(t)$ and $y_n(t)$ represent probabilities for each $m$ and $n$, they are confined to the interval $[0, 1]$ and must satisfy normalization conditions
\begin{align}
	 & \quad \quad \quad 0\leq x_{m}(t)\leq1,\ \forall t\ \mbox{ and } \forall m\in\mathcal{S}_{\mathsf{e}} , \nonumber    \\
	 & \quad \quad \quad 0\leq y_{n}(t)\leq1,\ \ \forall t\ \mbox{ and }  \forall n\in\mathcal{S}_{\mathsf{i}} ,\nonumber    \\
	 & \sum_{m=0}^{\arc{\tau}{e}{}{}{}}x_{m}(t)=1,\ \forall t                          \quad , \quad
	  \sum_{n=0}^{\arc{\tau}{i}{}{}{}}y_{n}(t)=1,\ \forall t
     \label{eq:probability_restriction}
\end{align}
With these restrictions we can define $x_{0}(t)$ and $y_{0}(t)$
as
\begin{align}
	x_{0}(t)=  1-\sum_{m=1}^{\tau_{\mathsf{e}}}x_{m}(t) \ , \ 
	y_{0}(t)=  1-\sum_{n=1}^{\tau_{\mathsf{i}}}y_{n}(t). 
    \label{eq:x0} 
\end{align}

Consequently, Eqs. \eqref{eq:MFT_x1}, \eqref{eq:MFT_y1}, and \eqref{eq:MFT_xiyi}, together with the constraints in \eqref{eq:probability_restriction}, form a set of coupled nonlinear dynamical maps that define the GCBM MF approximation. What remains is the computation of the transition probabilities $W_{01}^{\mathsf{n_0}}(\tilde{z}_{\mathsf{n}_{0}}(t))$. These depend on the specific motif under consideration, as will be demonstrated in the following section.

\subsection{Dictionary of Mean-Field Approximations}






In general, computing the MF equations requires evaluating the transition probabilities for all possible configurations of neighboring neurons. However, in this section, we will show that this is not necessary. Instead, it is sufficient to compute the transition probability for the case where all neighboring neurons are active, that is $\overline{W_{01}^{\mathsf{n_0}}}$ in Eq. \eqref{eq:Highest_W}. (Note we use $\mathsf{n_{0}}$ since we are referring to MF approximations.) Then,  to get from this transition probability to the MF equations it is necessary to apply a substitution rule, such that the transmission probabilities are multiplied by the probability that the corresponding neighboring neuron is active. 
This procedure establishes a direct dictionary between any network motif and its corresponding MF equations.

More explicitly, to derive the MF equations, Eqs.~\eqref{eq:MFT_x1} and~\eqref{eq:MFT_y1}, we begin by evaluating the term $\overline{W_{01}^{\mathsf{n_0}}}$ in Eq.~\eqref{eq:Highest_W}, which provides a compact starting expression.
The next step is to substitute the variables defined in Eqs.~\eqref{eq:E_m_term}, \eqref{eq:S_e_term}, \eqref{eq:I_m_term}, and~\eqref{eq:S_i_term}, followed by explicitly specifying the probabilities of the relevant events from Eqs.~\eqref{eq:Probability_events}.
This procedure yields the transition probability for the configuration in which all neighboring neurons are active, expressed in terms of the transmission probabilities. Although these are written with general indices, in the MF approximation we restrict to $\mathsf{e_0}$ and $\mathsf{i_0}$.
The final step is to apply the following substitution rules to $\overline{W_{01}^{\mathsf{e_0}}}$ and $\overline{W_{01}^{\mathsf{i_0}}}$ to obtain $F$ and $G$, respectively, 
\begin{align}
    &\overline{W_{01}^{\mathsf{e_0}}}(p_{s\mathsf{e}_{0}},\arc{P}{e}{\mu}{e}{0},\arc{P}{i}{\nu}{e}{0},\arcin{P}{e}{\mu}{e}{0}{\nu},\arcin{P}{}{}{e}{0}{\nu}, \ldots)\rightarrow F=\nonumber \\ &\overline{W_{01}^{\mathsf{e_0}}}(p_{s\mathsf{e}_{0}},\arc{P}{e}{\mu}{e}{0}x_{1}(t),\arc{P}{i}{\nu}{e}{0}y_{1}(t),   \arcin{P}{e}{\mu}{e}{0}{\nu}y_{1}(t),\arcin{P}{}{}{e}{0}{\nu}y_{1}(t),\ldots) \nonumber \\
    &\overline{W_{01}^{\mathsf{i_0}}}(p_{s\mathsf{i}_{0}},\arc{P}{e}{\mu}{i}{0},\arc{P}{i}{\nu}{i}{0}, \ldots)\rightarrow G=\nonumber \\ &\overline{W_{01}^{\mathsf{i_0}}}(p_{s\mathsf{i}_{0}},\arc{P}{e}{\mu}{i}{0}x_{1}(t),\arc{P}{i}{\nu}{i}{0}y_{1}(t),\ldots), \quad \quad \quad\forall\mathsf{e_{\mu},i_{\nu}}.\label{eq:substitutionrule}
\end{align}

To demonstrate the validity of these substitution rules, we first consider a simplified setting in which arc-arc interactions are neglected. The general case, including all arc-arc interaction terms, is presented in Appendix~\ref{App:Dictionary}.

Starting from the MF equations, and substituting the transition probabilities $W_{01}^{\mathsf{e_0}}(\tilde{z}_{\mathsf{e}_{0}}(t))$ and $W_{01}^{\mathsf{i_0}}(\tilde{z}_{\mathsf{i}_{0}}(t))$ using Eq.~\eqref{eq:transition_probability}, it can be shown that $F$ and $G$ can be expressed as 
\begin{align}
	F &=  \sum_{\forall\tilde{Z}_{\mathsf{e}_{0}}} U(\tilde{Z}_{\mathsf{e_0}})\cdot p_{s\mathsf{e}_{0}} \cdot  \sum_{\arc{z}{e}{1}{}{}(t)=0}^{\tau _{\mathsf{e}}
	}P(\arc{Z}{e}{1}{e}{0}\mid \arc{z}{e}{1}{}{}(t))\cdot \nonumber \\& x_{\arc{z}{e}{1}{}{}(t)} \cdots \sum_{\arc{z}{i}{1}{}{}(t)=0}^{\tau _{\mathsf{i}}
	}P(\arc{Z}{i}{1}{e}{0}\mid \arc{z}{i}{1}{}{}(t))\cdot y_{\arc{z}{i}{1}{}{}(t)} \cdots,\label{eq:MFT_F_arranged}
\end{align}
\begin{align}
	G&=  \sum_{\forall\tilde{Z}_{\mathsf{i}_{0}}} U(\tilde{Z}_{\mathsf{i_0}})\cdot p_{s\mathsf{i}_{0}} \cdot  \sum_{\arc{z}{e}{1}{}{}(t)=0}^{\tau _{\mathsf{e}}
	}P(\arc{Z}{e}{1}{i}{0}\mid \arc{z}{e}{1}{}{}(t))\cdot  \nonumber \\& x_{\arc{z}{e}{1}{}{}(t)} \cdots\sum_{\arc{z}{i}{1}{}{}(t)=0}^{\tau _{\mathsf{i}}
	}P(\arc{Z}{i}{1}{i}{0}\mid \arc{z}{i}{1}{}{}(t))\cdot y_{\arc{z}{i}{1}{}{}(t)} \cdots .\label{eq:MFT_G_arranged}
\end{align}

After some algebraic manipulations, and using the arc activation functions in Eq.~\eqref{eq:ArcProbabilityFxnSite}, the functions $F$ and $G$ can be written as
\begin{align}
	F &=  \sum_{\forall\tilde{Z}_{\mathsf{e}_{0}}} U(\tilde{Z}_{\mathsf{e_0}})\cdot  p_{s\mathsf{e}_{0}} \cdot \prod _{\mu=1}^{\numberneighbors{e}{e}}\big ( (1-\arc{P}{e}{\mu}{e}{0}x_{1}(t))\delta_{0,\arc{Z}{e}{\mu}{e}{0}}+\nonumber \\&\arc{P}{e}{\mu}{e}{0}x_{1}(t)\delta_{1,\arc{Z}{e}{\mu}{e}{0}}\big ) \cdot \prod _{\nu=1}^{\numberneighbors{i}{e}}\big ( (1-\arc{P}{i}{\nu}{e}{0}y_{1}(t))\delta_{0,\arc{Z}{i}{\nu}{e}{0}}+\nonumber \\&\arc{P}{i}{\nu}{e}{0}y_{1}(t)\delta_{1,\arc{Z}{i}{\nu}{e}{0}}\big ) ,\label{eq:MFT_F_explicit}
\end{align}
\begin{align}
	G&=  \sum_{\forall\tilde{Z}_{\mathsf{i}_{0}}} U(\tilde{Z}_{\mathsf{i_0}})\cdot p_{s\mathsf{i}_{0}} \cdot  \prod _{\mu=1}^{\numberneighbors{e}{i}}\big ( (1-\arc{P}{e}{\mu}{i}{0}x_{1}(t))\delta_{0,\arc{Z}{e}{\mu}{i}{0}}+\nonumber \\&\arc{P}{e}{\mu}{i}{0}x_{1}(t)\delta_{1,\arc{Z}{e}{\mu}{i}{0}}\big ) \cdot \prod _{\nu=1}^{\numberneighbors{i}{i}}\big ( (1-\arc{P}{i}{\nu}{i}{0}y_{1}(t))\delta_{0,\arc{Z}{i}{\nu}{i}{0}}+\nonumber \\&\arc{P}{i}{\nu}{i}{0}y_{1}(t)\delta_{1,\arc{Z}{i}{\nu}{i}{0}}\big ) .\label{eq:MFT_G_explicit}
\end{align}

Starting with Eq. \eqref{eq:transition_probability} one can write the cases when all the neighbors are active $\overline{W_{01}^{\mathsf{e_0}}}$ and $\overline{W_{01}^{\mathsf{i_0}}}$ as
\begin{align}
	\overline{W_{01}^{\mathsf{e_0}}} &=  \sum_{\forall\tilde{Z}_{\mathsf{e}_{0}}} U(\tilde{Z}_{\mathsf{e_0}})\cdot p_{s\mathsf{e}_{0}} \cdot  \prod _{\mu=1}^{\numberneighbors{e}{e}}\big ( (1-\arc{P}{e}{\mu}{e}{0})\delta_{0,\arc{Z}{e}{\mu}{e}{0}}+\nonumber \\&\arc{P}{e}{\mu}{e}{0}\delta_{1,\arc{Z}{e}{\mu}{e}{0}}\big ) \cdot \prod _{\nu=1}^{\numberneighbors{i}{e}}\big ( (1-\arc{P}{i}{\nu}{e}{0})\delta_{0,\arc{Z}{i}{\nu}{e}{0}}+\nonumber \\&\arc{P}{i}{\nu}{e}{0}\delta_{1,\arc{Z}{i}{\nu}{e}{0}}\big ) ,\label{eq:MFT_E_transition}
\end{align}
\begin{align}
	\overline{W_{01}^{\mathsf{i_0}}}&=  \sum_{\forall\tilde{Z}_{\mathsf{i}_{0}}} U(\tilde{Z}_{\mathsf{i_0}})\cdot p_{s\mathsf{i}_{0}} \cdot \prod _{\mu=1}^{\numberneighbors{e}{i}}\big ( (1-\arc{P}{e}{\mu}{i}{0})\delta_{0,\arc{Z}{e}{\mu}{i}{0}}+\nonumber \\&\arc{P}{e}{\mu}{i}{0}\delta_{1,\arc{Z}{e}{\mu}{i}{0}}\big ) \cdot \prod _{\nu=1}^{\numberneighbors{i}{i}}\big ( (1-\arc{P}{i}{\nu}{i}{0})\delta_{0,\arc{Z}{i}{\nu}{i}{0}}+\nonumber \\&\arc{P}{i}{\nu}{i}{0}\delta_{1,\arc{Z}{i}{\nu}{i}{0}}\big ) ,\label{eq:MFT_I_transition}
\end{align}
expressions that differ from $F$ and $G$ by the fact that, in the latter, each transmission probability is multiplied by the probability that the corresponding neighboring neuron is active. This means that Eq.~\eqref{eq:Highest_W} together with the substitution rules in Eq.~\eqref{eq:substitutionrule} provides a complete dictionary for constructing the MF equations associated with each network motif. 
An explicit example illustrating this methodology is presented in the following section.

\section{Example Motifs with Probability of Activation Computation}
\label{section4}

\begin{figure}[htbp]
\includegraphics[width=8cm]{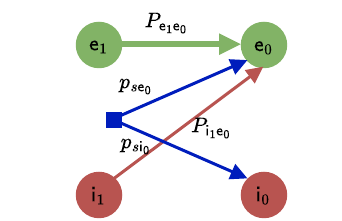}
\caption{
This motif has site inhibition with a number of excitatory neighboring neurons $\numberneighbors{e}{e}$=1, a number of inhibitory neighboring neurons $\numberneighbors{i}{e}=1$, and no neighbors for the target inhibitory neuron $\arc{z}{i}{0}{}{}$. In this figure, $\mathsf{e}_1$ and $\mathsf{i}_1$ represent $\mathsf{\bar e}$ and $\mathsf{\bar {i}}$ the mean neighborhood shown in Fig. \ref{fig:motif}. }
 \label{fig:simplemotif}
\end{figure}

For pedagogical reasons, we start with the simple motif shown in Fig.~\ref{fig:simplemotif} which has $\numberneighbors{e}{e}=\numberneighbors{i}{e}=1$, and $\numberneighbors{e}{i}=\numberneighbors{i}{i}=0$.
For this example, we set $\arc{\tau}{e}{}{}{} = \arc{\tau}{i}{}{}{} = 2$, which yields $(1+1)^2 = 4$ possible configurations for the motif.
For this example, $\mathsf{e}_1$ and $\mathsf{i}_1$ represent $\mathsf{\bar e}$ and $\mathsf{\bar {i}}$ the mean neighborhood shown in Fig. \ref{fig:motif}.

The first transition probability function, $W^{\sf{n_0}}_{01}(\tilde{z}_{\mathsf{n}_{0}}(t))$,  we will compute corresponds to the case where $\tilde{z}_{\mathsf{e_{1}}}=1,\tilde{z}_{\mathsf{i_{1}}}=1$; in other words, it is $\overline{W_{01}^{\mathsf{n_0}}}=W_{01}^{\sf{n_0}}((1;1))$.  Here, we assume $p_{s\mathsf{e}_{\mu}}=p_{s\mathsf{i}_{\nu}}=p_{s}$. We start computing the transition probability $\overline{W_{01}^{\mathsf{e_0}}}$ using Eq.~\eqref{eq:Highest_W}
\begin{align}
    \label{eq:W01eHighest_Motif1}
     \overline{W_{01}^{\mathsf{e_0}}}=
     & \, E_1^{\sf e_0} I_{[0]}^{\sf e_0}+E_2^{\sf e_0} I_{[1]}^{\sf e_0} \\
     =& (p_{s}+(1-p_s)\arc{P}{e}{1}{e}{0})(1-\arc{P}{i}{1}{e}{0})+    p_{s} \arc{P}{e}{1}{e}{0}  \arc{P}{i}{1}{e}{0} . \nonumber
\end{align}
Similarly, 
\begin{align}
    \label{eq:W01iHighest_Motif1}
	\overline{W_{01}^{\mathsf{i_{0}}}} & =E_{1}^{\sf i_0}
	I_{[0]}^{\sf i_0}=p_s         .                            
\end{align}
In Appendix~\ref{App:W11equations}, we provide detailed steps showing how to compute $\overline{W_{01}^{\mathsf{e_0}}}$ and $\overline{W_{01}^{\mathsf{i_0}}}$. 

Next, we compute $F$ and $G$ by applying the substitution rule, Eqs. \eqref{eq:substitutionrule}, to $\overline{W_{01}^{\mathsf{n_0}}}$. The resulting dynamical map $x_{1,2}(t), y_{1,2}(t)$ becomes
\begin{eqnarray}
    x_{1}(t+1) &=& x_0(t)((p_{s}+(1 \!-\!p_{s})\arc{P}{e}{1}{e}{0} x_1(t))(1 \!-\!\arc{P}{i}{1}{e}{0} y_1(t))\nonumber \\&& \quad \ \ \ \ \ \ + p_{s}\arc{P}{e}{1}{e}{0} x_1(t)\arc{P}{i}{1}{e}{0} y_1(t)) \nonumber \\
    x_{2}(t+1)&=& x_1(t) \nonumber \\
    y_1(t+1)&=&y_0(t) \ p_{s} \nonumber \\
    y_2(t+1)&=&y_1(t).
    \label{eq:MFmaps_Motif1}
\end{eqnarray}
The validity of the substitution rule for this example is demonstrated explicitly in Appendix~\ref{App:motif1x1y1}, where we show how to compute {\it all} transition probabilities $W^{\sf{n_0}}_{01}(\tilde{z}_{\mathsf{n}_{0}}(t))$. 

\begin{figure}
    \centering
    \includegraphics[width=8cm]{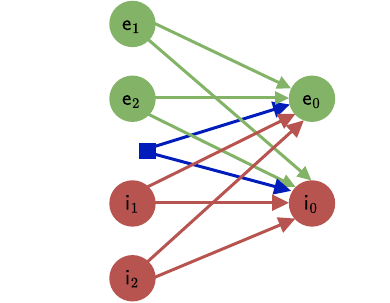}
    \caption{Representative motif selected for in-depth analysis. 
        The network motif consists of a balanced configuration with 
        $\numberneighbors{e}{e} = \numberneighbors{e}{i} = \numberneighbors{i}{e} = \numberneighbors{i}{i} = 2$. 
    In this figure, ($\mathsf{e}_1$, $\mathsf{e}_2$) and ($\mathsf{i}_1$, $\mathsf{i}_2$) represent $\mathsf{\bar e}$ and $\mathsf{\bar {i}}$ the mean neighborhood shown in Fig. \ref{fig:motif}.}
    \label{fig:motif2}
\end{figure}

Now that we have established how the transition probabilities and maps are established, we turn to the more representative motif shown in Fig.~\ref{fig:motif2}, which will be used in subsequent sections to describe the solutions of the GCBM MF equations. We will apply the same procedure to compute the maps in Eqs.~\eqref{eq:MFT_x1} and \eqref{eq:MFT_y1} for this motif.
The representative motif is chosen such that $\numberneighbors{e}{e}=\numberneighbors{i}{e}=\numberneighbors{e}{i}=\numberneighbors{i}{i}=2$. We refrain from assigning fixed values to $\arc{\tau}{e}{}{}{}$ and $\arc{\tau}{i}{}{}{}$, as these parameters will be varied in subsequent examples.
For this example, $(\mathsf{e}_1,\mathsf{e}_2)$ and $(\mathsf{i}_1,\mathsf{i}_2)$ represent $\mathsf{\bar e}$ and $\mathsf{\bar i}$, the mean neighborhoods shown in Fig.~\ref{fig:motif}.
Starting with computing the transition probability when all neighbors are active $\overline{W_{01}^{\mathsf{n_0}}}=W_{01}^{\sf{n_0}}((1,1;1,1))$ 
\begin{eqnarray}
    \label{motif07_We}
   \overline{W_{01}^{\mathsf{e_0}}}&=&E_1^{\sf e_0} I_{[0]}^{\sf e_0}+E_2^{\sf e_0} I_{[1]}^{\sf e_0}+ E_3^{\sf e_0}I_{[2]}^{\sf e_0} \\&
   =&(p_{s}+(1-p_s)(\arc{P}{e}{1}{e}{0}+\arc{P}{e}{2}{e}{0}-\arc{P}{e}{1}{e}{0} \arc{P}{e}{2}{e}{0}) )\nonumber \\ &&    
 ((1-\arc{P}{i}{1}{e}{0}) (1-\arc{P}{i}{2}{e}{0}) ) + \nonumber \\&&
 ( p_{s} (\arc{P}{e}{1}{e}{0} +\arc{P}{e}{2}{e}{0})+ (1 -2 p_{s}) \arc{P}{e}{1}{e}{0} \arc{P}{e}{2}{e}{0}) \nonumber \\ &&(\arc{P}{i}{1}{e}{0} \!+\!\arc{P}{i}{2}{e}{0}\!-\!2\arc{P}{i}{1}{e}{0} \arc{P}{i}{2}{e}{0} )
  \! + \! p_{s} \arc{P}{e}{1}{e}{0} \arc{P}{e}{2}{e}{0} \arc{P}{i}{1}{e}{0} \arc{P}{i}{2}{e}{0} . \nonumber
\end{eqnarray}
\begin{eqnarray}
    \label{motif07_Wi}
   \overline{W_{01}^{\mathsf{i_0}}}&=&E_1^{\sf i_0} I_{[0]}^{\sf i_0}+E_2^{\sf i_0} I_{[1]}^{\sf i_0}+ E_3^{\sf i_0}I_{[2]}^{\sf i_0} \\&
   =&(p_{s}+(1-p_s)(\arc{P}{e}{1}{i}{0}+\arc{P}{e}{2}{i}{0}-\arc{P}{e}{1}{i}{0} \arc{P}{e}{2}{i}{0}) )\nonumber \\ &&    ((1-\arc{P}{i}{1}{i}{0}) (1-\arc{P}{i}{2}{i}{0}) ) + \nonumber \\&&
 ( p_{s} (\arc{P}{e}{1}{i}{0} +\arc{P}{e}{2}{i}{0})+ (1 -2 p_{s}) \arc{P}{e}{1}{i}{0} \arc{P}{e}{2}{i}{0}) \nonumber \\ &&(\arc{P}{i}{1}{i}{0} \!+\!\arc{P}{i}{2}{i}{0}\!-\!2\arc{P}{i}{1}{i}{0} \arc{P}{i}{2}{i}{0} )
  \! + \! p_{s} \arc{P}{e}{1}{i}{0} \arc{P}{e}{2}{i}{0} \arc{P}{i}{1}{i}{0} \arc{P}{i}{2}{i}{0} . \nonumber
\end{eqnarray}
The steps for computing $\overline{W_{01}^{\mathsf{n_0}}}$ are ommitted due to the extensive length. 
Applying the substitution rule, Eq.~\eqref{eq:substitutionrule}, to the equation above leads to the dynamical map expressed as polynomial sums
\begin{eqnarray}
x_{1}(t+1)&=&x_{0}(t)  \sum_{c=0,d=0}^{2} a_{cd} \, x_{1}(t)^{c} y_{1}(t)^{d} ,\nonumber \\
   x_m(t+1) &=& x_{m-1}(t) , \quad  \mbox{ for } 2 \le m \le \tau_{\sf e}
 \label{dynamicmapx}
\end{eqnarray}
and 
\begin{eqnarray}
 y_{1}(t+1)&=&y_{0}(t)  \sum_{c=0,d=0}^{2} b_{cd} \, x_{1}(t)^{c} y_{1}(t)^{d} , \nonumber \\
   y_n(t+1) &=& y_{n-1}(t) , \quad  \mbox{ for } 2 \le n \le \tau_{\sf i}
 \label{dynamicmapy}
\end{eqnarray}
where the coefficients $a_{cd}$ and $b_{cd}$ are given explicitly in Appendix \ref{App:motif2x1y1}. The substitution rule was confirmed to be true, but it is omitted due to its length. The following section investigates this motif's dynamical equations in depth. 
Comparisons between the time series obtained from the MF and many-body GCBM simulations are presented in Appendix \ref{App:ValidateMF}.

\section{How Inhibition affects the dynamics}
\label{section5}

The role of inhibitory neurons in shaping brain dynamics is an active field of research \cite{Hattori_Kuchibhotla_Froemke_Komiyama_2017, Importance_of_Inhbitory_Neurons}.
Here, we investigate whether inhibitory neurons fundamentally alter the dynamics of our model or exert only a modulatory effect.
If the system is quasi-critical, we expect its core features to persist with some modification.
Using Eqs. \eqref{eq:MFT_x1} and \eqref{eq:MFT_y1}, we construct phase diagrams to examine how inhibitory parameters influence stability of the various phases.
While inhibition does not introduce qualitative changes, it expands the model’s stable region.
As inhibitory strength increases, the system becomes more stable, average activity decreases, and dynamical susceptibility is reduced.
In the unstable regime, we observe chaotic dynamics, which inhibition can suppress.
The model follows a distinctive route to chaos, characterized by marginally stable orbits, period doubling, and occasional seemingly intermittent bursts of chaos.
When $y_1$ acts as an external periodic input, strong inhibition induces unstable periodic behavior in excitatory neurons with the same period but a seemingly $-\pi/2$ phase shift.

\subsection{Establishing Phase Diagrams}
We begin by computing the Jacobian of the dynamical map (MF equations), at the fixed points, and by analyzing its eigenvalues.
A fixed point is stable if all eigenvalues have magnitude less than one, and unstable if any eigenvalue exceeds one in magnitude.
The Jacobian terms are defined as 
\begin{align*}
    J_{ij}=\frac{\partial h_i (t+1)}{\partial h_j (t)} ,
\end{align*}
where $h_i$ and $h_j \in  \{h \mid h =(x_1,\ldots,x_{\tau_{\mathsf{e}}},y_1,\ldots,y_{\tau_{\mathsf{i}}})\}$.

The fixed points of $x_{1}(t)$, $y_{1}(t)$, and Eqs.~\eqref{eq:x0} are computed with the system of equations
\begin{eqnarray}
  x_1^*&=&(1-\tau_{\sf {e}}x_1^*)\ F[ x_1^*, y_1^*,p_{s},...] , \nonumber \\
   y_1^*&=&(1-\tau_{\sf {i}}y_1^*)\ G[ x_1^*, y_1^*,p_{s},...] ,
   \label{fixed-pointeq}
\end{eqnarray}
Note that $1-\sum_{i=1}^{\tau_{\sf {n}}}h_{i}^{*} \rightarrow (1-\tau_{\sf n} h_{1}^{*})$ since $h_1^*=h_{i}^{*}$ for a fixed point, so should be constant.
An example Jacobian $\mathbb{J}$ for $\tau_{\sf{e}}=3$ and $\tau_{\sf{i}}=3$, which has the same form for every motif, is represented by
\begin{widetext}
\begin{equation}
\mathbb{J} = \left(\begin{array}{cccccc}
    -F+\frac{\partial F}{\partial x_1}\cdot x_{0}(t) & -F & -F & \frac{\partial F}{\partial y_1}\cdot x_{0}(t) & 0 & 0\\
    1 & 0 & 0 & 0 & 0 & 0\\
    0 & 1 & 0 & 0 & 0 & 0\\
    \frac{\partial G}{\partial x_1}\cdot y_{0}(t) & 0 & 0 & -G+\frac{\partial G}{\partial y_1}\cdot y_{0}(t) & -G & -G\\
    0 & 0 & 0 & 1 & 0 & 0\\
    0 & 0 & 0 & 0 & 1 & 0 \\
\end{array}\right) \ .
\label{eq:Jacobian-generalForm-Example}
\end{equation}

\end{widetext}

We will use the motif from Fig.~\ref{fig:motif2} to draw our phase diagrams. We are choosing that inhibitor neurons always inhibit.

\subsection{Nonequilibrium Phase Diagram and the Widom Line}

\begin{figure}
    \includegraphics[width=0.95\linewidth]{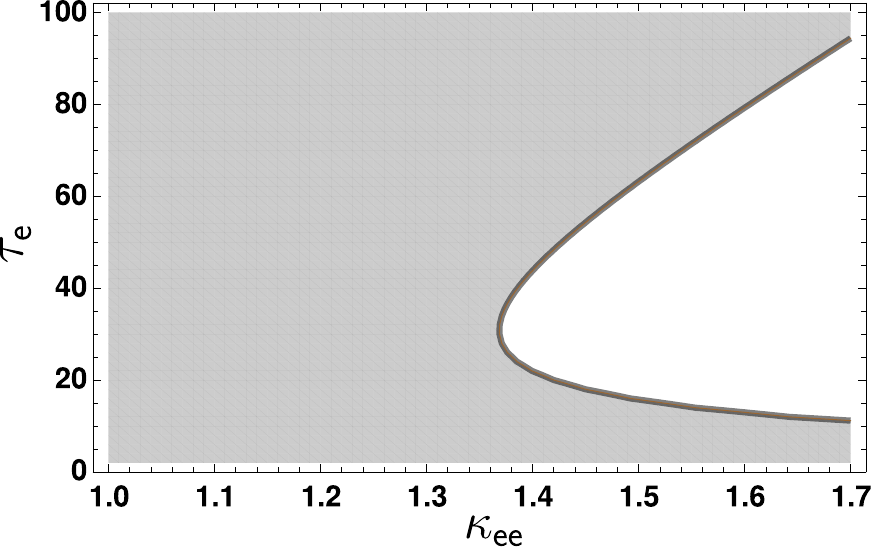}
    \caption{
        Stability of solutions in the representative motif.
        A cross-section of the phase diagram showing the stability of the solutions of the excitatory activity $x_{1}(t)$ and the inhibitory activity $y_{1}(t)$ for the representative motif in Fig.~\ref{fig:motif2}. 
        The x-axis represents the coupling strength between excitatory neurons ($\arc{\kappa}{e}{}{e}{}$), and the y-axis represents the refractory period of excitatory neurons ($\arc{\tau}{e}{}{}{}$). 
        The gray region corresponds to a stable nonzero fixed point, while the white region indicates an oscillatory phase with unstable fixed points. 
        The solid black line marks the stability phase boundary, which is non-analytical. 
        The remaining parameters are $\arc{\kappa}{ii}{}{}{} =\arc{\kappa}{ie}{}{}{} = \arc{\kappa}{ei}{}{}{}=1.6$,  $p_{s} = 10^{-3}$, $\arc{\tau}{i}{}{}{} = 5$, and
        $\numberneighbors{e}{e} =  
        \numberneighbors{e}{i} = 
        \numberneighbors{i}{e} =  
        \numberneighbors{i}{i} = 2$.
    }
    \label{fig:Representative_Phase_Diagram}
\end{figure}

Our phase diagram is a graphical representation that shows how the state of our dynamical system changes as a function of parameters such as $\arc{\kappa}{\mathsf{e}}{}{ \mathsf{e}}{}$, $\arc{\kappa}{\mathsf{e}}{}{ \mathsf{i}}{}$, $\arc{\kappa}{\mathsf{i}}{}{ \mathsf{i}}{}$, $\arc{\kappa}{\mathsf{i}}{}{ \mathsf{e}}{}$, $\arc{\tau}{\mathsf{e}}{}{}{}$, $\arc{\tau}{\mathsf{i}}{}{}{}$, $p_s$, and the neighborhood sizes $\numberneighbors{e}{e}$, $\numberneighbors{i}{e}$, and $\numberneighbors{e}{i}$, $\numberneighbors{i}{i}$. For clarity and comparison with the purely excitatory model, we focus on a two-dimensional cross-section of this space, using $\kappa_{\mathsf{ee}}$ on the x-axis (excitatory-to-excitatory coupling strength) and $\tau_{\mathsf{e}}$ on the y-axis (excitatory refractory period).
To explore how inhibition affects the dynamics, we vary the inhibitory parameters $\arc{\kappa}{ie}{}{}{}$, $\arc{\kappa}{ei}{}{}{}$, $\arc{\kappa}{ii}{}{}{}$, $\arc{\tau}{\mathsf{i}}{}{}{}$, and $p_s$, allowing us to examine their role in modulating stability and their relationship to the quasi-critical hypothesis.

We select parameters within the physical regime by ensuring that all probabilities are well-defined, with the upper bound given in Eq. \eqref{kappamax}.
Using the representative motif in Fig. \ref{fig:motif2}, we solve Eq. \eqref{fixed-pointeq} to find fixed points and assess their stability.

The representative phase diagram in Fig. \ref{fig:Representative_Phase_Diagram} shows the stability of the solutions to Eqs. \eqref{dynamicmapx} and  \eqref{dynamicmapy} for the motif in Fig.~\ref{fig:motif2}.
It reveals a phase transition across a boundary separating stable and unstable fixed points. 
While qualitatively similar to the purely excitatory case, the stability boundary is shifted rightward, toward higher values of $\arc{\kappa}{ee}{}{}{}$, indicating that inhibition expands the stable region of the phase space.
Inhibitory neurons thus serve as regulators, enhancing stability and allowing a broader range of configurations to remain dynamically stable. 
Varying the inhibitory parameters confirms that, in most cases, stronger inhibition shifts the stability boundary toward higher $\arc{\kappa}{ee}{}{}{}$ values and reduces the mean excitatory activity.
This effect is more pronounced at lower $\arc{\tau}{\mathsf{e}}{}{}{}$ and higher $\arc{\kappa}{ee}{}{}{}$ values.
An exception occurs at low refractory periods (e.g., $\arc{\tau}{\mathsf{e}}{}{}{} = 2$), where increasing inhibition can shift the stable region leftward to lower $\arc{\kappa}{ee}{}{}{}$ values, even as excitatory activity continues to decline.
\begin{figure}[htb]
     \includegraphics[width=0.95\linewidth]{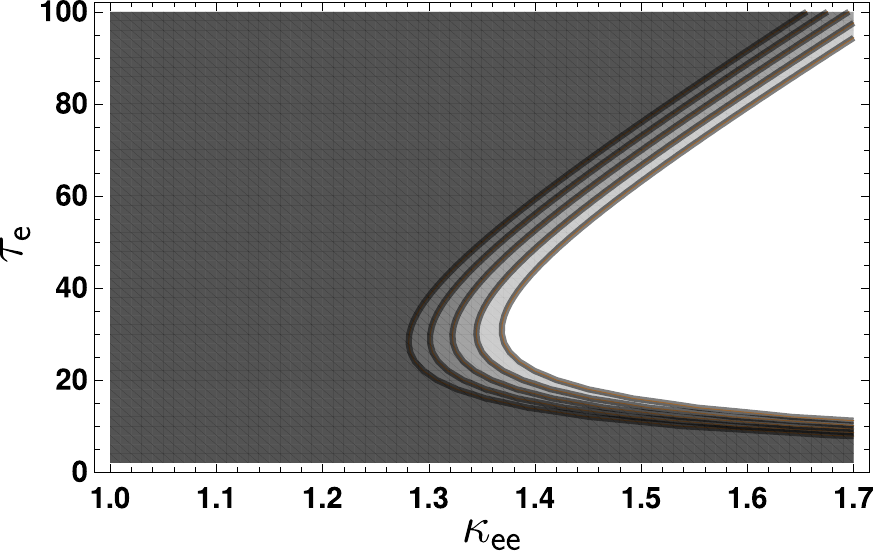}
        \caption{
        Effect of inhibitory-to-excitatory coupling on phase stability.
        A cross-section of the phase diagram showing the stability of the solutions of the excitatory activity $x_{1}(t)$ and the inhibitory activity $y_{1}(t)$ for the representative motif in Fig.~\ref{fig:motif2}, as the coupling strength from inhibitory to excitatory neurons ($\arc{\kappa}{ie}{}{}{}$) increases from 0 to 1.6 in steps of .4 (left to right, darker to lighter gray). 
        The gray regions indicate stable fixed-point phases, while the white region corresponds to an oscillatory phase with unstable fixed points. 
        The x-axis represents the coupling strength between excitatory neurons ($\arc{\kappa}{e}{}{e}{}$), and the y-axis represents the refractory period of excitatory neurons ($\arc{\tau}{e}{}{}{}$). 
        Solid black lines denote stability phase boundaries. 
        The other parameters are $\arc{\kappa}{ii}{}{}{} =\arc{\kappa}{ei}{}{}{} = 1.6$, $p_{s} = 10^{-3}$, $\arc{\tau}{i}{}{}{} = 5$, and 
        $\numberneighbors{e}{e} =  
        \numberneighbors{e}{i} = 
        \numberneighbors{i}{e} =  
        \numberneighbors{i}{i} = 2$.
    }
        \label{fig:kie}

\end{figure}

We begin by exploring how varying $\arc{\kappa}{ie}{}{}{}$ affects the phase diagram, as shown in Fig.~\ref{fig:kie}.
As $\arc{\kappa}{ie}{}{}{}$ increases, the stability boundary shifts toward higher $\arc{\kappa}{ee}{}{}{}$, indicating increased stability. 
Stronger inhibition reduces excitatory activity, especially at low $\arc{\tau}{\mathsf{e}}{}{}{}$, expanding the stable region.
On average, higher $\arc{\kappa}{ie}{}{}{}$ reduces inhibitory activity due to fewer active excitatory neurons.
An exception occurs at $\arc{\tau}{\mathsf{e}}{}{}{} = 2$, where the boundary shifts leftward, though excitatory activity still decreases.

Importantly, shifts in the phase transition boundaries only occur when both $\arc{\kappa}{ei}{}{}{} \neq 0$ and $\arc{\kappa}{ie}{}{}{} \neq 0$.
If either is zero, the Jacobian's upper-left and lower-right submatrices decouple, preventing inhibitory-excitatory interaction and eliminating the effect on stability.

\begin{figure}[htb]
\includegraphics[width=0.95\linewidth]{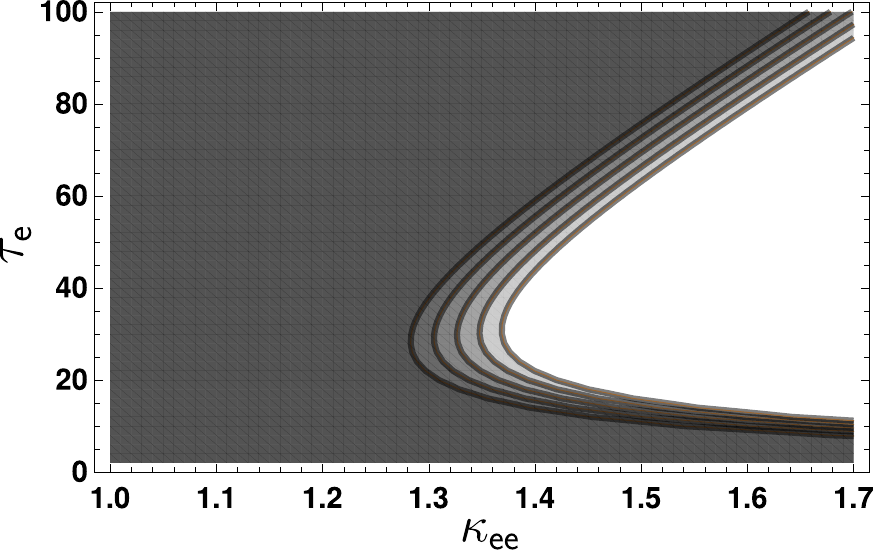}
\caption{
    Effect of excitatory-to-inhibitory coupling on phase stability.
    A cross-section of the phase diagram showing the stability of the solutions of the excitatory activity $x_{1}(t)$ and the inhibitory activity $y_{1}(t)$ for the representative motif in Fig.~\ref{fig:motif2}, as the coupling strength from excitatory to inhibitory neurons ($\arc{\kappa}{ei}{}{}{}$) increases from 0 to 1.6 in steps of .4 (left to right, darker to lighter gray). 
    The gray regions correspond to stable fixed-point phases, while the white region indicates an oscillatory phase with unstable fixed points. 
    The x-axis represents the coupling strength between excitatory neurons ($\arc{\kappa}{e}{}{e}{}$), and the y-axis represents the refractory period of excitatory neurons ($\arc{\tau}{e}{}{}{}$). 
    Solid black lines denote the stability phase boundaries. 
    The other parameters are $\arc{\kappa}{ii}{}{}{} =\arc{\kappa}{ie}{}{}{} = 1.6$, $p_{s} = 10^{-3}$, $\arc{\tau}{i}{}{}{} = 5$, and 
    $\numberneighbors{e}{e} =  
        \numberneighbors{e}{i} = 
        \numberneighbors{i}{e} =  
        \numberneighbors{i}{i} = 2$.
}

         \label{fig:kei}
\end{figure}

Next, we examine how varying $\arc{\kappa}{ei}{}{}{}$ affects the phase diagram, shown in Fig.~\ref{fig:kei}. 
The diagrams resemble those for varying $\arc{\kappa}{ie}{}{}{}$, but changes in $\arc{\kappa}{ei}{}{}{}$ shift the stability boundary less to the right for values up to about 1.6, where the shifts become comparable.
Increasing $\arc{\kappa}{ei}{}{}{}$ expands the stable region by enhancing excitatory input to inhibitory neurons, thereby increasing inhibitory activity and suppressing excitatory activity.
This increase in average inhibitory activity produces a more pronounced and wider shift across $\arc{\tau}{\mathsf{e}}{}{}{}$ values compared to varying $\arc{\kappa}{ie}{}{}{}$.
At $\arc{\tau}{\mathsf{e}}{}{}{} = 2$, the stability boundary shifts leftward with decreased excitatory activity.
Fixing either $\arc{\kappa}{ie}{}{}{}$ or $\arc{\kappa}{ei}{}{}{}$ at a larger value (here, 1.6) amplifies the effect of increasing the other parameter on the phase transition boundary.

\begin{figure}[htb]
\includegraphics[width=.95\columnwidth]{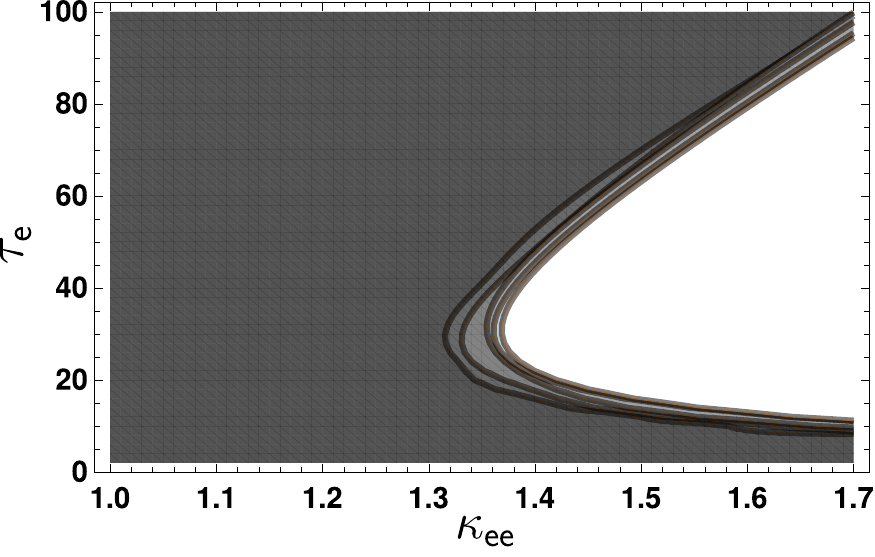}
\caption{
    Effect of inhibitory refractory period on phase stability.
    A cross-section of the phase diagram showing the stability of the solutions of the excitatory activity $x_{1}(t)$ and the inhibitory activity $y_{1}(t)$ for the representative motif in Fig.~\ref{fig:motif2}, as the inhibitory refractory period $\arc{\tau}{i}{}{}{}$ decreases from 100, 50, 25, 10, to 5 (left to right, darker to lighter gray). 
    The gray regions indicate stable fixed-point phases, while the white region corresponds to an oscillatory phase with unstable fixed points. 
    The x-axis represents the coupling strength between excitatory neurons ($\arc{\kappa}{e}{}{e}{}$), and the y-axis represents the refractory period of excitatory neurons ($\arc{\tau}{e}{}{}{}$). 
    Solid black lines denote the stability phase boundaries. 
    The other parameters are $\arc{\kappa}{ii}{}{}{} =\arc{\kappa}{ie}{}{}{} = \arc{\kappa}{ei}{}{}{} = 1.6$, $p_{s} = 10^{-3}$, and
    $\numberneighbors{e}{e} =  
        \numberneighbors{e}{i} = 
        \numberneighbors{i}{e} =  
        \numberneighbors{i}{i} = 2$.
}

    \label{fig:taui} 
\end{figure}

As $\arc{\tau}{\mathsf{i}}{}{}{}$ increases, the phase diagram shifts left toward lower $\arc{\kappa}{ee}{}{}{}$ values, indicating reduced stability Fig.~\ref{fig:taui}.
This occurs because a longer inhibitory refractory period reduces the number of active inhibitory neurons, leading to increased excitation and diminished inhibitory regulation, which limits stability.
An exception appears at $\arc{\tau}{\mathsf{e}}{}{}{} = 2$, where stability shifts rightward toward higher $\arc{\kappa}{ee}{}{}{}$ values, reflecting increased stability.
For the inhibitory self-coupling parameter $\arc{\kappa}{ii}{}{}{}$, values of $\arc{\tau}{\mathsf{e}}{}{}{} > 3$ show minimal impact on the stability boundary, so the corresponding plot is omitted.
However, at $\arc{\tau}{\mathsf{e}}{}{}{} = 2,3$, increasing $\arc{\kappa}{ii}{}{}{}$ shifts the stable region rightward, enhancing stability. 
This is accompanied by a slight increase in average excitation and a decrease in average inhibition.
Overall, increasing inhibitory effectiveness generally expands the stable region of the phase diagram.

\begin{figure}[htb]
\includegraphics[width=0.95\linewidth]{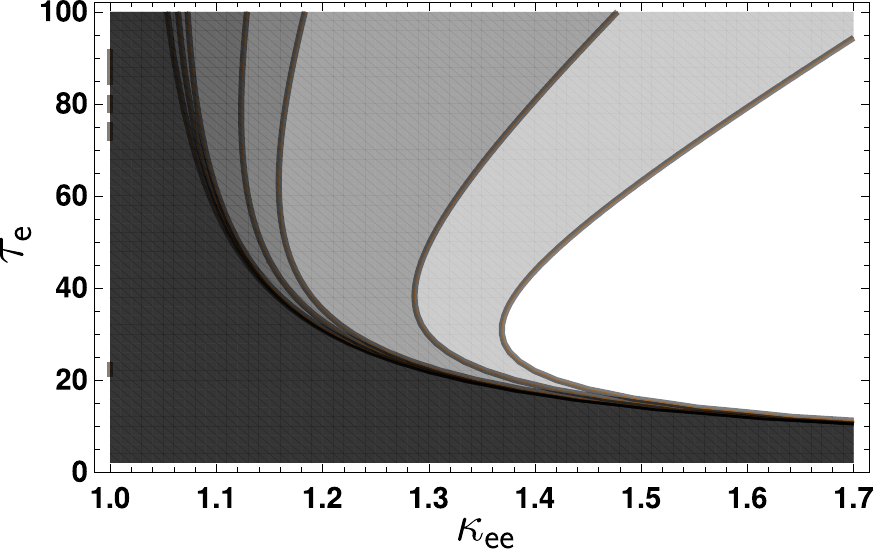}
\caption{
    Effect of spontaneous activation probability $p_{s}$ on phase stability. A cross-section of the phase diagram showing the stability of the solutions of the excitatory activity $x_{1}(t)$ and the inhibitory activity $y_{1}(t)$ for the representative motif in Fig.~\ref{fig:motif2}, as $p_{s}$ increases uniformly from left to right: $0$, $0.5 \times 10^{-5}$, and $1 \times 10^{-3}$ (darker to lighter gray). 
    The gray regions represent stable fixed-point phases, while the white region corresponds to an oscillatory phase with unstable fixed points. 
    The x-axis denotes the coupling strength between excitatory neurons ($\arc{\kappa}{e}{}{e}{}$), and the y-axis denotes the refractory period of excitatory neurons ($\arc{\tau}{e}{}{}{}$). 
    Solid black lines indicate the stability phase boundaries. 
    The other parameters are $\arc{\kappa}{ii}{}{}{} = \arc{\kappa}{ie}{}{}{} = \arc{\kappa}{ei}{}{}{} = 1.6$, $\arc{\tau}{i}{}{}{} = 5$, and 
    $\numberneighbors{e}{e} =  
        \numberneighbors{e}{i} = 
        \numberneighbors{i}{e} =  
        \numberneighbors{i}{i} = 2$.
}

    \label{fig:ps} 
\end{figure}

For completeness, we show in Fig.~\ref{fig:ps} that increasing $p_{s}$ shifts the stability boundary toward higher values of $\arc{\kappa}{ee}{}{}{}$, particularly at larger $\arc{\tau}{\mathsf{e}}{}{}{}$.
At low $\arc{\kappa}{ee}{}{}{}$ values, the stability boundary remains unchanged. 
This indicates that higher $p_{s}$ is needed to maintain stability as $\arc{\tau}{\mathsf{e}}{}{}{}$ increases, consistent with observations in the purely excitatory case \cite{williams2014quasicritical}. 
Additionally, increasing $p_{s}$ elevates both excitatory and inhibitory activity levels.

For Eqs. \eqref{dynamicmapx} and  \eqref{dynamicmapy}, there are four solution pairs for the fixed points for $x_1^*$ and $y_1^*$. The valid pair varies across the domain, except at $p_s=0$, where two physical pairs exist: one vanishing stable fixed point for $\arc{\kappa}{e}{}{e}{}<1$, and one unstable non-vanishing fixed point for $\arc{\kappa}{e}{}{e}{}\ge 1$. Thus, $\rho_{1}^{\mathsf {e}}(t)$ behaves like a Landau order parameter with $\arc{\kappa}{e}{}{e}{}$ as the control parameter.

\begin{figure}[htb]
    \centering
     \includegraphics[width=1.0\linewidth]{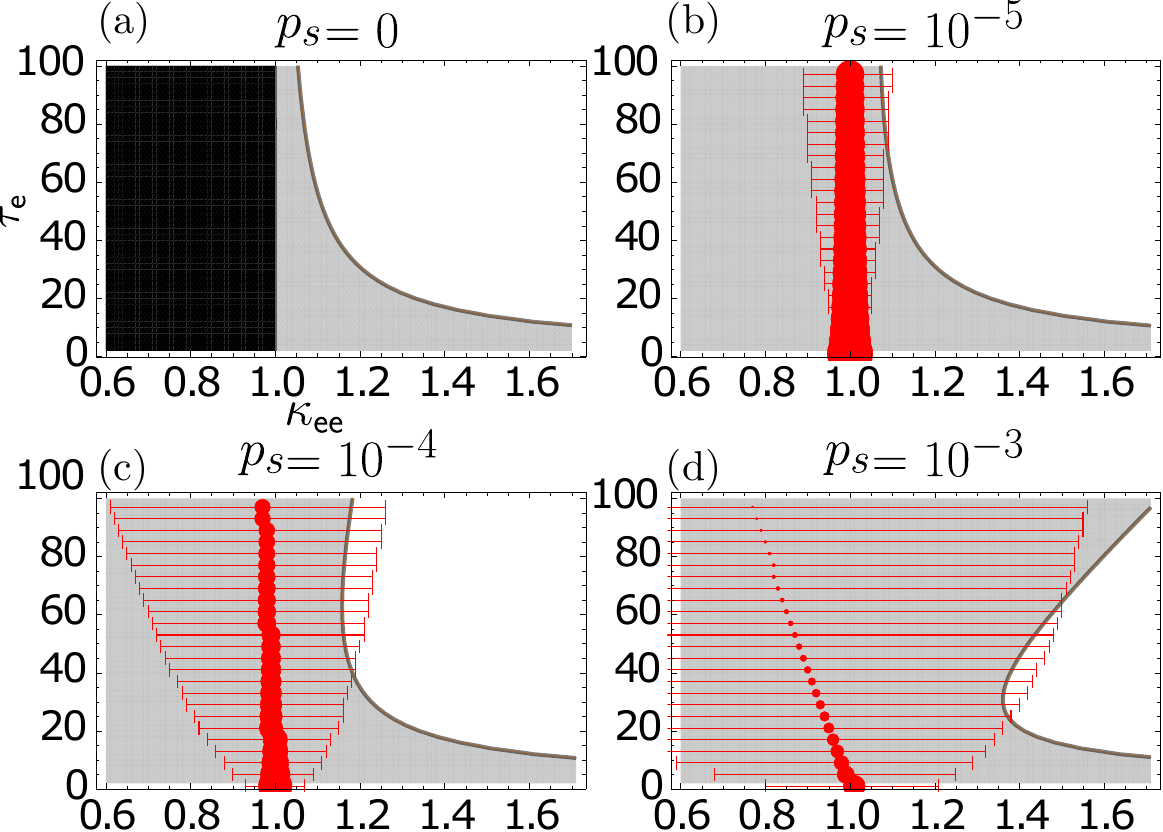}
    \caption{(Color online)
    Nonequilibrium MF phase diagram with maximum susceptibility for select values of the probability of spontaneous activation $p_{s}$.
    The black region represents a subcritical disordered phase characterized by a vanishing fixed point. The gray regions correspond to a nonzero stable fixed-point phase, while the white region denotes a quasiperiodic oscillatory phase with unstable fixed points. 
    The solid black line indicates a stability boundary. Red dots mark the locations of maximum susceptibility across $\arc{\kappa}{ee}{}{}{}$ for each value of $\arc{\tau}{e}{}{}{}$; the size of each dot is scaled logarithmically with the magnitude of the dynamical susceptibility. 
    Red whiskers represent the full width at half maximum. 
    The other parameters are $\arc{\kappa}{ii}{}{}{} = \arc{\kappa}{ie}{}{}{} = \arc{\kappa}{ei}{}{}{} = 1.6$, $\arc{\tau}{i}{}{}{} = 5$, and 
    $\numberneighbors{e}{e} =  
        \numberneighbors{e}{i} = 
        \numberneighbors{i}{e} =  
        \numberneighbors{i}{i} = 2$.
}

    \label{fig:pdandsus}
\end{figure}

There is a disordered phase for $\arc{\kappa}{ee}{}{}{} \leq 1$ and an ordered phase for $\arc{\kappa}{ee}{}{}{} \geq 1$, separated by a second-order phase transition (Fig.~\ref{fig:pdandsus}).
We have $\rho_{1}^{\mathsf {e}}(t) = 0$ for $\arc{\kappa}{e}{}{e}{} \leq 1$ and $\rho_{1}^{\mathsf {e}}(t) > 0$ for $\arc{\kappa}{e}{}{e}{} > 1$, with critical point $\arc{\kappa}{ee,c}{}{}{} = 1$.
Expanding $x_1^*$ near this critical point and $p_s=0$ via a two-variable Taylor series shows a critical exponent $\beta=1$: $x_1^* \propto (\arc{\kappa}{e}{}{e}{}-\arc{\kappa}{ee,c}{}{}{})^\beta$ for $\arc{\kappa}{e}{}{e}{} >1$.
Using $x_1^* \propto (\arc{\kappa}{e}{}{e}{}-\arc{\kappa}{ee,c}{}{}{})^1$ we found $\beta$ =1: $y_1^* \propto (\arc{\kappa}{e}{}{e}{}-\arc{\kappa}{ee,c}{}{}{})^\beta$ for $\arc{\kappa}{e}{}{e}{} >1$.
This exponent $\beta$ matches that of the excitatory-only model \cite{williams2014quasicritical} and coincides with the MF directed percolation universality class.
We also note that, in the many-body GCBM, the network topology influences  the effective critical exponents.

We can compute the dynamical susceptibility for the GCBM as follows 
\begin{align*}
    \arc{\chi}{n}{}{}{} &= \lim_{p_{s}\to 0 } \dfrac{\partial h_{1}^{*}}{\partial p_{s}} , \\
    \dfrac{\partial h_{1}^{*}}{\partial p_{s}} &=\dfrac{\partial}{\partial p_{s}}(1-\tau _{\sf n} h_{1}^{*})\ R[x_1^*, y_1^*,p_{s},...] ,
\end{align*}
where $R[x_1^*, y_1^*,p_{s},...]$=$F[x_1^*,y_1^*,p_{s},...]$ for $\arc{\chi}{e}{}{}{}$, and $R[x_1^*, y_1^*,p_{s},...]$=$G[x_1^*,y_1^*,p_{s},...]$ for $\arc{\chi}{i}{}{}{}$. After computing the susceptibility, and using the critical exponents of $x_1^*$ and $y_1^*$, the susceptibility diverges at $\arc{\kappa}{ee,c}{}{}{}$ with exponent $\gamma' = 1$ for $\arc{\kappa}{e}{}{e}{} < 1$ as $\arc{\chi}{e,i}{}{}{} \propto (\arc{\kappa}{ee,c}{}{}{} - \arc{\kappa}{e}{}{e}{})^{-\gamma'}$, and for $\arc{\kappa}{e}{}{e}{} > 1$ with exponent $\gamma = 1$ as $\arc{\chi}{e,i}{}{}{} \propto (\arc{\kappa}{e}{}{e}{} - \arc{\kappa}{ee,c}{}{}{})^{-\gamma}$. 
These exponents match MF directed percolation and agree with the excitatory-only model for both $\arc{\chi}{e}{}{}{}$ and $\arc{\chi}{i}{}{}{}$ \cite{williams2014quasicritical}.
Calculations used a simplified motif with one incoming connection per neuron type ($\numberneighbors{e}{e} = \numberneighbors{i}{e} = \numberneighbors{i}{i} = 1$) due to complexity; for purely excitatory neurons, increasing neighbors does not change exponents, as expected.

\begin{figure}[htb]
    \centering
    \includegraphics[width=0.95\linewidth]{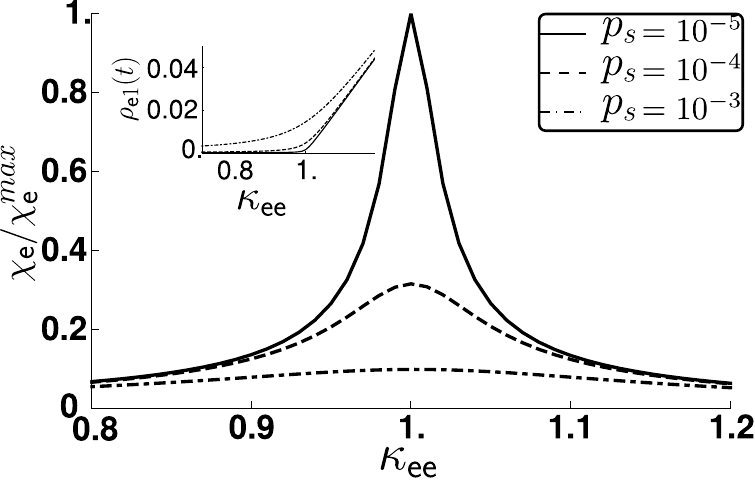}
     \caption{
    Dynamical susceptibility $\arc{\chi}{e}{}{}{}$ of the excitatory neurons activity $\rho_{1}^{\mathsf {e}}$,  as a function of the excitatory coupling strength $\arc{\kappa}{ee}{}{}{}$, normalized by the maximum value of $\arc{\chi}{e}{}{}{}$. The inset shows the average firing density of excitatory neurons, $\rho_{1}^{\mathsf {e}}$, as a function of the excitatory coupling strength $\arc{\kappa}{ee}{}{}{}$. The dynamical susceptibility is normalized to the quasi-critical point at $\arc{\kappa}{e}{}{e}{} = 1$ for $p_{s} = 10^{-5}$. 
    The other parameters are $\arc{\kappa}{ii}{}{}{} = \arc{\kappa}{ie}{}{}{} = \arc{\kappa}{ei}{}{}{} = 1.6$, $\arc{\tau}{i}{}{}{} = 2$, $\arc{\tau}{e}{}{}{} = 2$, and 
    $\numberneighbors{e}{e} =  
        \numberneighbors{e}{i} = 
        \numberneighbors{i}{e} =  
        \numberneighbors{i}{i} = 2$.
}

    \label{fig:criticalexponents}
\end{figure}

For fixed $\arc{\tau}{e}{}{}{}$ and nonzero $p_{s}$, the peak susceptibility defines a nonequilibrium Widom line in the $\arc{\tau}{e}{}{}{}$–$\arc{\kappa}{e}{}{e}{}$ plane. 
Fig. \ref{fig:pdandsus} shows susceptibility curves for various $p_{s}$ and $\arc{\tau}{e}{}{}{}$, with red dots marking maximum susceptibility (log-scaled size) and red whiskers indicating full width at half maximum.
For $\arc{\tau}{e}{}{}{}=2$, the susceptibility peaks at a quasi-critical point $\arc{\kappa}{e}{}{e}{}=1$, defining a Widom line in the $p_{s}$–$\arc{\kappa}{e}{}{e}{}$ plane (inset Fig. \ref{fig:criticalexponents}). 
The maximum susceptibility decreases as $p_{s}$ increases, analogous to a magnetic field in the Ising model.
Compared to the CBM, the GCBM shows reduced susceptibility peaks but similar shape, indicating correlation functions decrease with added inhibition \cite{GerardoBook}.
This reduction broadens the full width at half max, theoretically widening the susceptibility peak, 
thereby enlarging the quasi-critical region \cite{GriffithsPhase}.

\subsection{Chaotic Dynamics }

Using the CBM with only excitatory neurons and the simplest parameters ($\numberneighbors{e}{e}=1$, $\arc{\tau}{e}{}{}{}=2$), a period-doubling route to chaos emerges at the instability line as $\arc{\kappa}{e}{}{e}{}$ increases, starting with period-four dynamics \cite{Williams_Garc_a_2022}. 
However, these chaotic dynamics occur in the non-physical regime where $\arc{\kappa}{e}{}{e}{} > \arc{\kappa}{e}{}{e}{}^{\sf max}$ (Eq. \eqref{kappamax}). 
For $\arc{\tau}{e}{}{}{} \geq 2$, $\numberneighbors{e}{e} \geq 1$, or with inhibition, a {\it marginally stable region} appears beyond the stability boundary.
Increasing $\arc{\kappa}{e}{}{e}{}$ eventually leads to period-doubling bifurcations and chaos. 
Although chaos typically lies outside the physical regime, sufficiently high $\numberneighbors{e}{e}$ and $\arc{\tau}{e}{}{}{}$ can bring it into the physical regime with or without inhibition, though adding inhibition generally delays the onset of chaos, allowing more non-chaotic dynamics.
These biologically realistic parameters are computationally costly, but we provide examples demonstrating physical chaos.
Such chaos under realistic conditions is relevant for understanding epileptic seizures, where chaotic transitions may underlie abrupt neural changes \cite{Sackellares2000EPILEPSYW, Chaos_Theory_and_Epilepsy}.
Incorporating inhibitory neurons and additional parameters confirms a richer chaotic dynamics consistent with this route to chaos.

Generally, determining whether a system's dynamics is chaotic is challenging, as no universal definition exists.
A commonly used method for identifying chaos is to test sensitivity to initial conditions by computing the largest Lyapunov exponent (LLE), $\lambda (\Vec{x}_0)$. 
Specifically, $\lambda (\Vec{x}_0) < 0$ indicates periodic behavior, $\lambda (\Vec{x}_0) = 0$ corresponds to a marginally stable orbit, and $\lambda (\Vec{x}_0) > 0$ suggests instability or chaos \cite{WOLF1985285}.
This approach was used in the previous analysis for the excitatory-only case.
The calculation of the LLE for $\Vec{x}(t+1)$ is detailed in Appendix \ref{App:lypexp}.

\begin{figure*}
\centering
\includegraphics[width=1\textwidth]{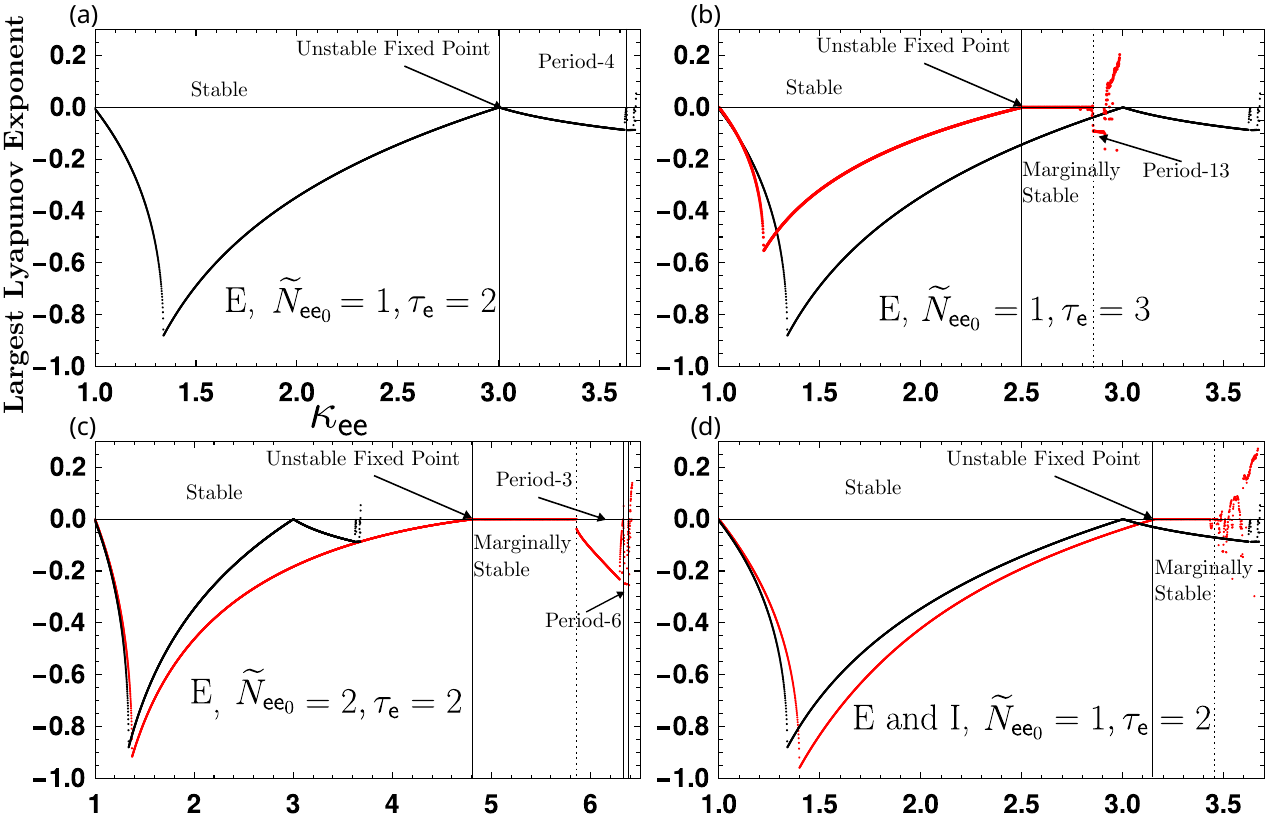}
\caption{(Color Online)
        Largest Lyapunov exponent (LLE) $\lambda(\Vec{x}_0)$ of the excitatory activity $x_{1}(t)$ as a function of the excitatory coupling strength $\arc{\kappa}{ee}{}{}{}$. Panel (a) shows the basic excitatory model (E) with the number of excitatory neighbors of the representative excitatory neuron $\numberneighbors{e}{e} = 1$, the refractory period of the representative excitatory neuron $\arc{\tau}{e}{}{}{} = 2$, and the probability of spontaneous activity $p_{s} = 0$ (black). Panels (b), (c), and (d) contrast different variations of the excitatory-only model (E) in red, with the excitatory-only model (black) for comparison:  Panel (b) shows the effect of increasing refractory period $\arc{\tau}{e}{}{}{}$ from 2 to 3, Panel (c) shows the effect of increasing number of excitatory neighbors $\numberneighbors{e}{e}$ from 1 to 2, Panel (d) shows the inclusion of inhibition (I) in the model.
    Black vertical lines indicate the locations of unstable fixed points, followed by bifurcation points labeled with the emerging period. The dashed black line marks the approximate end of the marginally stable region.
}
    \label{fig:LLEchanges}
\end{figure*}

Our goal is to explore new dynamics by incorporating biologically realistic parameters like inhibition and provide further evidence for chaos using additional methods.
Besides computing the LLE, we visualize attractors via time-delay embedding of $x_{1}(t)$, compute its discrete Fourier transform, and analyze the full Lyapunov spectrum \cite{Eckmann1985}. 
This enables estimating the Lyapunov dimension, approximating the fractal dimension of strange attractors \cite{FARMER1983153}, implemented in Mathematica \cite{Sandri1996} as detailed in Appendix \ref{App:lypexp}.
While the LLE is computed across parameters, we focus on $\arc{\kappa}{ee}{}{}{}$ for comparison with the basic CBM (Fig.~\ref{fig:LLEchanges} (a)). Near unstable fixed points, dynamics become computationally challenging, appearing to slowly converge to periodic behavior—for example, at $\arc{\kappa}{ee}{}{}{}=3$, values stabilize to $10^{-6}$ after $10^{8}$ iterations.

\begin{figure}[htb]
\centering
\includegraphics[width=.95\columnwidth]{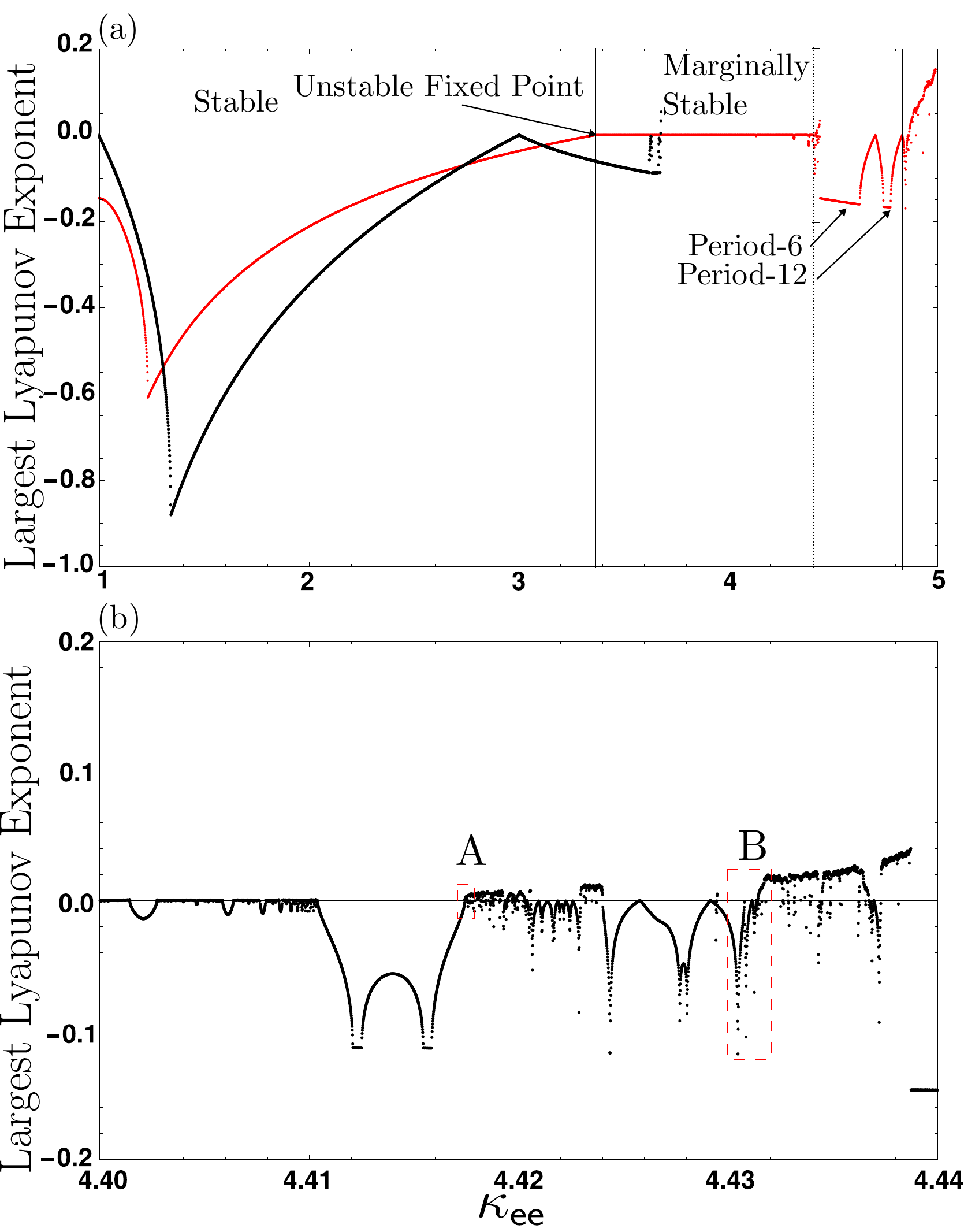}
\caption{(Color Online) 
(a) Largest Lyapunov exponent (LLE) $\lambda(\Vec{x}_0)$ of the excitatory activity $x_{1}(t)$ as a function of $\arc{\kappa}{ee}{}{}{}$ with parameters: $p_{s} = 10^{-3}$, $\arc{\tau}{e}{}{}{} = 3$, $\arc{\tau}{i}{}{}{} = 2$, $\numberneighbors{e}{e} =   \numberneighbors{e}{i} = 
        \numberneighbors{i}{e} =  
        \numberneighbors{i}{i} = 2$. The solid black vertical lines mark, in sequence, the onset of instability (unstable fixed point) and subsequent bifurcations. The dashed black line indicates the approximate end of the marginally stable region. The black box in panel (a), starting at $\arc{\kappa}{e}{}{e}{}=4.40$ shows the region magnified in panel (b). Label A marks a possible abrupt transition out of periodic behavior. Label B highlights a region consistent with period-doubling en route to chaos.}

    \label{fig:LLEkin2ps3tx3}
\end{figure}

We tested a non-exhaustive set of parameters, including excitatory-only networks, added inhibition, neighborhood connectivities ($\numberneighbors{e}{e}=1$ or $2$), delays ($\arc{\tau}{e}{}{}{}=2$ or $3$), and synaptic strengths ($p_{s}$), summarized in Figs. \ref{fig:LLEchanges} and \ref{fig:LLEkin2ps3tx3}. 
Using more biologically realistic parameters—excitatory and inhibitory neurons with $\numberneighbors{e}{e}=\numberneighbors{e}{i}=\numberneighbors{i}{e}=\numberneighbors{i}{i}=2$, $\arc{\tau}{e}{}{}{}=3$, and $p_{s}=10^{-3}$, we observed the LLE shown in Fig. \ref{fig:LLEkin2ps3tx3}.
The first region where the LLE nears zero corresponds to the second-order phase transition (e.g., $\arc{\kappa}{ee}{}{}{}=1$, $p_{s}=0$), with further zeros indicating marginal stability, bifurcations, and chaos onset in Fig.~\ref{fig:LLEchanges} (a).
In Figs.~\ref{fig:LLEchanges} and \ref{fig:LLEkin2ps3tx3}, solid black lines mark marginal stability points (bifurcations or onset of marginally stable regions), while dashed lines approximate their boundaries.

Figure~\ref{fig:LLEchanges} (b) illustrates the basic excitatory model with $\numberneighbors{e}{e}=1$ and $p_{s}=0$, comparing $\arc{\tau}{e}{}{}{}=2$ (black) to $\arc{\tau}{e}{}{}{}=3$ (red).
Increasing $\arc{\tau}{e}{}{}{}$ introduces a marginally stable region rather than a single marginally stable point. 
For $\arc{\tau}{e}{}{}{}=3$, the first marginally stable point occurs at $\arc{\kappa}{ee}{}{}{}=2.5$, compared to $\arc{\kappa}{ee}{}{}{}=3$ for $\arc{\tau}{e}{}{}{}=2$.
Additionally, chaotic dynamics emerge earlier for $\arc{\tau}{e}{}{}{}=3$. 
For marginally stable orbits, we find that the LLE is generally positive but converges to zero with increasing iterations, requiring at least $10^6$ iterations in the studied cases.
Figure~\ref{fig:LLEchanges} (c) shows the impact of increasing $\numberneighbors{e}{e}=1$ (black) to $\numberneighbors{e}{e}=2$ (red).
This modification introduces a marginally stable region for higher values of $\arc{\kappa}{ee}{}{}{}$, along with new chaotic regions.
Figure~\ref{fig:LLEchanges} (d) incorporates inhibition into the CBM model, revealing that while the dynamics occur at similar values of $\arc{\kappa}{ee}{}{}{}$, the unstable fixed point occurs later, and chaotic dynamics emerges earlier.
Including inhibition also introduces a marginally stable region, similar to the excitatory-only cases.
Interestingly, for $p_{s}<10^{-3}$, the effect on marginally stable points is minimal, except at $\arc{\kappa}{ee}{}{}{}=1$, where a non-zero $p_{s}$ disrupts the phase transition.

Incorporating biologically realistic parameters still yields marginally stable regions and chaotic dynamics at higher $\arc{\kappa}{ee}{}{}{}$ values (Fig. \ref{fig:LLEkin2ps3tx3}).
With added parameters, doubling routes to chaos—beginning with finite-period windows—remain visible but occur after the marginally stable region rather than immediately following the unstable fixed point, as shown in Fig. \ref{fig:LLEkin2ps3tx3} (a).
Additional parameters also introduce new doubling routes with varying $\arc{\kappa}{ee}{}{}{}$ widths, such as region B in Fig.~\ref{fig:LLEkin2ps3tx3} (b).
In contrast, region A shows an abrupt transition from periodic to chaotic behavior without a clear route to chaos. 
Though narrow period-doubling windows may exist, they are undetectable within $10^{-6}$ precision in $\arc{\kappa}{ee}{}{}{}$. 
Small negative dips in the LLE within marginally stable regions quickly return to zero.

\begin{figure}[htb]
\centering
\includegraphics[width=1.00\columnwidth]{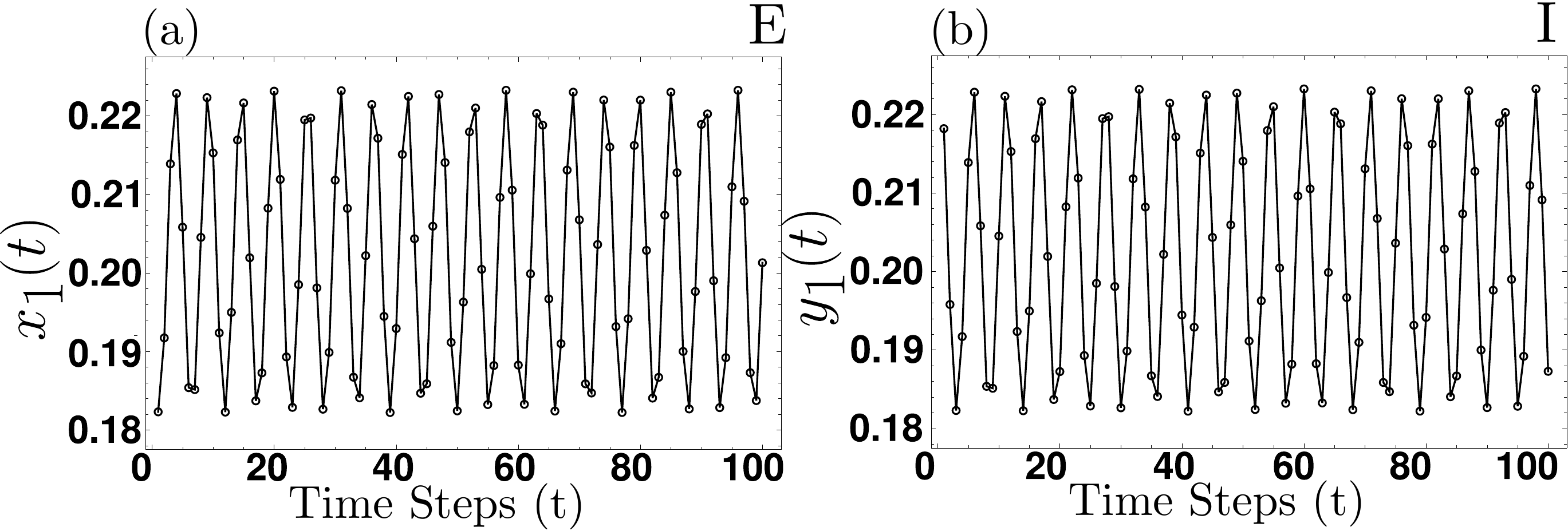}
\caption{Time series solutions in the marginally stable region of (a) the excitatory activity $x_{1}(t)$ and (b) the inhibitory activity $y_{1}(t)$, shown for the last 200 of $10^6$ iterations. Parameters are: $\arc{\kappa}{ee}{}{}{} = 3.37$, $\arc{\kappa}{ii}{}{}{} =\arc{\kappa}{ie}{}{}{} =\arc{\kappa}{ei}{}{}{} = 1$, $p_{s} = 10^{-3}$, $\arc{\tau}{e}{}{}{} = 3$, $\arc{\tau}{i}{}{}{} = 2$, $B = 0.5$, and $\numberneighbors{e}{e} =  
        \numberneighbors{e}{i} = 
        \numberneighbors{i}{e} =  
        \numberneighbors{i}{i} = 2$.
}
\label{fig:mapEIkin2}
\end{figure}

\begin{figure*}[htb]
    \includegraphics[width=1.0\textwidth]{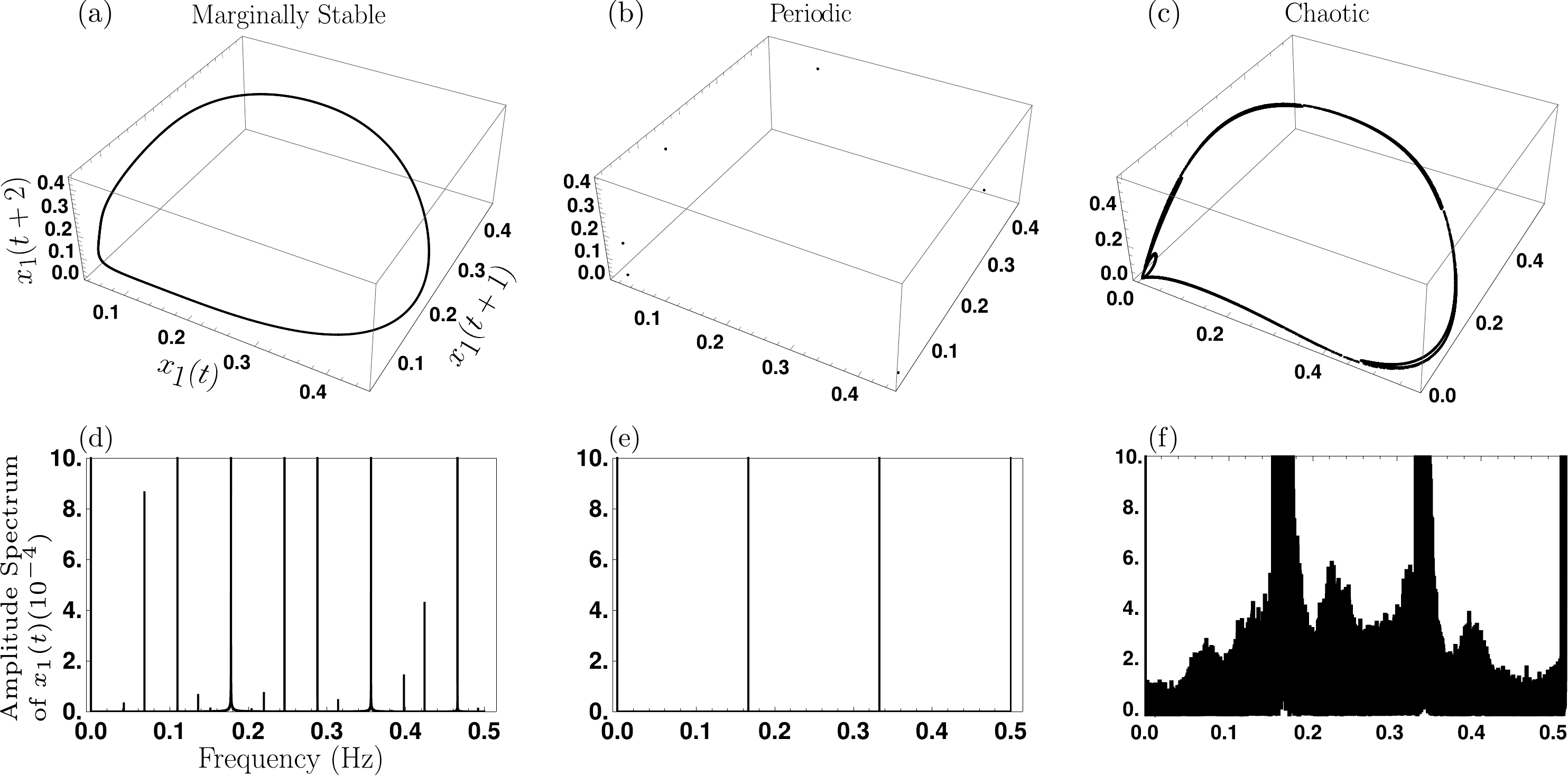}
    \caption{Time-delay embeddings and Fourier spectra of ${\arc{x}{1}{}{}{}(t)}$. Upper row (a-c) is the time delay embedding of $\arc{x}{1}{}{}{}(t)$ and lower row (d-f) is the discrete Fourier transform of $\arc{x}{1}{}{}{}(t)$ with the single-sided amplitude being the y-axis. The columns correspond to the marginally stable region (a, d, $\arc{\kappa}{ee}{}{}{}=4.0$), periodic region (b, e, $\arc{\kappa}{ee}{}{}{}=4.5$), and chaotic region (c, f, $\arc{\kappa}{ee}{}{}{}=4.95$).  The other parameters are the same as Fig.~\ref{fig:LLEkin2ps3tx3}: $\kappa_{\sf{ii}}=\kappa_{\sf{ie}}=\kappa_{\sf{ei}}=1$, $\arc{\kappa}{e}{}{e}{}$=1.6, $p_{s}=10^{-3}$, $\tau_{\sf e}=3$, and $\tau_{\sf i}=2$, $\numberneighbors{e}{e}=\numberneighbors{e}{i}=\numberneighbors{i}{e}=\numberneighbors{i}{i}=2$.}
    \label{fig:timedelayembEIkin2tx3ps3}
\end{figure*}

The marginally stable region appears nearly periodic in the solution plots of $x_{1}(t)$ and $y_{1}(t)$ from the map (Fig. \ref{fig:mapEIkin2}), though the trajectory never exactly repeats, raising the question of chaos.
To better distinguish marginally stable, periodic, and chaotic regimes, we computed time-delay embeddings and discrete Fourier transforms of $x_{1}(t)$, shown in Fig. \ref{fig:timedelayembEIkin2tx3ps3} using the same parameters as Fig.~\ref{fig:LLEkin2ps3tx3}.
The left column (a–c) shows time-delay embeddings, and the right column (d–f) shows the single-sided amplitude spectra, corresponding respectively to the marginally stable ($\arc{\kappa}{ee}{}{}{}=4.0$), periodic ($\arc{\kappa}{ee}{}{}{}=4.5$), and chaotic ($\arc{\kappa}{ee}{}{}{}=4.95$) regimes. 
Together, these visualizations offer a complementary view beyond the LLE for characterizing system dynamics.

The marginally stable region shows a deformed elliptical time-delay embedding and broad peaks in the Fourier spectrum (Fig. \ref{fig:LLEkin2ps3tx3} (a,d)).
The periodic regime displays a finite number of points in the time-delay embedding and half as many sharp, nonzero peaks in the one-sided Fourier spectrum; for example, Fig. \ref{fig:LLEkin2ps3tx3} (b,e) shows a period-6 orbit with 6 distinct points in the embedding and 3 peaks at 1/6, 1/3, and 1/2 Hz.
In contrast, the chaotic regime is characterized by either a self-wrapping deformed ellipse or an infinite sequence of bifurcations in the embedding, along with a broadband Fourier spectrum (Fig.~\ref{fig:LLEkin2ps3tx3} (c,f)).
While time-delay embeddings for marginally stable and chaotic regimes may appear similar, their Fourier spectra differ markedly.
Within the marginally stable region, increasing $\arc{\kappa}{ee}{}{}{}$ introduces additional frequencies in the Fourier spectrum of $x_{1}(t)$, consistent with a period-doubling route to chaos.

The delay embedding in Fig. \ref{fig:timedelayembEIkin2tx3ps3} (c) forms a strange attractor, as evidenced by its fractional Lyapunov dimension \cite{Fractals}.
The Lyapunov dimension $\mathit{D}_L$ is computed from the Lyapunov spectrum of $\Vec{x}(t+1)$ using the procedure described in Appendix \ref{App:lypexp}, following Ref.~\cite{Sandri1996}.
According to the Kaplan–Yorke conjecture, the Lyapunov dimension is defined as
\begin{equation}
    \mathit{D}_L=j+\frac{\sum_{i=1}^{j}\lambda_{i}}{|\lambda_{j+1}|}.
\end{equation}
With the Lyapunov exponents ordered from largest to smallest $\lambda_1\ge\lambda_2\ge \dots \ge \lambda_n$.
Then, $j$ is defined by the largest index such that the following conditions hold
\begin{align*}
    &\sum_{i=1}^j\lambda_i\ge 0 , \ \mbox{ and }\ 
    \sum_{i=1}^{j+1}\lambda_i < 0 .
\end{align*}

For Fig.~\ref{fig:timedelayembEIkin2tx3ps3} (c) $\mathit{D}_L=1.02433$ with $j=1$, which is a fractional dimension, meaning it is a strange attractor.
In general, $1\leq\mathit{D}_L \leq 2$ for each variation computed.

So far, our analysis of chaos has focused on the non-physical regime, where $\arc{\kappa}{e}{}{e}{} > \arc{\kappa}{e}{}{e}{}^{\sf max}$.
However, by fixing $B = 0.5$ and choosing $\arc{\tau}{e}{}{}{} = 40$, $\numberneighbors{e}{e} = 5$, $\numberneighbors{e}{i} = \numberneighbors{i}{e} = \numberneighbors{i}{i} = 1$, $\arc{\kappa}{ii}{}{}{} = \arc{\kappa}{ie}{}{}{} = \arc{\kappa}{ei}{}{}{} = 0.8$, $\arc{\tau}{i}{}{}{} = 2$, and $p_{s} = 0$, we observe chaos within the physical regime, where all coupling strengths satisfy $\arc{\kappa}{n}{}{n}{} < \arc{\kappa}{n}{}{n}{}^{\sf max}$.
Among these parameters, $\arc{\tau}{e}{}{}{}$ and $\numberneighbors{e}{e}$ appear to be the most critical for enabling chaos in the physical regime when $B$ is held fixed.
Notably, inhibition is not necessary for the emergence of chaos in this regime.
We find that increasing $\arc{\tau}{e}{}{}{}$ to a sufficiently high value, 40 in this example, lowers the minimal value of $\arc{\kappa}{ee}{}{}{}$ required for chaos to appear in the physical regime, with chaos observed around $\arc{\kappa}{ee}{}{}{} = 2$. 
Additionally, setting $\numberneighbors{e}{e} = 5$ ensures that $\arc{\kappa}{e}{}{e}{}^{\sf max} > 2.33$, thus maintaining the system within the physical regime.
Interestingly, the range of Lyapunov exponents remains largely unchanged for this parameter set.
These values are illustrative rather than minimal, and merely demonstrate that chaos is indeed possible within the physical regime under suitable conditions.

\subsubsection{Admissible Initial Conditions}

It is necessary to find permissible initial conditions such that the maps define valid probabilities.
Two criteria must be met: the probabilities for all possible states of $x_m(t)$ and $y_n(t)$ sum to one at all times, and each probability remains bounded between zero and one for all time shown in Eqs.~\eqref{eq:probability_restriction}.
To determine the permissible initial conditions required to express Eqs.~\eqref{eq:probability_restriction} in terms of the initial values only, we use Eqs.~\eqref{eq:MFT_x1} and~\eqref{eq:MFT_y1} together with the definitions of $x_{0}(t)$ and $y_{0}(t)$ given in Eq.~\eqref{eq:x0}, along with the following equations, which must be applied repeatedly

\begin{align}
    &\sum_{m=1}^{\arc{\tau}{e}{}{}{}} x_{m}(t)= \sum_{l=2}^{t} x_{1}(l)+ \sum_{m=1}^{\arc{\tau}{e}{}{}{}-(t-1)} x_{m}(1) ,\nonumber \\& \ \,
    \sum_{n=1}^{\arc{\tau}{i}{}{}{}} y_{n}(t)= \sum_{l=2}^{t} y_{1}(l)+ \sum_{n=1}^{\arc{\tau}{i}{}{}{}-(t-1)} y_{n}(1).
    \label{eq:simplifedIC}
\end{align}

\begin{figure}[htb]
    \centering
\includegraphics[width=1.0\columnwidth]{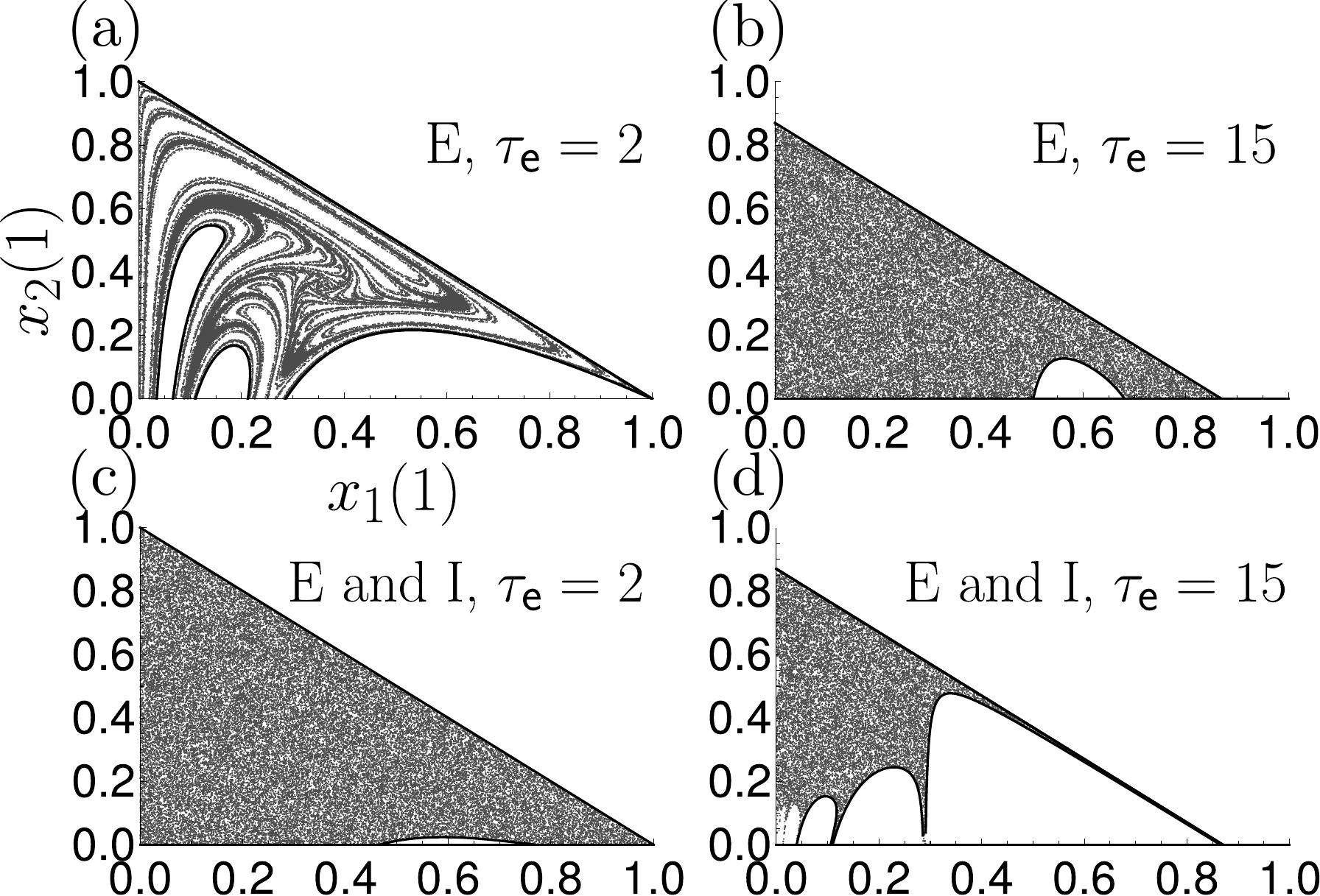}
    \caption{Scatter plots of permissible initial conditions for $\arc{\kappa}{ee}{}{}{} = 3.5$ after $10^2$ iterations. The black lines indicate analytical constraints derived from Eqs.~\eqref{eq:probability_restriction}, evaluated up to $t = 4$. Panels (a) and (b) show the CBM with $\arc{\tau}{e}{}{}{} = 2$ and $\arc{\tau}{e}{}{}{} = 15$, respectively. Panels (c) and (d) show the GCBM with the same corresponding $\arc{\tau}{e}{}{}{}$ values. All plots use $p_{s} = 0$, $\numberneighbors{e}{e} = 2$, and initial conditions $x_{3,\ldots,\arc{\tau}{e}{}{}{}}(1) = 0.01$. The bottom row additionally includes: $\numberneighbors{i}{e} = \numberneighbors{e}{i} = \numberneighbors{i}{i} = 2$, $\arc{\tau}{i}{}{}{} = 2$, $y_{1,2}(1) = 0.01$, $\arc{\kappa}{ii}{}{}{} = \arc{\kappa}{ie}{}{}{} = \arc{\kappa}{ei}{}{}{} = 1.6$.}
    \label{fig:IC}
\end{figure}

We plotted cross-sections of the permissible initial conditions for different motifs with $\arc{\kappa}{ee}{}{}{} = 3.5$ and $\numberneighbors{e}{e} = 2$, comparing excitatory-only (E) neurons and networks including inhibitory (I) neurons with refractory periods $\arc{\tau}{e}{}{}{} = 2$ or $15$, as shown in Fig. \ref{fig:IC}.
The permissible initial conditions were found by randomly sampling $10^5$ points uniformly in the unit square $x_1(1), x_2(1) \in [0,1]$, then plotting those (gray dots) that satisfy Eq. \eqref{eq:probability_restriction} up to time $t=10^2$.
We analytically solved Eqs. \eqref{eq:probability_restriction} and  \eqref{eq:simplifedIC} up to $t=4$, shown as black lines in Fig. \ref{fig:IC}; computations beyond $t=4$ were neither feasible nor insightful.
A strict upper bound constraint from $0 \leq \sum_{m=1}^{\arc{\tau}{e}{}{}{}} x_{m}(t) \leq 1$ at $t=1$, restricts the permissible space maximally to the lower left triangle since $ x_{2}(1) \le 1-x_{1}(1)-\sum_{m=3}^{\arc{\tau}{e}{}{}{}} x_{m}(1) $.
Additional black lines appear sequentially at later times, further constraining the permissible initial conditions.
Where appropriate we fixed $p_{s}=0$, $\numberneighbors{i}{e}=\numberneighbors{e}{i}=\numberneighbors{i}{i}=2$, $x_3(1)=x_4(1)=...=x_{\arc{\tau}{e}{}{}{}}(1)=.01$, $\arc{\tau}{i}{}{}{}=2$, $y_1(1)=y_{2}(1)=.01$, $\arc{\kappa}{i}{}{i}{}=\arc{\kappa}{i}{}{e}{}=\arc{\kappa}{e}{}{i}{}$=1.6.  Note that relaxing the constraints on $x_3(1)$ and $y_{1,2}(1)$ over $[0,1]$ fills the previously inadmissible region of the lower-left triangle when projected onto the same axes.

\subsection{CBM with an External Periodic Inhibitory Source}

\begin{figure}[htb]
    \centering
    \includegraphics[width=0.95\columnwidth]{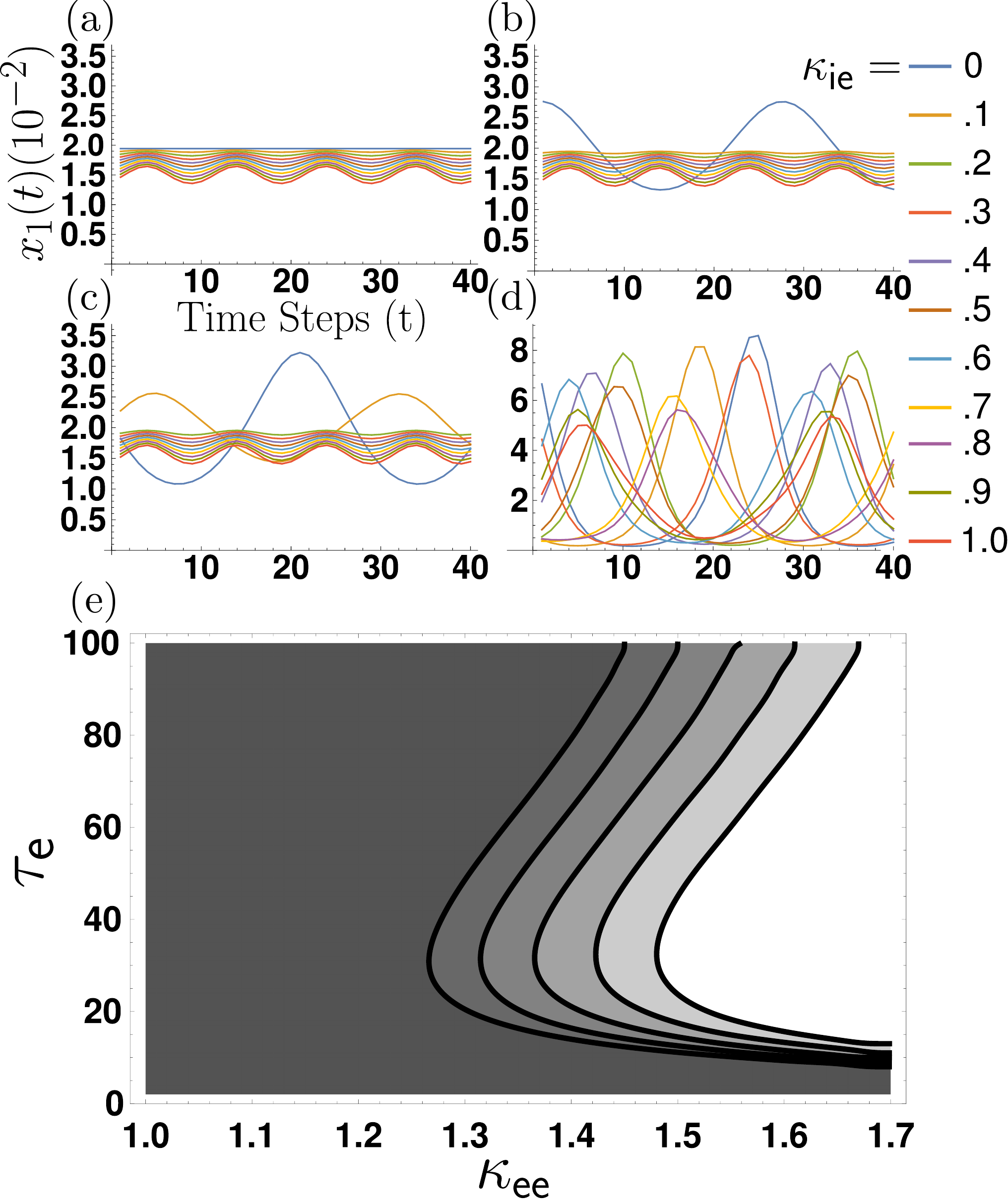}
\caption{(Color Online)
Periodicity enslavement by an external periodic inhibitory source.
(a–d) Time series plots of excitatory activity $x_1(t)$ with increasing $\arc{\kappa}{ie}{}{}{}$ from 0 to 1 (as indicated in the legend) and $p_{s} = 10^{-3}$, $\arc{\tau}{e}{}{}{} = 15$. Each panel corresponds to a different excitatory coupling strength: $\arc{\kappa}{ee}{}{}{} = \{1.36, 1.37, 1.39, 1.60\}$ in panels (a), (b), (c), and (d), respectively. In panel (a), when $\arc{\kappa}{ie}{}{}{} = 0$, $x_1(t)$ settles into a stable fixed point; for $\arc{\kappa}{ie}{}{}{} > 0$, $x_1(t)$ becomes periodic and adopts the periodicity of the external inhibitory input $y_1(t)$. In panel (b), $x_1(t)$ starts in a marginally stable quasiperiodic regime and transitions to periodicity as $\arc{\kappa}{ie}{}{}{}$ increases. In panel (c), $x_1(t)$ remains in a marginally stable quasiperiodic regime for $\arc{\kappa}{ie}{}{}{} \leq 0.1$ but transitions to periodic behavior for $\arc{\kappa}{ie}{}{}{} > 0.1$, enslaved by $y_1(t)$. In panel (d), all values of $\arc{\kappa}{ie}{}{}{}$ maintain marginally stable quasiperiodic behavior. In panel (e), a plot distinguishing when $x_{1}(t)$ is periodic (gray) and when it is non-periodic (white) generally being marginally stable, as a function of $\arc{\kappa}{ee}{}{}{}$ and $\arc{\kappa}{ie}{}{}{}$. $\arc{\kappa}{ie}{}{}{}$ varies from 0 to 1.6 in steps of 0.4, with darker shades representing lower values. All panels use: $\numberneighbors{e}{e} = 2$, $\numberneighbors{e}{i} = 0$, $\numberneighbors{i}{e} = 1$, and $\numberneighbors{i}{i} = 0$.
}

\label{fig:periodicy}
\end{figure}

Experiments seem to indicate that inhibitory neurons exhibit periodic activation \cite{PeriodicInhibition}.
We can impose a periodic form on $y_1(t)$ in the MF approximation as
\begin{equation}
    y_{1}(t)=D+A \sin ^2 \left (\frac{\pi t}{T}\right ),
\end{equation}
choosing the square of the sine function to keep all values positive. 
This reduces the excitatory dynamics to that of a CBM driven by an external periodic inhibitory input.
When $\kappa_{\sf{ie}}=0$, the periodic forcing of $y_1(t)$ does not affect the periodicity of $x_1(t)$’s solution, leaving the many-body CBM dynamics unchanged. 
We examine how imposing this periodic $y_1(t)$ alters $x_1(t)$’s dynamics by plotting $x_1(t)$ for varying $\arc{\kappa}{ie}{}{}{}$ and fixed $\arc{\kappa}{ee}{}{}{} = {1.36, 1.37, 1.39, 1.60}$, each exhibiting distinct behaviors (Fig.~\ref{fig:periodicy}).
For illustration, we choose $D=0.05$, $A=0.1$, and $T=10$. The dynamics of $x_1(t)$ depend sensitively on these parameters governing the periodic inhibitory input.

For $\arc{\kappa}{ee}{}{}{} = 1.36$, $x_1(t)$ starts in a stable fixed point region at $\arc{\kappa}{ie}{}{}{} = 0$; once $\arc{\kappa}{ie}{}{}{} > 0$, $x_1(t)$ transitions to a periodic phase (Fig. \ref{fig:periodicy} (a)), becoming entrained to $y_1(t)$ with the same period. 
This periodic behavior occurs for small amplitudes of the periodic function $y_1(t)$ and small coupling $\arc{\kappa}{ie}{}{}{}$.
At $\arc{\kappa}{ee}{}{}{} = 1.37$, $x_1(t)$ begins in a marginally stable phase when $\arc{\kappa}{ie}{}{}{}=0$ and again shifts to the periodic phase with the same period as $y_1(t)$ for $\arc{\kappa}{ie}{}{}{} > 0$ (Fig. \ref{fig:periodicy} (b)).
For $\arc{\kappa}{ee}{}{}{} = 1.39$, values of $\arc{\kappa}{ie}{}{}{} \leq 0.1$ correspond to a marginally stable phase for $x_1(t)$, while $\arc{\kappa}{ie}{}{}{} > 0.1$ leads to periodic behavior matching $y_1(t)$’s period (Fig. \ref{fig:periodicy} (c)).
Finally, at $\arc{\kappa}{ee}{}{}{} = 1.6$, $x_1(t)$ remains marginally stable for $0 \leq \arc{\kappa}{ie}{}{}{} \leq 1.5$, transitioning to the periodic phase for $\arc{\kappa}{ie}{}{}{} > 1.5$ (Fig. \ref{fig:periodicy} (d)).
Figure~\ref{fig:periodicy} (e) summarizes these transitions by showing periodic regions in gray and non-periodic in white across values of $\arc{\kappa}{ee}{}{}{}$ and $\arc{\tau}{e}{}{}{}$ as $\arc{\kappa}{ie}{}{}{}$ increases from 0 to 1.6.

This diagram differs from the phase diagrams since it only distinguishes whether the dynamics are periodic, not whether fixed points are stable. 
In the special case $\arc{\kappa}{ie}{}{}{} = 0$, it coincides with the CBM phase diagram for  $\numberneighbors{e}{e}= 2$. 
Originally, it is a phase diagram because the nonanalytic line separates stable fixed points (period-1 functions) from unstable ones.
Notably, regions where $x_1(t)$ would normally be unstable fixed points become periodic when $y_1(t)$ is forced to be periodic. 

\section{Discussion}
\label{section6}

By incorporating inhibitory neurons into branching-process models, the GCBM opens new avenues for further research.
While the GCBM exhibits activity suppression consistent with integrate-and-fire models \citep{brunel_dynamics_2000}, other connections remain to be established. 
Additionally, the interplay between driving and relaxation timescales—and the role of causality in the presence of inhibitory neurons—which constitutes a key motivation for the quasicriticality hypothesis \citep{williams2014quasicritical}, requires further investigation \citep{williams-garcia_unveiling_2017,zierenberg_description_2020}. 
The quasicriticality hypothesis may provide a unifying framework emphasizing the role of emergent avalanche distributions in artificial neural networks, with potential applications to computing systems \citep{cramer_control_2020,safavi_signatures_2024}, learning in artificial intelligence \citep{ghavasieh_toward_2025}, and training in in vitro neuronal networks \citep{habibollahi_critical_2023}.

We developed a methodology for generating low-dimensional representations of functional networks using MF approximations. This framework was applied to a minimal neural dynamics model incorporating both excitatory and inhibitory neurons, in agreement with predictions of the quasi-criticality hypothesis.

We constructed a dictionary of fundamental rules for deriving MF equations across a range of network motifs. Our results demonstrate that the inclusion of inhibitory neurons generally promotes stability within the system. Additionally, we characterized chaotic dynamics in detail within the non-physical regime and identified examples within the physical regime. In both cases, we observed the emergence of a marginally stable region that transitions to chaos either via a period-doubling route or through an abrupt onset. These findings highlight the stabilizing influence of inhibition and elucidate the mechanisms underlying the transition to complex dynamics in neural networks.

Our findings support the quasi-criticality hypothesis by demonstrating that the system’s maximum susceptibility decreases as the strength of the external drive increases, even with the inclusion of inhibitory neurons. Despite this modulation, the system remains within the directed percolation universality class, exhibiting a disorder–order phase transition when the probability of spontaneous activation vanishes. Inhibitory connections reshape the phase diagram by expanding the stable region: increasing inhibitory-to-excitatory (I$\rightarrow$E) or excitatory-to-inhibitory (E$\rightarrow$I) coupling strengths, weakening inhibitory-to-inhibitory (I$\rightarrow$I) interactions, or shortening the inhibitory refractory period all enhance network stability. An exception occurs at very small values of the excitatory refractory period, $\arc{\tau}{e}{}{}{}$, where the stable region contracts for reasons that remain unclear.
The inclusion of one additional arc-to-arc inhibitory interaction acting on arcs originating from the excitatory neuron primarily produces quantitative shifts in the phase boundaries rather than qualitatively new dynamical regimes. In particular, when $\arc{\tau}{e}{}{}{}=2$ (Fig.~\ref{fig:kie}), the addition of this interaction shifts the phase transition to lower values of $\arc{\kappa}{ee}{}{}{}$ while preserving the qualitative structure of the phase diagram.

Notably, these changes consistently reduce the average excitatory activity, highlighting the role of inhibition in maintaining the network near the quasi-critical regime. Furthermore, the addition of inhibitory neurons broadens the region of high susceptibility while lowering its peak values, suggesting that inhibition enables the system to remain sensitive across a wider parameter range. Although our framework does not capture temporal mutual information due to the absence of time-resolved activity, these results emphasize the critical role of inhibitory dynamics in stabilizing network activity and supporting quasi-critical behavior.

We have improved the classification of unstable regions exhibiting chaotic dynamics by incorporating more biologically realistic parameters into our model. Across most network configurations, beyond the simplest excitatory-only case, we consistently observed a sequence of dynamic regimes: a stable region, followed by a marginally stable region, which often transitions into chaos via a period-doubling route. Although chaos was primarily analyzed in the non-physical regime due to the computational cost of simulations with physical parameters, we found no significant qualitative differences between the two. In some cases, we also observed a direct transition from marginal stability to chaos, though this pathway requires further investigation. To characterize chaotic behavior, we employed multiple analytical techniques, including the computation of the largest Lyapunov exponent, time-delay embedding, and discrete Fourier analysis of the average excitatory activity. A critical aspect of this analysis was the careful selection of permissible initial conditions to ensure accurate identification of chaotic regimes. Together, these results offer new insights into the rich dynamics of neural networks composed of both excitatory and inhibitory populations.


Our findings further support Ref. \cite{williams2014quasicritical}'s hypothesis that oscillations in the quasiperiodic region may be linked to synchronizations observed during epileptic seizures \cite{williams2014quasicritical, SeizureSynchronization}. The presence of chaotic dynamics with more biologically realistic parameters in the unstable fixed-point region strengthens the potential connection between the unstable regime and epileptic seizures, which can also exhibit chaotic behavior \cite{Low_Dimensional_Chaos, Chaos_Theory_and_Epilepsy}. We found that forcing the inhibitory neuron activation density to be periodic induces periodic behavior in excitatory neurons with the same period and an approximate $-\pi/2$ phase shift, but only when the connection strength between inhibitory and excitatory neurons is sufficiently high. This threshold increases as the strength of the excitatory to excitatory connection grows. These findings may provide insights into how neurons could synchronize during an epileptic seizure.

A key motivation for our work is to identify a homeostatic mechanism that maintains the system within the quasi-critical regime, potentially mediated by inhibitory neurons. In addition, we aim to develop an experimentally measurable biomarker that indicates the system’s proximity to chaotic dynamics, offering insight into transitions toward instability. We envision that our model could help guide the design of such a biomarker, leveraging controlled manipulations of excitatory and inhibitory connectivity to assess network susceptibility and stability. Of particular interest is the application of this biomarker to real-time seizure prediction, providing clinical value for individuals with epilepsy, a direction we are actively pursuing.

\appendix

\section{LIST OF SYMBOLS AND NOTATION USED}
\label{App:Notation Guide}

$N_{\sf e}$: Number of Excitatory Neurons

$N_{\sf i}$: Number of Inhibitory Neurons

$\mathsf{e}_{\mu}$: Excitatory Neurons

$\mathsf{i}_{\nu}$: Inhibitory Neurons

$\mu$: Index of Excitatory Neurons

$\nu$: Index of Inhibitory Neurons

$\mathsf{n}$: Either Excitatory or Inhibitory Neuron

$\arc{z}{e}{\mu}{}{}(t)$: Excitatory Dynamical States Variable

$\arc{z}{i}{\nu}{}{}(t)$: Inhibitory Dynamical States Variable

$t$: Time Index of State Variables

$\mathcal{S} = \{\mathcal{S}_{\sf e}, \mathcal{S}_{\sf i}\}$:  Set of Node States of Excitatory and \hspace*{30pt} Inhibitory Neurons

$\arc{\tau}{e}{}{}{}$: Refractory Period of Excitatory Neurons

$\arc{\tau}{i}{}{}{}$: Refractory Period of Inhibitory Neurons

$z(t)$: Sequence of Individual Neuron States

$C$: Configuration Space Neuron's States

$Z$: States of the Arcs (Signals Transmission)

$Z_{\mathsf{n}}$: State of Directed Edges for Spontaneous \hspace*{30pt} Activation

$Z_{\sf nn'}$: State of Directed Edges Two Vertices

$Z^{{\sf i}_\nu}_{\sf ee'}$: State of Directed Edges Three Vertices

$Z^{{\sf i}_\nu}_{{\sf e}}$: State of Directed Edges Two Vertices with \hspace*{30pt} External source

$P$: Transmission Probabilities

$\arc{P}{n}{}{n'}{}$: Probability of Transmission Two Vertices

$P_{{\sf n}{\sf n}'}^{\sf i_{\nu}}$: Probability of Transmission Three Vertices

$p_{s\mu}$: Probability of Spontaneous Activation  for \hspace*{30pt} Excitatory Neurons

$p_{s\nu}$: Probability of Spontaneous Activation  for \hspace*{30pt} Inhibitory Neurons

$p_{s}$: Probability of Spontaneous Activation when  \hspace*{30pt}$p_{s\mu}=p_{s\nu}$


$k_{\mathsf{n}\mathsf{n'}}$: In Degree of Nodes

$p^{\mathsf{n}\mathsf{n'}}_{\rm r}$: Exponential Branching Decay Factor

$\arc{\kappa}{n}{}{n'}{} \in \{\kappa_{\sf{ee}}, \kappa_{\sf{ei}}, \kappa_{\sf{ie}}, \kappa_{\sf{ii}} \}$: Branching Parameter

$\arc{\kappa}{n}{}{n'}{}^{\sf max}$: Maximum Physical $\kappa$ Value 

$B$: Bias Parameter

$\mathcal{N}_{\mathsf{n}}$: Neighborhood of Postsynaptic Neuron $\mathsf{n}$  

$\widetilde N_{\mathsf{e}\mathsf{n}}$: Number of Excitatory Neighbors of \\ \hspace*{30pt} Postsynaptic Neuron $\mathsf{n}$  

$\widetilde N_{\mathsf{i}\mathsf{n}}$: Number of Inhibitory Neighbors of \\ \hspace*{30pt} Postsynaptic Neuron $\mathsf{n}$  

$\arc{\tilde{Z}}{e}{\mu'}{}{}$: Sequence of All Excitatory Arc States of \hspace*{30pt} Postsynaptic Neuron $\mathsf{e}$

$\arc{\tilde{Z}}{i}{\nu'}{}{}$: Sequence of All Inhibitory Arc States of \hspace*{30pt} Postsynaptic Neuron $\mathsf{e}$

$U(\tilde{Z}_{\mathsf{e'}})$: Excitatory Site Activation Function

$U(\tilde{Z}_{\mathsf{i'}})$: Inhibitory Site Activation Function

$\tilde{z}_{\sf n'}(t)$: The Set of States of the Neighboring Neurons \hspace*{30pt} of Postsynaptic Neuron $\mathsf{n'}$  

$W^{\sf{n}}_{01}(\tilde{z}_{\mathsf{n}_{0}}(t))$: Transition Probability from Quiescent (0) \hspace*{30pt} to  Active (1) State 

$\overline{W_{01}^{\mathsf{n}}}$ Transition Probability when All Neighbors are \hspace*{30pt} Active 

$A_{0}^{\arc{}{n}{0}{}{}}(t), A_{\mu}^{\arc{}{n}{0}{}{}}(t), B_{\nu}^{\arc{}{n}{0}{}{}}(t)$: Event Notation \\ \hspace*{30pt} for the Dictionary

$P(A_{0}^{\arc{}{n}{0}{}{}}(t)), P(A_{\mu}^{\arc{}{n}{0}{}{}}(t)), P(B_{\nu}^{\arc{}{n}{0}{}{}}(t))$: Probability of the \\ \hspace*{30pt} Events for the Dictionary

$E_m^{\sf n}$: Probability of at least $m$ Excitatory Arcs, targeted at $\mathsf{n}$,  are Simultaneously Active

$I_{[m]}^{\sf n}$: Probability of exactly $m$ Inhibitory Arcs, targeted at $\mathsf{n}$, are Simultaneously Active

$\rho^{\sf{e}}_{1}(t)$: Excitatory Density of Activation

$\rho^{\sf{i}}_{1}(t)$: Inhibitory Density of Activation

$\Bar{\rho}^{\sf{n}}_{1}$: Time Average Densities

$N_{T}$: Number of Iterations of the Map

$\arc{\chi}{n}{}{}{}$: Zero Field Dynamical Susceptibility

$\mathsf{e}_{0}$: Representative Excitatory Neuron

$\mathsf{i}_{0}$: Representative Inhibitory Neuron

$z_{\mathsf{n}_{0}}(t)$: State of Representative Neuron

$\tilde{z}_{\mathsf{n}_{0}}(t)$: States of Neighborhood Neurons of \\ \hspace*{30pt} Representative Neurons $z_{\mathsf{n}_{0}}(t)$

$x_{m}(t)$: Probability for an Excitatory Neuron to be in \hspace*{30pt} State $m$

$y_{n}(t)$: Probability for an Inhibitory Neuron to be in \hspace*{30pt} State $n$

$J_{ij}$: Jacobian Matrix

$x_1^*$: Fraction of Active Excitatory Neurons' Fixed Point

$y_1^*$: Fraction of Active Inhibitory Neurons' Fixed Point

$\arc{\kappa}{e}{}{e,c}{}$: Critical $\arc{\kappa}{e}{}{e}{}$ Value

$\beta$: Critical exponent of Activation Density and $\arc{\kappa}{e}{}{e}{}$

$\gamma$: Critical exponent of Susceptibility and $\arc{\kappa}{e}{}{e}{}$

$\lambda (\Vec{x}_0)$: largest Lyapunov exponent (LLE)

$\mathit{D}_L$: Lyapunov Dimension

\section{COMPLETE DICTIONARY}
\label{App:Dictionary}
We have previously established the MF dictionary for the case of arcs, but not for hyperarcs, which we now present here. We begin by explicitly defining the arc-activation probability function for each arc and hyperarc as
\begin{align}
	P(Z_{\mathsf{nn'}}|z_{\mathsf{n}}(t))&=\delta_{1,\arc{z}{n}{0}{}{}(t)}((1-P_{\mathsf{nn'}})\delta_{0,Z_{\mathsf{nn'}}}+ \nonumber  \\ \nonumber &P_{\mathsf{nn'}}\delta_{1,Z_{\mathsf{nn'}}})+ (1-\delta_{1,z_{\mathsf{n}_{0}}(t)})(1-Z_{\mathsf{nn'}}(t)),  \nonumber\\
    P(\arcin{Z}{e}{}{e'}{}{}|z_{\mathsf{e}}(t))&=\delta_{1,\arc{z}{i}{0}{}{}(t)}((1-\arcin{P}{e}{}{e'}{}{})\delta_{0,\arcin{Z}{e}{}{e'}{}{}}+   \nonumber \\&\arcin{P}{e}{}{e'}{}{}\delta_{1,\arcin{Z}{e}{}{e'}{}{}})+ (1-\delta_{1,z_{\mathsf{i}_{0}}(t)})(1-\arcin{Z}{e}{}{e'}{}{}(t)), \nonumber \\ 
    P(\arcin{Z}{e}{}{}{}{}|z_{\mathsf{e}}(t))&=\delta_{1,\arc{z}{i}{0}{}{}(t)}((1-\arcin{P}{e}{}{}{}{})\delta_{0,\arcin{Z}{e}{}{}{}{}}+   \nonumber \\ &\arcin{P}{e}{}{}{}{}\delta_{1,\arcin{Z}{e}{}{}{}{}})+ (1-\delta_{1,z_{\mathsf{i}_{0}}(t)})(1-\arcin{Z}{e}{}{}{}{}(t)).
    \label{eq:ArcProbabilityFxnActiveOnlyFull} 
\end{align} 

Starting from the MF equations, and substituting the transition probabilities $W_{01}^{\mathsf{e_0}}(\tilde{z}_{\mathsf{e}_{0}}(t))$ and $W_{01}^{\mathsf{i_0}}(\tilde{z}_{\mathsf{i}_{0}}(t))$ using Eq.~\eqref{eq:transition_probability}, one can show that 
\begin{align}
    F &=  \sum_{\forall\tilde{Z}_{\mathsf{e}_{0}}} U(\tilde{Z}_{\mathsf{e_0}})\cdot p_{s\mathsf{e}_{\mu}} \cdot \hspace{-.5em} \sum_{\arc{z}{e}{1}{}{}(t)=0}^{\tau _{\mathsf{e}}
    }\hspace{-.5em}P(\arc{Z}{e}{1}{e}{0}\mid \arc{z}{e}{1}{}{}(t))\cdot x_{\arc{z}{e}{1}{}{}(t)} \cdots \nonumber \\&  \sum_{\arc{z}{i}{1}{}{}(t)=0}^{\tau _{\mathsf{i}}
    }P(\arcin{Z}{e}{1}{e}{0}{1}\mid \arc{z}{i}{1}{}{}(t))\cdot y_{\arc{z}{i}{1}{}{}(t)}\cdots \nonumber \\& \sum_{\arc{z}{i}{1}{}{}(t)=0}^{\tau _{\mathsf{i}}
    }P(\arcin{Z}{}{}{e}{0}{1}\mid \arc{z}{i}{1}{}{}(t))\cdot y_{\arc{z}{i}{1}{}{}(t)} \cdots \nonumber \\& \sum_{\arc{z}{i}{1}{}{}(t)=0}^{\tau _{\mathsf{i}}
    }P(\arc{Z}{i}{1}{e}{0}\mid \arc{z}{i}{1}{}{}(t))\cdot y_{\arc{z}{i}{1}{}{}(t)}\cdots,\label{eq:MFT_F_arranged_Full}
\end{align}
\begin{align}
    G&=  \sum_{\forall\tilde{Z}_{\mathsf{i}_{0}}} U(\tilde{Z}_{\mathsf{i_0}})\cdot p_{s\mathsf{i}_{\nu}}\cdot \hspace{-.5em}\sum_{\arc{z}{e}{1}{}{}(t)=0}^{\tau _{\mathsf{e}}
    }\hspace{-.5em}P(\arc{Z}{e}{1}{i}{0}\mid \arc{z}{e}{1}{}{}(t))\cdot x_{\arc{z}{e}{1}{}{}(t)} \cdots \nonumber \\&\sum_{\arc{z}{i}{1}{}{}(t)=0}^{\tau _{\mathsf{i}}
    }P(\arc{Z}{i}{1}{i}{0}\mid \arc{z}{i}{1}{}{}(t))\cdot y_{\arc{z}{i}{1}{}{}(t)} \cdots .\label{eq:MFT_G_arranged_Full}
\end{align}

With some algebra, and using the arc activation functions in Eq.~\eqref{eq:ArcProbabilityFxnActiveOnlyFull}, the functions $F$ and $G$ can be written as follows
\begin{align}
    F &=  \sum_{\forall\tilde{Z}_{\mathsf{e}_{0}}} U(\tilde{Z}_{\mathsf{e_0}})\cdot p_{s\mathsf{e}_{\mu}} \cdot \prod _{\mu=1}^{\numberneighbors{e}{e}}\big ( (1-\arc{P}{e}{\mu}{e}{0}x_{1}(t))\delta_{0,\arc{Z}{e}{\mu}{e}{0}}+\nonumber \\&\arc{P}{e}{\mu}{e}{0}x_{1}(t)\delta_{1,\arc{Z}{e}{\mu}{e}{0}}\big ) \cdot \hspace{-1em}\prod _{\forall \mathsf{ e_{\mu}}, \mathsf{ i_{\nu}} \in\numberneighbors{i}{e}}\hspace{-1em}\big ( (1-\arcin{P}{e}{\mu}{e}{0}{\nu}y_{1}(t))\delta_{0,\arcin{Z}{e}{\mu}{e}{0}{\nu}}+\nonumber \\&\arc{P}{i}{\nu}{e}{0}y_{1}(t)\delta_{1,\arc{Z}{i}{\nu}{e}{0}}\big ) \cdot \prod _{\forall \mathsf{ i_{\nu}} \in\numberneighbors{i}{e}}\big ( (1-\arcin{P}{}{}{e}{0}{\nu}y_{1}(t))\delta_{0,\arcin{Z}{}{}{e}{0}{\nu}}+\nonumber \\&\arc{P}{i}{\nu}{e}{0}y_{1}(t)\delta_{1,\arc{Z}{i}{\nu}{e}{0}}\big ) \cdot \prod _{\nu=1}^{\numberneighbors{i}{e}}\big ( (1-\arc{P}{i}{\nu}{e}{0}y_{1}(t))\delta_{0,\arc{Z}{i}{\nu}{e}{0}}+\nonumber \\&\arc{P}{i}{\nu}{e}{0}y_{1}(t)\delta_{1,\arc{Z}{i}{\nu}{e}{0}}\big ),\label{eq:MFT_F_explicit_Full}
\end{align}
\begin{align}
    G&=  \sum_{\forall\tilde{Z}_{\mathsf{i}_{0}}} U(\tilde{Z}_{\mathsf{i_0}})\cdot p_{s\mathsf{i}_{\nu}} \cdot  \prod _{\mu=1}^{\numberneighbors{e}{i}}\big ( (1-\arc{P}{e}{\mu}{i}{0}x_{1}(t))\delta_{0,\arc{Z}{e}{\mu}{i}{0}}+\nonumber \\&\arc{P}{e}{\mu}{i}{0}x_{1}(t)\delta_{1,\arc{Z}{e}{\mu}{i}{0}}\big ) \cdot \prod _{\nu=1}^{\numberneighbors{i}{i}}\big ( (1-\arc{P}{i}{\nu}{i}{0}y_{1}(t))\delta_{0,\arc{Z}{i}{\nu}{i}{0}}+\nonumber \\&\arc{P}{i}{\nu}{i}{0}y_{1}(t)\delta_{1,\arc{Z}{i}{\nu}{i}{0}}\big ) ,\label{eq:MFT_G_explicit_Full}
\end{align}

Starting with Eq. \eqref{eq:transition_probability} can write the cases when all the neighbors are active $\overline{W_{01}^{\mathsf{e_0}}}$ and $\overline{W_{01}^{\mathsf{i_0}}}$ as
\begin{align}
    \overline{W_{01}^{\mathsf{e_0}}} &=  \sum_{\forall\tilde{Z}_{\mathsf{e}_{0}}} U(\tilde{Z}_{\mathsf{e_0}})\cdot p_{s\mathsf{e}_{\mu}} \cdot \prod _{\mu=1}^{\numberneighbors{e}{e}}\big ( (1-\arc{P}{e}{\mu}{e}{0})\delta_{0,\arc{Z}{e}{\mu}{e}{0}}+\nonumber \\&\arc{P}{e}{\mu}{e}{0}\delta_{1,\arc{Z}{e}{\mu}{e}{0}}\big ) \cdot \hspace{-1em}\prod _{\forall \mathsf{ e_{\mu}}, \mathsf{ i_{\nu}} \in\numberneighbors{i}{e}}\hspace{-1em}\big ( (1-\arcin{P}{e}{\mu}{e}{0}{\nu})\delta_{0,\arcin{Z}{e}{\mu}{e}{0}{\nu}}+\nonumber \\&\arc{P}{i}{\nu}{e}{0}\delta_{1,\arc{Z}{i}{\nu}{e}{0}}\big ) \cdot \prod _{\forall \mathsf{ i_{\nu}} \in\numberneighbors{i}{e}}\big ( (1-\arcin{P}{}{}{e}{0}{\nu})\delta_{0,\arcin{Z}{}{}{e}{0}{\nu}}+\nonumber \\&\arc{P}{i}{\nu}{e}{0}\delta_{1,\arc{Z}{i}{\nu}{e}{0}}\big ) \cdot \prod _{\nu=1}^{\numberneighbors{i}{e}}\big ( (1-\arc{P}{i}{\nu}{e}{0})\delta_{0,\arc{Z}{i}{\nu}{e}{0}}+\nonumber \\&\arc{P}{i}{\nu}{e}{0}\delta_{1,\arc{Z}{i}{\nu}{e}{0}}\big ) ,\label{eq:MFT_E_transition_Full}
\end{align}
\begin{align}
    \overline{W_{01}^{\mathsf{i_0}}}&=  \sum_{\forall\tilde{Z}_{\mathsf{i}_{0}}} U(\tilde{Z}_{\mathsf{i_0}})\cdot p_{s\mathsf{i}_{\nu}} \cdot  \prod _{\mu=1}^{\numberneighbors{e}{i}}\big ( (1-\arc{P}{e}{\mu}{i}{0})\delta_{0,\arc{Z}{e}{\mu}{i}{0}}+\nonumber \\&\arc{P}{e}{\mu}{i}{0}\delta_{1,\arc{Z}{e}{\mu}{i}{0}}\big ) \cdot \prod _{\nu=1}^{\numberneighbors{i}{i}}\big ( (1-\arc{P}{i}{\nu}{i}{0})\delta_{0,\arc{Z}{i}{\nu}{i}{0}}+\nonumber \\&\arc{P}{i}{\nu}{i}{0}\delta_{1,\arc{Z}{i}{\nu}{i}{0}}\big ) ,\label{eq:MFT_I_transition_Full}
\end{align}
when the substitution rules in Eq.~\eqref{eq:substitutionrule} are applied to $\overline{W_{01}^{\mathsf{e_0}}}$ and $\overline{W_{01}^{\mathsf{i_0}}}$, they yield expressions that are equivalent to $F$ and $G$, respectively.

\section{TRANSITION PROBABILITIES $\overline{W_{01}^{\mathsf{n_0}}}$ FOR THE MOTIF IN FIG.~\ref{fig:simplemotif}}
\label{App:W11equations}

In Section~\ref{section4}, the solution of the transition probability $\overline{W_{01}^{\mathsf{e_0}}}$ for the motif in Fig.~\ref{fig:simplemotif} was already given, and now we will provide further details. This will be followed by showing the solutions for the other transition probabilities, starting with $\overline{W_{01}^{\mathsf{e_0}}}$.

First, $\overline{W_{01}^{\mathsf{e_{0}}}}$ is calculated from Eq. \eqref{eq:Highest_W}
\begin{align*}
	\overline{W_{01}^{\mathsf{e_{0}}}} & =E_{1}^{\sf e_0}I_{[0]}^{\sf e_0}+E_{2}^{\sf e_0}I_{[1]}^{\sf e_0} ,
\end{align*}
where $E_{1}^{\sf e_0} =S_{1}^{\mathsf{ee_0}}-S_{2}^{\mathsf{ee_0}}$ and 
$E_{2}^{\sf e_0} =S_{2}^{\mathsf{ee_0}}$ are derived from Eq. \eqref{eq:E_m_term}. Equation \eqref{eq:S_e_term} provides 
expressions for $S_{1}^{\mathsf{ee_0}} =P\left(A_{0}^{\mathsf{e_0}}\right)+P\left(A_{1}^{\mathsf{e_0}}\right)
=p_{s}+P_{\mathsf{e_{1}e_{0}}}$ and $S_{2}^{\mathsf{ee_0}}  =P\left(A_{0}^{\mathsf{e_0}}\cap A_{1}^{\mathsf{e_0}}\right)      =p_{s}P_{\mathsf{e_{1}e_{0}}}$. Similarly, the terms $I_{[0]}^{\sf e_0} =S_{0}^{\mathsf{ie_0}}-S_{1}^{\mathsf{ie_0}}$ and $I_{[1]}^{\sf e_0} =S_{1}^{\mathsf{ie_0}}$, with $S_{0}^{\mathsf{ie_0}}=1$ and $S_{1}^{\mathsf{ie_0}} =P\left(B_{1}^{\mathsf{e_0}}\right) =P_{\mathsf{i_{1}e_{0}}}$, are obtained from Eqs. \eqref{eq:I_m_term} and \eqref{eq:S_i_term}. Putting all the pieces together leads to Eq. \eqref{eq:W01eHighest_Motif1}.

One can follow a similar procedure to compute $\overline{W_{01}^{\mathsf{i_{0}}}}$ 
\begin{align*}
	\overline{W_{01}^{\mathsf{i_{0}}}} & =E_{1}^{\sf i_0}
	I_{[0]}^{\sf i_0} ,\end{align*}
where $E_{1}^{\sf i_0} =S_{1}^{\mathsf{ei_0}}=P\left(A_{0}^{\mathsf{i_0}}\right)=p_{s}$ and $I_{[0]}^{\sf i_0}=S_{0}^{\mathsf{ii_0}}=1$, with the result $\overline{W_{01}^{\mathsf{i_{0}}}}=p_s$.  

\section{DYNAMICAL EQUATIONS OF THE MOTIF IN FIG.~\ref{fig:simplemotif}: AN EXAMPLE PROOF OF THE SUBSTITUTION RULE}
\label{App:motif1x1y1}

To demonstrate the validity of the substitution rule for the example motif, it is necessary to compute the remaining transition probabilities $W_{01}^{\sf{e_0}} \ne \overline{W_{01}^{\mathsf{e_{0}}}}$. These can be obtained from $\overline{W_{01}^{\mathsf{e_{0}}}}$ by setting to zero the transmission–probability variables corresponding to inactive neighbors.


From Eq. \eqref{eq:W01eHighest_Motif1}
\begin{align}
    \overline{W_{01}^{\mathsf{e_{0}}}}=& \, (p_{s}+(1-p_s)\arc{P}{e}{1}{e}{0})(1-\arc{P}{i}{1}{e}{0})+ p_{s} \arc{P}{e}{1}{e}{0}  \arc{P}{i}{1}{e}{0},\nonumber
\end{align}
one obtains
\begin{eqnarray}
    W_{01}^{\sf{e_0}}((1;\cancel{1}))&=& p_{s}+(1-p_s)\arc{P}{e}{1}{e}{0} ,\nonumber \\
    W_{01}^{\sf{e_0}}((\cancel{1};1))&=&p_{s}(1-\arc{P}{i}{1}{e}{0}),\nonumber \\
    W_{01}^{\sf{e_0}}((\cancel{1};\cancel{1}))&=&p_{s}.\nonumber
\end{eqnarray}
We can now explicitly write $x_1(t+1)$ using Eqs.~\eqref{eq:MFT_x1} and  \eqref{eq:x0}, together with the transition probabilities above, and after some algebraic manipulations obtain
\begin{eqnarray}
    \frac{x_{1}(t+1)}{x_{0}(t)} &=& \ W_{01}^{\sf{e_0}}((1;1)) \ {x_1(t)} {y_1(t)} \nonumber \\  && +W_{01}^{\sf{e_0}}((1;\cancel{1})) \  x_1(t) (1-y_1(t))\nonumber \\  && +W_{01}^{\sf{e_0}}((\cancel{1};1)) \ (1-x_1(t)) y_1(t)\nonumber \\ &&+W_{01}^{\sf{e_0}}((\cancel{1};\cancel{1})) \ (1-x_1(t)) (1-y_1(t))\nonumber \\
    &=& (p_{s}+(1-p_{s})\arc{P}{e}{1}{e}{0} x_1(t)) (1-\arc{P}{i}{1}{e}{0} y_1(t)) \nonumber \\&& \ \ + p_{s}\arc{P}{e}{1}{e}{0} x_1(t)\arc{P}{i}{1}{e}{0} y_1(t). 
\end{eqnarray}
\normalsize

Similarly, for $W_{01}^{\sf{i_0}}$ the only neighbor is $p_{s}$ so
\begin{equation}
    W_{01}^{\sf{i_0}}=p_{s} .
\end{equation}

We can now explicitly write $y_1(t+1)$ using Eq. \eqref{eq:MFT_y1}
\begin{eqnarray}
    y_{1}(t+1)=y_{0}(t) \ p_{s} .
\end{eqnarray}
We see that these dynamical equations match exactly those obtained via the substitution rule for the non-trivial equations in Eqs.~\eqref{eq:MFmaps_Motif1}, thereby providing an explicit example of the validity of the substitution rule.

\section{DYNAMICAL EQUATIONS OF THE MOTIF IN FIG.~\ref{fig:motif2}}
\label{App:motif2x1y1}

The dynamic maps expressed as a polynomial sum for the motif in Fig.~\ref{fig:motif2} were given in Section~\ref{section4}, and here we provide the corresponding coefficients.
The substitution rule used to compute these dynamical maps was verified to hold, but the full derivation is omitted here due to its length. The final expression for excitation is
\begin{align}
 x_{1}(t+1)=x_{0}(t)\sum_{c=0,d=0}^{\numberneighbors{e}{e},\numberneighbors{i}{e}} a_{cd} \, x_{1}(t)^{c} y_{1}(t)^{d} ,
\end{align}
where
\begin{align}
    &a_{00}=p_{s} \nonumber\\&
    a_{01}=-p_s(\arc{P}{i}{1}{e}{0}+\arc{P}{i}{2}{e}{0})\nonumber\\&
    a_{02}=p_{s}\arc{P}{i}{1}{e}{0}\arc{P}{i}{2}{e}{0}\nonumber\\&
    a_{10}=(1-p_{s})(\arc{P}{e}{1}{e}{0}+\arc{P}{e}{2}{e}{0})\nonumber\\&
    a_{11}=(-1+2p_{s})(\arc{P}{e}{1}{e}{0}+\arc{P}{e}{2}{e}{0})(\arc{P}{i}{1}{e}{0}+\arc{P}{i}{2}{e}{0})
    \nonumber\\&
    a_{12}=(1-3 p_{s})(\arc{P}{e}{1}{e}{0}+\arc{P}{e}{2}{e}{0})\arc{P}{i}{1}{e}{0}\arc{P}{i}{2}{e}{0}\nonumber\\&
    a_{20}=(p_{s}-1)\arc{P}{e}{1}{e}{0}\arc{P}{e}{2}{e}{0}\nonumber\\&
    a_{21}=(2-3p_{s})\arc{P}{e}{1}{e}{0}\arc{P}{e}{2}{e}{0}(\arc{P}{i}{1}{e}{0}+\arc{P}{i}{2}{e}{0})\nonumber\\&
    a_{22}=(2 p_{s}-1)3\arc{P}{e}{1}{e}{0}\arc{P}{e}{2}{e}{0}\arc{P}{i}{1}{e}{0}\arc{P}{i}{2}{e}{0} , \label{eq:term_acd_Motif2}
\end{align}
and, for inhibition, 
\begin{align}
 y_{1}(t+1)=y_{0}(t)\sum_{c=0,d=0}^{\numberneighbors{e}{i},\numberneighbors{i}{i}} b_{cd} \, x_{1}(t)^{c} y_{1}(t)^{d} ,
\end{align}
where
\begin{align}
    &b_{00}=p_{s} \nonumber\\&
    b_{01}=-p_s (\arc{P}{i}{1}{i}{0}+\arc{P}{i}{2}{i}{0})\nonumber\\&
    b_{02}=p_{s}\arc{P}{i}{1}{i}{0}\arc{P}{i}{2}{i}{0}\nonumber\\&
    b_{10}=(1-p_{s})(\arc{P}{e}{1}{i}{0}+\arc{P}{e}{2}{i}{0})\nonumber\\&
    b_{11}=(-1+2p_{s})(\arc{P}{e}{1}{i}{0}+\arc{P}{e}{2}{i}{0})(\arc{P}{i}{1}{i}{0}+\arc{P}{i}{2}{i}{0})
    \nonumber\\&
    b_{12}=(1-3 p_{s})(\arc{P}{e}{1}{i}{0}+\arc{P}{e}{2}{i}{0})\arc{P}{i}{1}{i}{0}\arc{P}{i}{2}{i}{0}\nonumber\\&
    b_{20}=(p_{s}-1)\arc{P}{e}{1}{i}{0}\arc{P}{e}{2}{i}{0}\nonumber\\&
    b_{21}=(2-3p_{s})\arc{P}{e}{1}{i}{0}\arc{P}{e}{2}{i}{0}(\arc{P}{i}{1}{i}{0}+\arc{P}{i}{2}{i}{0})\nonumber\\&
    b_{22}=(2 p_{s}-1)3\arc{P}{e}{1}{i}{0}\arc{P}{e}{2}{i}{0}\arc{P}{i}{1}{i}{0}\arc{P}{i}{2}{i}{0} . \label{eq:term_bcd_Motif2}
\end{align}

\section{VALIDATION OF MF MODELS AGAINST MANY-BODY SIMULATIONS}
\label{App:ValidateMF}

\begin{figure}
    \includegraphics[width=8.5cm]{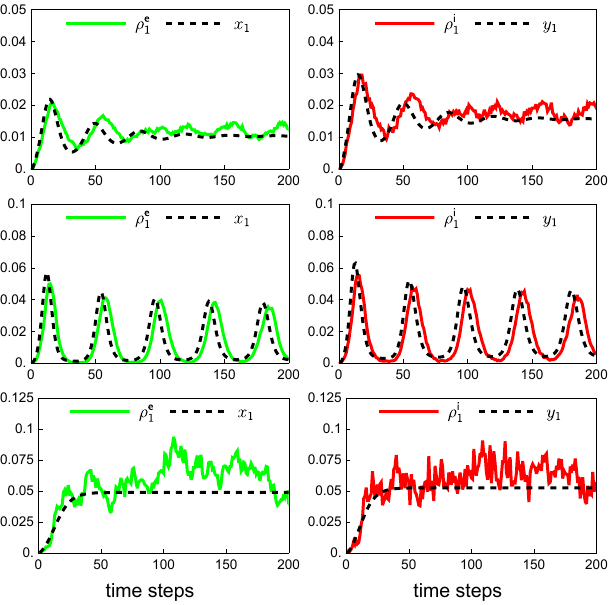}

	\caption{ Time series obtained from the MF and many-body simulations (with system size $N_{\mathsf{e}}=N_{\mathsf{i}}=2048$), averaged over an ensemble of networks, are shown as dashed and solid lines, respectively. The MF approximation qualitatively captures the dynamics observed in the many-body simulations within certain parameter regimes. The top row, corresponding to $\tau_{\mathsf{e}}=20$ and $\kappa_{\mathsf{ee}}=1.2$, shows the stable fixed-point regime, in which both the MF and many-body models exhibit an overshoot before relaxing to a fixed-point activity on the same time scale. However, the amplitudes in the many-body simulations are larger and have been rescaled to match the MF. The middle row, corresponding to $\tau_{\mathsf{e}}=30$ and $\kappa_{\mathsf{ee}}=1.5$,  shows the  quasiperiodic regime, where the simulation amplitudes have likewise been rescaled. Finally,   the bottom row, correspoinding to $\tau_{\mathsf{e}}=1$ and $\kappa_{\mathsf{ee}}=1.14$,  shows time series from a single network to illustrate fluctuations, highlighting departures from MF behavior and emphasizing the need for a many-body description. The rest of the simulation parameters are identical to those of the motif described in Fig.~\ref{fig:motif2}. 
}
     \label{fig:Comparison_MFT_MBS}

\end{figure}

To validate the MF approximations, we show that they qualitatively capture the dynamics of the many-body system in both the nonzero stable fixed-point regime and the quasiperiodic oscillatory phase (see Fig.~\ref{fig:Comparison_MFT_MBS}). 
The MF probabilities for a node to be active, $x_{1}(t)$ and $y_{1}(t)$, are compared with the simulated densities of active excitatory and inhibitory neurons, $\rho_{1}^{\mathsf{e}}(t)$ and $\rho_{1}^{\mathsf{i}}(t)$.
Simulations are performed on randomly connected networks with a fixed number of neighbors, and the corresponding MF approximations use identical parameter values. We use the motif shown in Fig.~\ref{fig:motif2}, in which the numbers of inhibitory and excitatory neurons are equal.  
While MF captures the qualitative behavior, it underestimates the amplitudes; therefore, the simulated densities are rescaled to compare the fluctuation structure.
The MF approximation breaks down in the fluctuation-dominated regime.
This breakdown is illustrated in the bottom row of Fig.~\ref{fig:Comparison_MFT_MBS}, emphasizing the importance of the many-body GCBM for extracting effective critical exponents and dynamical scaling relations.
The correspondence between the fluctuation regimes observed in many-body simulations —which may involve a variety of network topologies— and those predicted by MF approximations deserves further investigation.

\section{COMPUTATION OF THE SPECTRUM OF LYAPUNOV EXPONENTS}
\label{App:lypexp}

We calculate the  largest Lyapunov exponent for the GCBM map $\Vec{x}(t+1)=f(\Vec{x}(t))$
Where we relabel the map as
\begin{equation}
    \Vec{x}_{p+1}=f(\Vec{x}_p)
\end{equation}
Without loss of generality, we will assume $p$ is the period of orbit such that
\begin{equation}
    \Vec{x}_p=\Vec{x}_0\ \text{ with } \\ \Vec{x}_p=f^p(\Vec{x}_0)
\end{equation}
We will perturb in the direction of a tangent vector $\Vec{x}_0$
\begin{equation}
    \Vec{x}_{p+1}=Df(\Vec{x}_p)\cdot\Vec{x}_p
\end{equation}
Where $D f$ is the Jacobian Matrix.

Therefore, we can define
\begin{equation}
    \Vec{x}_p=Df^p(\Vec{x}_0)\cdot\Vec{x}_0 ,
\end{equation}
with
\begin{equation}
\hspace*{-0.2cm}
    Df^p(\Vec{x}_0)=Df^{p-1}(\Vec{x}_{p-1})\cdot Df^{p-2}(\Vec{x}_{p-2})\cdots Df(\Vec{x}_0) .
\end{equation}
We now can define the largest Lyapunov exponent as \cite{LyapunovExponents}
\begin{align}
    \lambda(\Vec{x}_0)=&\lim_{p\rightarrow \infty}\frac{1}{p} \ln(\frac{||\Vec{x}_p||}{||\Vec{x}_0||})\\
    \lambda(\Vec{x}_0)=&\lim_{p\rightarrow \infty}\frac{1}{p} \ln||Df^{p}(\Vec{x}_0)\cdot \frac{\Vec{x}_0}{||\Vec{x}_0||}|| \ .
\end{align}
To compute the Lyapunov spectrum 
$Df^p(\Vec{x}_0)$ one may use the QR decomposition method of Ref. \cite{Eckmann1985},
\begin{equation}
    Df(\Vec{x}_0)=Q_1 R_1 ,
\end{equation}
where $Q_1$ is an orthogonal matrix, and $R_1$ is an upper triangular with non-negative diagonal elements. 
The decomposition is unique if $ Df(\Vec{x}_0)$ is invertible.
We define 
\begin{equation*}
    Df^k=Df^{k-1}(\Vec{x}_{k-1})Q_{k-1} ,
\end{equation*}
which can be decomposed
\begin{equation*}
     Df^k=Q_k R_k ,
\end{equation*}
 allowing us to write 
\begin{equation*}
    Df^p(\Vec{x}_0)=Q_p R_p \cdots R_1 ,
\end{equation*}
since the matrices $Q_k$s are orthogonal.

The diagonal elements $\lambda_{ii}(\Vec{x}_0)^n$ of the upper right triangle of the product of $R_p \cdots R_1$ satisfy \cite{Johnson1987}
\begin{equation}
    \lim_{n\rightarrow \infty} \frac{1}{n} \log\lambda_{ii}(\Vec{x}_0)^n=\lambda_i(\Vec{x}_0) .
\end{equation}

\bibliography{main}

\end{document}